\documentclass[11pt, a4paper]{article}
\pdfoutput=1

\usepackage[height=8.85in,width=6.4in]{geometry}

\usepackage{ifpdf}
\ifpdf      
  \usepackage{graphicx, xcolor, rotating} 
  \usepackage[bookmarksopen,colorlinks=true,linkcolor=light_blue,
citecolor=light_pink,urlcolor=dark_red,linktoc=all]{hyperref}       
\else     
  \usepackage[dvipdfmx]{graphicx, xcolor,rotating}
  \usepackage[dvipdfmx, bookmarksopen,colorlinks=true,linkcolor=light_blue,
citecolor=light_pink,urlcolor=dark_red,linktoc=all]{hyperref}     
 \fi

\usepackage{amsmath}
\usepackage{mathtools}
\usepackage{amsfonts}
\usepackage{amssymb}
\usepackage{slashed}
\usepackage{braket}
\usepackage{bm}
\usepackage{cite}
\usepackage{comment}
\usepackage{xcolor}
\usepackage[utf8]{inputenc}
\usepackage[footnotesize]{caption}
\usepackage[normalem]{ulem}

\usepackage{fourier}

\newcommand{\vev}[1]{ \left\langle {#1} \right\rangle }
\newcommand{\dd}{\mathrm{d}}

\newcommand{\Tr}{\text{Tr\,}}

\newcommand{\GeV}{\,\mathrm{GeV}}

\newcommand{\del}{\partial}

\newcommand{\lmk}{\left(}  
\newcommand{\rmk}{\right)}

\newcommand{\diag}{{\rm diag}}
\def\eq#1{Eq.~(\ref{#1})}

\makeatletter
\@addtoreset{equation}{section}

\makeatother

\makeatletter
\def\maketag@@@#1{\hbox{\m@th\normalfont\normalsize#1}}
\makeatother

\usepackage{multicol}
\definecolor{dark_red}{rgb}{0.7, 0., 0.}
\definecolor{light_pink}{rgb}{1,0.4,0.4}
\definecolor{light_blue}{rgb}{0.284602,0.317763,0.963947}
\definecolor{cred}{RGB}{180,50,40} 
\definecolor{darkgreen}{RGB}{0, 100, 0}
\definecolor{desy_blue}{HTML}{009EE2}
\definecolor{desy_orange}{HTML}{FD8800}
\definecolor{forestgreen}{HTML}{228B22}
\definecolor{ochre}{HTML}{CCAA2B}

\begin{document}

\hypersetup{pageanchor=false}
\begin{titlepage}

\begin{center}

\hfill DESY 20-100\\
\hfill CERN-TH-2020-088\\
\hfill TU-1102\\

\vskip .5in

{
\Huge \bfseries Spontaneous Baryogenesis from Axions \\ with Generic Couplings\\
}
\vskip .7in

{\Large Valerie Domcke$^{\lozenge \, \square}$, Yohei Ema$^{\lozenge}$, Kyohei Mukaida$^\lozenge$, Masaki Yamada$^{\blacklozenge}$}

\vskip .3in
\begin{tabular}{ll}
$^\lozenge$&\!\!\!\! \emph{DESY, Notkestra{\ss}e 85, D-22607 Hamburg, Germany}\\[.3em]
$^\square$&\!\!\!\! \emph{Theoretical Physics Department, CERN, 1 Esplanade des Particules,}\\
& \!\!\!\!\! \emph{CH-1211 Geneva 23, Switzerland}\\[.3em]
$^\square$&\!\!\!\! \emph{Institute of Physics, Laboratory for Particle Physics and Cosmology, EPFL,}\\
&\!\!\!\!\! \emph{CH-1015 Lausanne, Switzerland}\\[.3em]
$^\blacklozenge$&\!\!\!\! \emph{Frontier Research Institute for Interdisciplinary Sciences, Tohoku University,}\\
& \!\!\!\!\! \emph{Sendai, Miyagi 980-8578, Japan}\\[.3em]
$^\blacklozenge$&\!\!\!\! \emph{Department of Physics, Tohoku University, Sendai, Miyagi 980-8578, Japan}
\end{tabular}

\end{center}
\vskip .5in

\begin{abstract}
\noindent Axion-like particles can source the baryon asymmetry of our Universe through spontaneous baryogenesis. Here we clarify that this is a generic outcome for essentially any coupling of an axion-like particle to the Standard Model, requiring only a non-zero velocity of the classical axion field while baryon or lepton number violating interactions are present in thermal bath. In particular, coupling the axions only to gluons is sufficient to generate a baryon asymmetry in the presence of electroweak sphalerons or the Weinberg operator. Deriving the transport equation for an arbitrary set of couplings of the axion-like particle, we provide a general framework in which these results can be obtained immediately. If all the operators involved are efficient, it suffices to solve an algebraic equation to obtain the final asymmetries. Otherwise one needs to solve a simple set of differential equations. This formalism clarifies some theoretical subtleties such as redundancies in the axion coupling to the Standard Model particles associated with a field rotation. We demonstrate how our formalism automatically evades potential pitfalls in the calculation of the final baryon asymmetry.

\newpage

\end{abstract}

\end{titlepage}

\tableofcontents
\thispagestyle{empty}
\renewcommand{\thepage}{\arabic{page}}
\renewcommand{\thefootnote}{$\natural$\arabic{footnote}}
\setcounter{footnote}{0}
\newpage
\hypersetup{pageanchor=true}

\section{Introduction}
\label{sec:intro}

The axion was introduced 
to solve the strong $CP$ problem~\cite{Peccei:1977hh,Peccei:1977ur,Weinberg:1977ma,Wilczek:1977pj}, but has since matured into a broader concept addressing many open questions in particle physics and cosmology. These axion-like particles\footnote{In the following, we will for brevity refer to all axion-like particles simply as `axions', using the term `QCD axion' to refer to the axion addressing the strong $CP$ problem.} are pseudoscalars which couple to the Standard Model (SM) gauge fields and fermions via (classically) shift-symmetric couplings mediated by dimension five operators. For example, in the context of cosmic inflation, this shift symmetry ensures a sufficiently flat direction in field space suitable to drive the exponential expansion of the very early Universe~\cite{Freese:1990rb}. In the context of dark matter, these small interactions with the SM ensure that an axion dark matter candidate is sufficiently long lived, while simultaneously providing an avenue for detection~\cite{PRESKILL1983127,ABBOTT1983133,DINE1983137}. In the context of string theory, axions are ubiquitous and typically arise as a result of the compactification~\cite{Banks:2003sx,Svrcek:2006yi,Ibanez:1986xy}.

Beyond all this, axions provide all the ingredients necessary to generate the matter antimatter asymmetry of our Universe via spontaneous baryogenesis: a non-vanishing velocity of a classical axion field spontaneously breaks $CPT$, which, in the presence of baryon number violating interactions, can generate a baryon asymmetry~\cite{Cohen:1987vi,Cohen:1988kt}.
This idea has been pursued \textit{e.g.}, in Refs.~\cite{Chiba:2003vp,Takahashi:2003db,Kusenko:2014uta,Ibe:2015nfa,Takahashi:2015waa,Jeong:2018jqe,Bae:2018mlv,Co:2019wyp}. There are two main points which differ among these works. Firstly, the motion of the axion may happen at any time between cosmic inflation or the electroweak phase transition, with correspondingly different physical processes responsible for triggering this motion. Secondly, different studies chose different couplings of the axion to the SM particle content, \textit{i.e.}, different linear combinations of the possible shift-symmetric operators.
   
\begin{figure}[t]
	\centering
 	\includegraphics[width=0.65\linewidth]{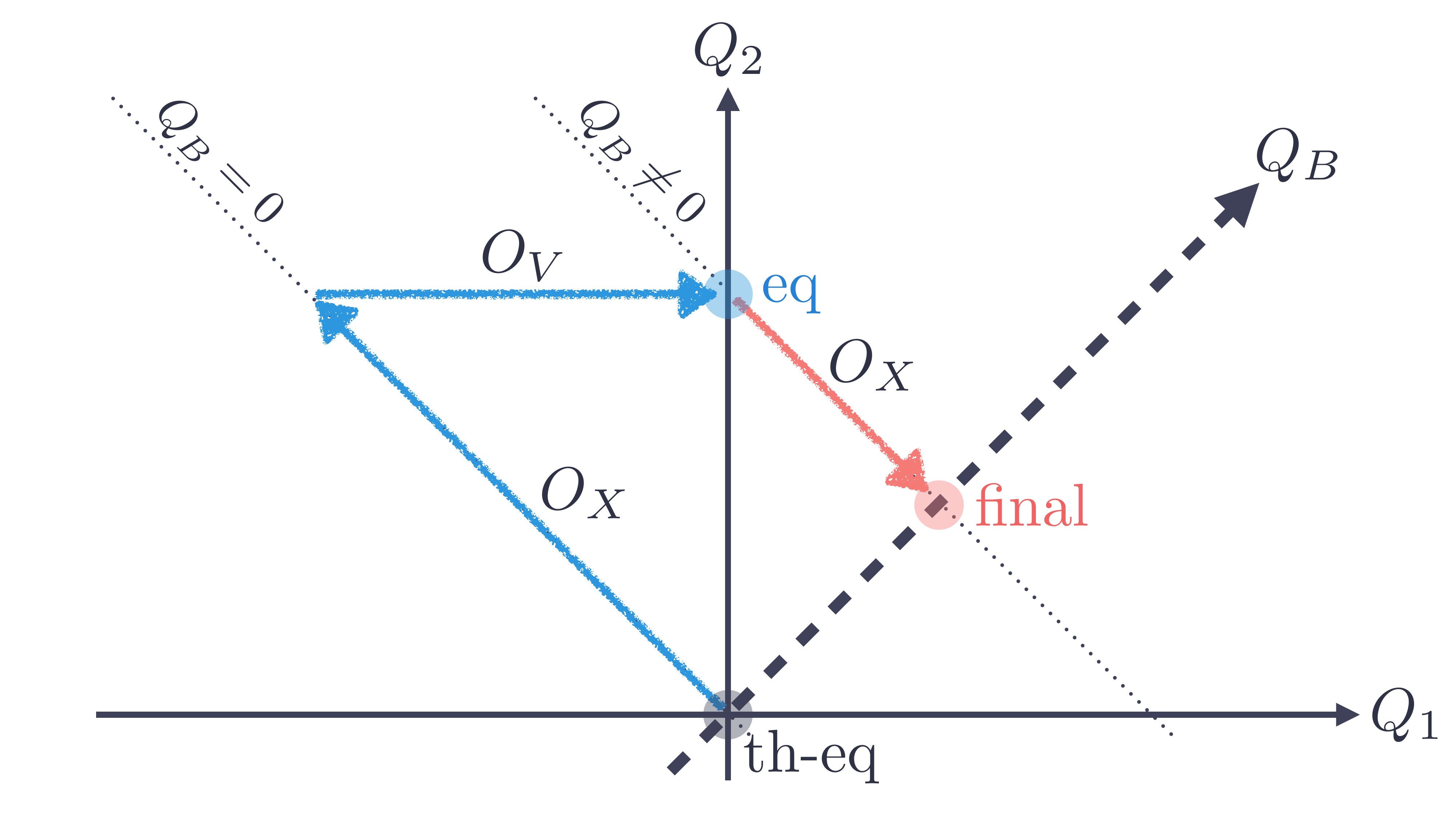}
	\caption{ 
	A schematic figure of our baryogenesis mechanism. Since the SM involves many particle species and interactions,
	here we consider a toy model composed of two species and two interactions as an illustration.
	Its current equations are $\partial_\mu J_1^\mu = - O_X + O_V$ and $\partial_\mu J_2^\mu = O_X$, and the degrees of freedom for $Q_1$ and $Q_2$ are assumed to be the same.
	We would like to generate $Q_B = Q_1 + Q_2$, which is violated by $O_V$ as $\partial_\mu J^\mu_B = O_V$.
	Conventional spontaneous baryogenesis introduces a direct coupling of a scalar field $a$ to the $Q_B$-violating operator as $a O_V$ or equivalently $a \partial_\mu J_B^\mu$.
	However, a coupling of $a O_X$ (or equivalently $a \partial_\mu J_2^\mu$) which is not directly related to the $J_B$-current is enough for baryogenesis.
	A non-vanishing velocity ($\dot a \neq 0$) biases $Q_1$ and $Q_2$ through $a O_X$ while $O_V$ tries to wash out $Q_1$. As a result, the new equilibrium solution ({\color[rgb]{0.147398,0.511420,0.836949}eq}) is different from the thermal equilibrium ({\color[rgb]{0.185048,0.192302,0.272102}th-eq}) and has a non-vanishing $Q_B$.
	After $O_V$ freezes-out, $Q_B$ becomes conserved, and we end up with $Q_B \neq 0$ ({\color[rgb]{0.942597,0.388320,0.387950}final}) when $\dot a = 0$. 
	The only way to \textit{avoid} generating $Q_B \neq 0$ is to couple the axion as $a(c_XO_X + c_V O_V)$ with $c_X+2c_V = 0$.
	Then the axion velocity only biases $Q_1 - Q_2$ violated by $-2 O_X + O_V$, which is orthogonal to $Q_B$.
	}
	\label{fig:schematic}
\end{figure}

In this paper we provide a simple formalism to study this class of models in a more systematic way. Starting from an axion $a$ coupling to an arbitrary combination of classically shift-symmetric operators (with coefficients encoded in the source charge vector $n_S$) we compute the final baryon asymmetry taking into account all the SM equilibration processes. 
A non-vanishing velocity of the axion biases the SM processes by acting as an effective chemical potential, thus modifying the equilibrium state of the system. As long as the baryon violating processes are involved in attaining this new equilibrium, the baryon asymmetry becomes non-zero and
its final value is conserved when the baryon violating processes freeze-out (see Fig.~\ref{fig:schematic} as an illustration).
Therefore this mechanism generically leads to a generation of a baryon asymmetry even if there is no direct coupling between the axion and any baryon or lepton number violating operator.\footnote{
A similar idea was discussed in Refs.~\cite{Chiba:2003vp, Takahashi:2003db}, 
where a baryon (and/or) lepton asymmetry is generated from a scalar field that is not coupled to the baryon nor lepton current. 
They consider the case of
an operator $O_V$ that violates both a global Peccei-Quinn symmetry U(1)$_{\rm PQ}$
and the baryon (and/or lepton) symmetry U(1)$_B$,
\textit{i.e.}, $\partial_\mu J_\text{PQ}^\mu = \Delta_\text{PQ} O_V$ and $\partial_\mu J_{B}^\mu = \Delta_{B} O_V$
with $\Delta_{\rm PQ}$ and $\Delta_B$ characterizing the amounts of violation of each symmetry.
An axion coupling of $a \partial_\mu J_\text{PQ}^\mu$ can generate the baryon (and/or lepton) asymmetry because one can rewrite $a \partial_\mu J_\text{PQ}^\mu$ as $a \partial_\mu J_B^\mu$ by performing a field rotation associated with $Q_\text{PQ} - (\Delta_\text{PQ}/\Delta_B) Q_B$.
In this paper, we will show that adding such an operator is not necessary for baryogenesis if we introduce an additional ingredient. See Fig.~\ref{fig:schematic}.
There we illustrate our idea with a toy model: $\partial_\mu J_B^\mu = \partial_\mu (J_1^\mu + J_2^\mu) = O_V$ and $\partial_\mu J_2^\mu = O_X$. As explained in the caption, a derivative coupling of $a \partial_\mu J_2^\mu$ can generate $Q_B$ although $J_2$ is not broken by $O_V$.
By applying this mechanism to a more realistic case, we show, for instance, the SU$(3)$ Chern-Simons coupling $a G \tilde G$ can source the baryon asymmetry, although it has nothing to do with baryon number violation.
}

We formulate this process by an algebraic matrix equation with the entries of the matrix encoding the various SM processes and the source vector $n_S$ corresponding to the axion coupling.
The only condition for baryogenesis is that the source charge vector $n_S$ should not be fully orthogonal to the charge vector of the baryon number violating process, \textit{i.e.}, a baryon number violating process needs to be involved either directly in the axion coupling or in the subsequent equilibration of the asymmetry. 
Our formalism correctly accounts for two important technical points: i) the transport equation, which describes the equilibration process, is independent of the choice of field basis related by field redefinitions and ii) the charge vectors of the involved processes are a priori not linearly independent.
In particular,  point i) implies that performing a  field rotation 
mapping the axion coupling to one operator (\textit{e.g.}, the electroweak sphaleron $a W \tilde W$) to another (such as the lepton current $a \partial_\mu J_L^\mu$) 
does not change the dynamics of baryogenesis. Such operations, if performed correctly, can therefore never change the condition for baryogenesis, and hence the resulting baryon asymmetry should be given in a form invariant under this transformation.\footnote{A similar point was noted in Ref.~\cite{Abel:2018fqg}. Our analysis extends this result to non-equilibrium situations, which in particular arise when marginally relevant processes are involved in the generation of the baryon asymmetry.}
Also, if marginally relevant processes are involved, we have to track the time-evolution of the baryon asymmetry in order to determine its final value, but our condition remains as a necessary condition for successful baryogenesis.

As a concrete example we apply this formalism to baryogenesis around the electroweak phase transition, invoking the original Peccei-Quinn axion and an Affleck-Dine type mechanism to trigger the axion motion (see Ref.~\cite{Co:2019wyp}). In this case, a notable subtlety arises because the charge vectors of the up-Yukawa, the down-Yukawa and the strong sphaleron and not linearly independent. As a consequence, the generated charge asymmetries in principle backreact on the axion equation of motion,\footnote{
	The backreaction to the axion is correctly taken into account in Refs.~\cite{McLerran:1990de,Co:2019wyp} in the case where the axion couples to the strong Chern-Simons term, $a G \tilde G$.
	Our formalism generalizes this to an arbitrary transport equation with an arbitrary axion couplings.
} though in the parameter space of interest this is not of phenomenological importance.

As a second example, we consider high-scale baryogenesis invoking the lepton-number violating Weinberg operator as well as a coupling to the lepton current or to $\tilde W W$ during reheating (see Ref.~\cite{Kusenko:2014uta}). Since the electroweak sphaleron comes into equilibrium only when the Weinberg operator drops out of equilibrium, the final baryon asymmetry (obtained by numerically solving the appropriate differential equation) is suppressed compared to the equilibrium solution (see also Ref.~\cite{Daido:2015gqa}). 
We point out that, in the presence of the lepton-number violating Weinberg operator, the couplings to the lepton current and to $\tilde W W$ are not equivalent. This is a consequence of the invariance of the transport equation under  field rotations.

In addition, by deriving a general condition for the axion coupling to trigger successful baryogenesis, we show that other couplings such as the coupling of the axion to gluons, $a \tilde G G$, which itself preserves baryon and lepton number, can account for the present baryon asymmetry both in electroweak-scale~\cite{Co:2019wyp} and high-scale baryogenesis.\footnote{
	Soon after we uploaded our paper on the arXiv, Ref.~\cite{Co:2020jtv} appeared, independently also pointing out that $a G \tilde G$ can source the $B-L$ asymmetry in the presence of the Weinberg operator.
}

The remainder of this paper is organized as follows. In Sec.~\ref{sec:transport} we derive the transport equation describing the time evolution of chemical potentials in the presence of an axion coupling to a set of operators (see Appendix~\ref{sec:derivation} for details). Without making any assumptions on the particle content and operator involved, we lay out the framework to compute the equilibrium solution and the final asymmetries. We explicitly demonstrate the invariance of the transport equation under field rotations which seemingly change the axion coupling
and discuss backreaction of the induced chemical potentials on the axion equation of motion (see Appendix~\ref{sec:br_pr} for details). In Sec.~\ref{sec:SM-transport} we specify the relevant Standard Model (SM) processes as well as their equilibration temperatures, extending the discussion in Ref.~\cite{Garbrecht:2014kda} by including the renormalization group running of the Yukawa couplings. Sections~\ref{sec:b+l} and \ref{sec:b-l} are dedicated to two concrete examples of baryogenesis around the electroweak phase transition and reheating, respectively. We conclude in Sec.~\ref{sec:conc}.

\section{Transport equation and basis independence}
\label{sec:transport}

\subsection{Transport equation}
\label{subsec:transport}

In this section, we discuss the general structure of the transport equation without specifying a particular system. We take a rather general attitude and derive several properties that hold for any transport equation in a homogeneous and isotropic system. Starting from the current equation and symmetry properties, which follow immediately from the Lagrangian of a given system, we invoke linear response theory to obtain a simple linear algebra system describing the equilibrium solution for all chemical potentials.
Some concrete examples will be considered in Secs.~\ref{sec:b+l} and \ref{sec:b-l}, with a particular focus on the resulting asymmetries in the total baryon number.
For the convenience of readers, we summarize our definitions of indices and symbols in Appendix.~\ref{sec:symbols}.

\paragraph{Current equation.}
Our starting point is the following operator
equation:
\begin{align}
	\partial_{\mu} J^{\mu}_{i} (x) = \sum_{\alpha} n^{\alpha}_{i} O_{\alpha} (x)\,.
	\label{eq:current-eq}
\end{align}
Here $J_{i}^{\mu}$ is the current corresponding to a particle species $i$ (with $i = 1, \cdots, N$) and the operator $O_{\alpha}$ encodes \textit{e.g.}, the anomalous contribution $F \tilde F$ or Yukawa interactions.
For each $O_\alpha$, there exists a vector $n_{i}^{\alpha}$ that specifies the charge of each species $i$ involved in the process of $O_{\alpha}$ (see Sec.~\ref{sec:SM-transport} for details on these operators in the SM).
For conserved currents, the right-hand side of Eq.~\eqref{eq:current-eq} vanishes.

\paragraph{Transport equation.}
We can derive a transport equation by taking the expectation value of both sides of this equation.
As usual, we assume that the chemical equilibration associated with the current equations is much slower than typical scatterings.
This justifies the approximation of kinetic equilibrium for the system, and hence the deviation from the chemical equilibrium can be characterized by slowly varying chemical potentials $\mu_{i}$ for each charge $J_{i}^{0}$.
Let $q_{i}$ be a charge density of species $i$ defined by
\begin{align}
	q_{i} (t) \equiv \frac{1}{\operatorname{vol} (\mathbb{R}^{3})} \int \dd^{3} x \,\vev{J_{i}^{0} (t, \bm{x})} = \vev{J_{i}^{0} (t, \bm{0})}\,.
	\label{eq:q_i}
\end{align}
Throughout this paper, we assume homogeneity and isotropy of the system.
We use this property in the second equality.
The connection to the chemical potential is given by
\begin{align}
	q_{i} (t) = g_{i} \mu_{i} (t) \frac{T^{2}}{6}\,,
	\label{eq:mu_i}
\end{align}
where $T$ is the temperature of the ambient plasma and the multiplicity is $g_{i}  = 1, 2$ for a chiral fermion and a complex scalar, respectively. 
We assume $\mu_i \ll T$ for all $i$ throughout this paper.
Note that one should introduce different chemical potentials for each species which are distinguishable by any of the relevant interactions.

The left-hand side of Eq.~\eqref{eq:current-eq} gives
\begin{align}
	\frac{1}{\operatorname{vol} (\mathbb{R}^{3})} \int \dd^{3} x\, \partial \cdot \vev{J_{i} (t, \bm{x})}
	= \dot q_{i} (t)\,.
\end{align}
One may evaluate the right-hand side {by computing the linear response of the system to a small perturbation $\mu_{i} / T \ll 1$.}
As can be seen from Eq.~\eqref{eq:current-eq}, an operator $\alpha$ involves $n^{\alpha}_{i}$ charges for each species $i$.
Therefore, the expectation value of $O_{\alpha}$ is given by
\begin{align}
	\frac{1}{\operatorname{vol} (\mathbb{R}^{3})} \int \dd^{3} x \, \vev{O_{\alpha} (t, \bm{x})} = -
	\Gamma_{\alpha} \sum_{j} n^{\alpha}_{j} \frac{\mu_{j}}{T}\,,
	\label{eq:transport-coef}
\end{align}
where
\begin{align}
	\Gamma_{\alpha} \equiv \left. \frac{T G_{\alpha}^{\rho} (\omega, \bm{0})}{2 \omega} \right|_{\omega = 0} \,, \quad 
	G_{\alpha}^{\rho} (\omega,\bm{p}) \equiv \int \dd^4 x\, e^{i\omega x^0 - i \bm{p} \cdot \bm{x}} \vev{ \left[ O_{\alpha} (x), O_{\alpha} (0) \right] }.
	\label{eq:Gamma_alpha}
\end{align}
at the linear response.\footnote{
This $\Gamma_\alpha$ is the linear response coefficient of interaction processes to a chemical potential. Regarding sphaleron processes, one may alternatively define $\Gamma$ by the diffusion constant per unit volume of Chern-Simons number. The latter one is twice larger than the former one by a fluctuation dissipation relation. The difference comes from the average between the forward and backward sphaleron rates~\cite{McLerran:1990de}. 
}
See Appendix.~\ref{sec:derivation} for the derivation and a more precise definition of correlators.
$\Gamma_{\alpha}$ represents the rate per unit time-volume for $O_{\alpha}$.
For later convenience, we also define the rate per unit time by
\begin{align}
	\gamma_{\alpha} \equiv \frac{\Gamma_{\alpha}}{T^{3}/6}\,.
	\label{eq:gamma_alpha}
\end{align}
Therefore the transport equation corresponding to the current equation \eqref{eq:current-eq} can be expressed as
\begin{align}
	\dot q_{i} = - \sum_{j} \Gamma_{ij} \frac{\mu_{j}}{T}\, , \qquad
	\Gamma_{ij} \equiv \sum_{\alpha} \Gamma_{\alpha} n_{i}^{\alpha} n_{j}^{\alpha}\, .
	\label{eq:transport}
\end{align}

\paragraph{Conserved quantities.}
In general, this matrix $\Gamma_{ij}$ can have vanishing eigenvalues.
Let $\{n_{i}^{A}\}$ be a set of eigenvectors with zero eigenvalues.
The presence of these eigenvectors indicates that, if one pumps up a chemical potential as $\sum_{i} n^{A}_{i} J^{0}_{i}$, it does not induce any chemical reactions.
Therefore this set corresponds to the conserved charges in this system.
One can see this by multiplying this vector from the left to both sides of Eq.~\eqref{eq:transport}, leading to
\begin{align}
	0 = \sum_{i} n^{A}_{i} \dot q_{i} \quad \longrightarrow \quad
	q_{A} = \text{const.}\,, \quad q_{A} \equiv \sum_{i} n^{A}_{i} q_{i}\, ,
	\label{eq:q_A}
\end{align}
which is equivalent to
\begin{align}
	\sum_{i} n^{A}_{i} g_{i} \frac{\mu_{i}}{T} = c_{A}\,,
	\label{eq:conservation}
\end{align}
with $c_A$ being a constant.
Here a constant $c_{A}$ sets the conserved charge $A$ of this system as
$q_{A} = c_{A} T^{3} / 6$.

\paragraph{Interaction basis.}
In general, some charge vectors can be expressed as a linear combination of the others.
One may choose a complete set of vectors $n^{\alpha}_{i}$ (associated with $O_{\alpha}$) that are linearly independent, which we denote as $\{ n_i^{\hat\alpha} \}$.
For later convenience, we denote the rest of the charge vectors as $n_i^{\alpha_\Delta}$ which can be expressed as a linear combination of $\{n_i^{\hat \alpha}\}$.
Now the set of $\{ n^{\hat\alpha}_{i}\}$ and $\{n^{A}_{i} \}$ forms a complete basis of the chemical potential space.
Note here that the vector spaces spanned by $\{ n_i^{\hat \alpha} \}$ and $\{ n_i^A \}$ are orthogonal because $0 = \sum_i n_i^{\hat \alpha} n_i^A$ for any $\hat \alpha$ and $A$.
Since the sets of basis vectors $\{ n_i^{\hat \alpha} \}$ and $\{ n_i^A \}$ are not orthonormal,
we define dual basis vectors $\{ \bar{n}_i^{\hat \alpha} \}$ and $\{ \bar{n}_i^A \}$ respectively such that $\sum_i \bar{n}_i^{\hat \alpha} n_i^{\hat\beta} = \delta_{\hat \alpha \hat \beta}$ and $\sum_i \bar{n}_i^A n_i^{B} = \delta_{AB}$.
Note that we also have $0 = \sum_i \bar{n}_i^{\hat \alpha} n_i^A = \sum_i \bar{n}_i^A n_i^{\hat \alpha} = \sum_i \bar{n}_i^{\hat \alpha} \bar{n}_i^{A}$ because the vector spaces spanned by $\{ n^{\hat\alpha}_{i}\}$ and $\{n^{A}_{i} \}$ are orthogonal.
For notational brevity, we introduce a collective notation $\{ n^{X}_{i} \} \equiv \{ n^{\hat\alpha}_{i}, n^{A}_{i} \}$ and $\{ \bar n^{X}_{i} \} \equiv \{ \bar n^{\hat\alpha}_{i}, \bar n^{A}_{i} \}$ with
\begin{align}
	\sum_{i} \bar n^{X}_{i} n^{Y}_{i} = \delta_{XY}\,, \quad
	\sum_{X = \hat\alpha, A} \bar n^{X}_{i} n^{X}_{j}  = \delta_{ij}\,.
	\label{eq:orthogonality}
\end{align}
We denote the number of the basis vectors $\{ n^{\hat\alpha}_{i} \}$ ($\{ n^{A}_{i} \}$) as $N_{\hat\alpha} (N_{A})$.
By definition, we have $N = N_{\hat\alpha} + N_{A}$. 
For later convenience, we further divide the basis vectors  $n_i^{\hat\alpha}$ into $\{ n_i^{\hat \alpha_\perp} \} \equiv \{ n_i^{\hat \alpha} | \sum_i \bar{n}_i^{\hat \alpha} n_i^{\alpha_\Delta} = 0~\text{for all}~\alpha_\Delta\}$ and $\{ n_i^{\hat \alpha_\parallel} \} \equiv \{ n_i^{\hat \alpha} | \sum_i \bar{n}_i^{\hat \alpha} n_i^{\alpha_\Delta} \neq 0 ~\text{for some}~\alpha_\Delta\}$. The latter set $\{n_i^{\hat\alpha_\parallel}\}$ involves a linear dependent relation with $n_i^{\alpha_\Delta}$ as $n_i^{\alpha_\Delta} = \sum_{\hat\alpha_\parallel} U^T_{\alpha_\Delta \hat\alpha_\parallel} n_i^{\hat\alpha_\parallel}$ with $U^T_{\alpha_\Delta \hat\alpha_\parallel} \neq 0$.
Note that the dual vector $\bar{n}_i^{\hat\alpha}$ is related to a conserved charge 
in the case where the interaction $\hat\alpha$ is turned off if $\hat \alpha \in \{ \alpha_\perp\}$. In this case the conserved charge
is $q_{\hat\alpha} = \sum_i \bar{n}_i^{\hat\alpha} q_i$ and the corresponding chemical potential is $\mu_{\hat\alpha} = \sum_i g_i \bar{n}_i^{\hat\alpha}$.\footnote{
These conserved charges provide the physical intuition behind distinguishing between linearly independent and dependent basis vectors, namely $\sum_i \bar{n}_i^{\hat\alpha} n_i^{\alpha_\Delta} = 0$ and $\sum_i \bar{n}_i^{\hat\alpha} n_i^{\alpha_\Delta}\neq 0$, respectively. A linearly dependent basis vector implies a reduced number of conserved charges when we turn off its corresponding interaction.}

\subsection{Transport equation including an axion}
\label{sec:basis-indep}

Now we shall include a coupling to an axion. In particular, we will provide a general transport equation in the presence of a non-vanishing velocity of the axion by assuming that the change of the axion velocity is much slower than the typical interactions in the ambient plasma (see Appendix.~\ref{sec:derivation} for the details of derivation and assumptions).
As an aside, we show that the resulting transport equation is invariant under field redefinitions associated with charges in the current equation, which seemingly change the coupling to the axion.

\paragraph{Coupling to the axion.}

Before discussing the coupling to the axion, let us briefly recall the derivation of~\eqref{eq:current-eq}.
Let $\{\Phi\}$ be a set of fields in the action $S$ and
consider a U$(1)_k$ transformation: $\{\Phi'\} = \{e^{i \theta_k Q_k} \Phi\}$.
The current equation follows if the path-integral fulfills 
\begin{align}
	\int \left[ \dd \mu' \right] F[\{\Phi'\}] e^{i S[\{ \Phi' \}]} 
	&= \int \left[ \dd \mu \right] F[\{\Phi\}] e^{i S[\{ \Phi \}] 
	+ \int \dd^4 x\, i \theta_k  \left( \partial \cdot J_k - \sum_\alpha n_k^\alpha O_\alpha  \right) + i R (\theta_k)} 
	\label{eq:derivation_current}
	\\
	&= \int \left[ \dd \mu \right] F[\{\Phi\}] e^{i S[\{ \Phi \}] }
	\left[ 1 + \int \dd^4 x \, i \theta_k  \left( \partial \cdot J_k - \sum_\alpha n_k^\alpha O_\alpha  \right) + \mathcal{O} (\theta_k^2) \right]
	\,,
\end{align}
with $[\dd \mu]$ being a measure of the path-integral and $F[\{\Phi\}]$ being a functional of $\{\Phi\}$ invariant under the U$(1)_k$ transformation.\footnote{
	We take $F[\{\Phi\}]$ invariant under the U$(1)_k$ transformation just for simplicity. If $F[\{\Phi\}]$ is charged under this, we get the anomalous Ward-Takahashi identity associated with the commutator $[Q_k,F]$.
}
Here $R$ in the first equation involves terms at a higher order in $\theta_k$.
Differentiating with respect to $\theta_k$ and taking $\theta_k = 0$, we obtain the current equation \eqref{eq:current-eq}.

Now we are ready to couple the current equation \eqref{eq:current-eq} to a homogeneous axion field $a(t)$ with a decay constant $f$.
There are two types of (classically) shift-symmetric couplings.
One is a direct coupling with the current:
\begin{align}
	\mathcal{L}_a = - \frac{a}{f} \partial \cdot J_k
	= \frac{\dot a}{f} J_{k}^{0} + (\text{total derivative})\,.
	\label{eq:current-coupling}
\end{align}
After integration by parts, this coupling can be regarded as an external chemical potential.
The other is an indirect coupling to the current with an operator $O_{\beta}$ appearing in the current equation:
\begin{align}
	\mathcal{L}_a = 
	- \frac{a}{f} O_{\beta}\,,
	\label{eq:operator-coupling}
\end{align}
at linear order in $a/f$.
The second coupling respects the classical shift symmetry of $a$,
if one can rewrite it as $(a/f) O_{\hat\beta} =  (a/f) \sum_{i} \bar n^{\hat \beta}_{i} \partial \cdot J_{i}$ by reversing the transformation in Eq.~\eqref{eq:derivation_current} at linear order in $a/f$\footnote{
	If we keep the nonlinear part appearing in the transformation~\eqref{eq:derivation_current},
	the equality should be $(a/f) O_{\hat\beta} =  \sum_{i}\bar n^{\hat\beta}_{i} (a/f) \partial \cdot J_{i} + R(\{\bar{n}_i^{\hat\beta} a/f\})$.
	However, throughout the main text, we assume that the axion mass originating from this coupling is negligible and the typical time scale of axion motion is much slower than $1/T$.
	Under these approximations, one may always rotate away a constant term in the axion, and also expand the resulting equations in $\dot a / (fT)$.
	These are the underlying reasons why we may use transport equations at linear order in the axion field.
	Since we restrict ourselves to this case in this paper anyway, we can drop the nonlinear part of $a/f$ in this discussion.
\label{fn:axion_mass}
}
or if it is a total derivative of some other operator $O_{\beta} = \partial \cdot K_{\beta}$ (\textit{e.g.}, $W \tilde W = \partial \cdot K_\text{WS}$).\footnote{
	The latter case could be broken quantum mechanically by the instanton.
	This is why we said ``classically'' shift symmetric.
}
More general couplings can be generated from a linear combination of the two couplings in Eqs.~\eqref{eq:current-coupling} and \eqref{eq:operator-coupling}, and hence these two are sufficient for our discussion.
These shift-symmetric couplings with the axion modify the transport equation as follows:
\begin{align}
	\dot q_{i} = - \sum_{j} \Gamma_{ij} \frac{\mu_{j}}{T} + \frac{\dot a / f}{T} S_{i}\,.
	\label{eq:transport-with-a}
\end{align}
In the following, we discuss the source term $S_{i}$  for each case.

\paragraph{Axion source terms.}
Let us start with the direct coupling to the current, given by Eq.~\eqref{eq:current-coupling}. As already mentioned, this coupling can be regarded as an external chemical potential of $(\dot a / f) J_{k}^{0}$.
Such an external chemical potential gives rise to a shift of $\mu_{k} \mapsto \mu_{k} - \dot a / f$, indicating $\mu_{k} = \dot a / f$ at equilibrium.
This observation implies the following form for the transport equation:
\begin{align}
	\dot q_{i} = - \sum_{j} \Gamma_{ij} \left( \frac{\mu_{j}}{T} - \frac{\dot a / f}{T} \delta_{jk} \right)
	\qquad \text{for the axion coupling} \quad \frac{\dot a}{f} J_{k}^{0}\,.
	\label{eq:transport-eq-current}
\end{align}
This means that the source term in Eq.~\eqref{eq:transport-with-a} is given by
\begin{align}
	S_{i} = \sum_{j} \Gamma_{ij} \delta_{jk}
	\qquad \text{for the axion coupling} \quad \frac{\dot a}{f} J_{k}^{0}\,.
	\label{eq:axion-chemical}
\end{align}
as expected.

Next we move on to the coupling with an operator $- (a / f) O_{\beta}$ [see Eq.~\eqref{eq:operator-coupling}]. 
In linear response, this interaction introduces a bias on the processes involving this operator.
Therefore, we expect
\begin{align}
	\frac{1}{\operatorname{vol}(\mathbb{R}^{3})} \int \dd^{3}x\, \vev{ O_{\alpha} (t, \bm{x}) |_{a/f} } = -\Gamma_{\alpha} \sum_{j} n_{j}^{\alpha} \frac{\mu_{j}}{T} 
	+ \delta_{\alpha \beta} \Gamma_{\alpha} \frac{\dot a / f}{T}\,,
	\label{eq:axion-op-lr}
\end{align}
where the expectation value with a superscript of $a / f$ implies the presence of the axion coupling.
We show that this relation indeed holds in Appendix.~\ref{sec:derivation} by means of linear response theory.
In the derivation of the transport equation from Eq.~\eqref{eq:current-eq}, the expectation value of the right-hand side should be replaced with Eq.~\eqref{eq:axion-op-lr}.
Hence, the transport equation becomes
\begin{align}
	\dot q_{i} =  - \sum_{\alpha} n^{\alpha}_{i} \Gamma_{\alpha} \left( 
	 \sum_{j} n^{\alpha}_{j} \frac{\mu_{j}}{T} - \frac{\dot a / f}{T} \delta_{\alpha \beta} 
	\right)  \qquad
	\text{for the axion coupling} \quad - \frac{a}{f} O_{\beta}\,,
	\label{eq:axion-op-transport}
\end{align}
implying that the corresponding source term in Eq.~\eqref{eq:transport-with-a} is given by
\begin{align}
	S_{i} = \sum_{\alpha}  n^{\alpha}_{i} \Gamma_{\alpha} \delta_{\alpha \beta}
	\qquad
	\text{for the axion coupling} \quad - \frac{a}{f} O_{\beta}\,.
	\label{eq:axion-op}
\end{align}

For a more general coupling, it is convenient to introduce a source vector $n_S^\alpha$ such that 
\begin{align}
 S_i \equiv \sum_\alpha \Gamma_\alpha n_i^\alpha n_S^\alpha. 
 \label{eq:sourcevec}
\end{align}
Then we obtain, \textit{e.g.}, $n_S^\alpha = \delta_{\alpha \beta}$ for the coupling of $-(a/f) O_\beta$, and
$n_S^\alpha = n_k^\alpha$ for $- (a / f) \sum_{\beta} n^{\beta}_{k} O_{\beta}$ or $- (a / f) \del \cdot J_k$. 
If the axion couples to a current $J_Q$ ($= \sum n_i^Q J_i$) where $n_i^Q$ is its charge vector, the source vector is given by $n_S^\alpha = \sum_i n_i^Q n_i^\alpha$.

\paragraph{Basis independence.}
So far, we have seen how the shift-symmetric couplings of the axion given in Eqs.~\eqref{eq:current-coupling} and \eqref{eq:operator-coupling} give rise to the source terms in the transport equation.
However, there are subtleties in this computation because the coupling to the axion has redundancies in its description owing to the current equations and the conserved charges.

As an illustration, let us consider a theory with $- (a/f) \partial \cdot J_{k}$.
One may compute the source term of this coupling and then obtain Eq.~\eqref{eq:axion-chemical}.
Instead, one may perform a field rotation associated with the charge $J_{k}^{0}$,
which yields $- (a/f) \sum_{i} n^{\alpha}_{k} O_{\alpha}$ at linear order in $a/f$ [see Eq.~\eqref{eq:derivation_current}].\footnote{Also, one may consider other transformations such as a field rotation associated with a conserved charge. Then, one can replace the coupling with $a \sum_{i\neq k} n^{A}_{i} \partial \cdot J_{i}$.
Moreover, one could perform a field rotation associated with other charges and then rewrite this coupling in a more complicated form.
All these transformations of a field basis (which we simply refer to as `field rotations' in this paper) give exactly the same transport equation.
}
The transport equation computed in this field basis should be the same as the original transport equation sourced by $- (a/f) \partial \cdot J_{k}$.
In the following, we directly confirm this \emph{basis independence} of the axion coupling in the transport equation.
In other words, we will prove that the source vector $n_S^\alpha$ does not depend on these redundancies of the axion-coupling related to a field rotation.

The fundamental building block of the independence under such basis transformations is an equivalence between $- a \partial \cdot J_{k}$ and $- a \sum_{\alpha} n^{\alpha}_{k} O_{\alpha}$.
Once we show this, other more complicated transformations are just given by considering linear combinations of these operators.
Hence, a proof for these two couplings is sufficient.
For the coupling of $-(a/f) \partial \cdot J_{k}$, the source term is given in Eq.~\eqref{eq:axion-chemical}:
\begin{align}
	S_{i} = \sum_{j} \Gamma_{ij} \delta_{jk} = \sum_{\alpha} \Gamma_{\alpha} n^{\alpha}_{i} n^{\alpha}_{k}\,.
	\label{eq:basis1}
\end{align}
On the other hand, the source term for $- (a / f) \sum_{\beta} n^{\beta}_{k} O_{\beta}$ can be obtained from Eq.~\eqref{eq:axion-op} by multiplying $n^{\beta}_{k}$ and summing over $\beta$.
This in the end gives
\begin{align}
	S_{i} = \sum_{\beta} n^{\beta}_{k} \sum_{\alpha} n^{\alpha}_{i} \Gamma_{\alpha} \delta_{\alpha \beta} = \sum_{\alpha} \Gamma_{\alpha} n^{\alpha}_{i} n^{\alpha}_{k} \,,
\end{align}
which coincides with Eq.~\eqref{eq:basis1}.
Therefore these two couplings yield exactly the same transport equations as expected.
This proves that the transport equation is invariant under such field rotations, \textit{i.e.}, the phenomenology of the axion couplings is basis independent.

\subsection{Asymmetry generation}
\label{sec:asymmetry_generation}
In this section, we discuss how the axion induces a non-vanishing asymmetry.
Assuming that the equilibration is faster than the axion dynamics, we first sketch how to obtain an equilibrium solution of chemical potentials for a given set of conserved charges $\{c_{A}\}$ in the presence of non-vanishing $\dot a$.
Then, we derive a condition for the couplings to the axion so that the non-vanishing velocity $\dot a$ yields an asymmetry for a specific charge.
We finally discuss how this dynamics gives rise to a friction term for the axion,
and point out a special case where this friction term vanishes identically.

\paragraph{Equilibrium solution.}
Now we sketch how to obtain the equilibrium solution for a given set of conserved charges $\{c_{A}\}$ in the presence of a source term.
An equilibrium solution is defined by $\dot q_{i} = 0$ for all $i$.
By multiplying $\bar n^{\hat\alpha}_{i}$ from the left to both sides of Eq.~\eqref{eq:transport-with-a} we find the following set of equations :
\begin{align}
	\sum_i \bar{n}_i^{\hat\alpha} S_i \frac{\dot a / f}{T} = \sum_{i,j} \bar{n}_i^{\hat\alpha} \Gamma_{ij} \frac{\mu_j}{T} \quad 
	\longrightarrow
	\quad
	\sum_{\beta} S_{\hat\alpha \beta} n_S^{\beta} \frac{\dot a / f}{T}
	= \sum_{\hat\beta} \Gamma_{\hat\alpha \hat\beta} \sum_{j}  n_j^{\hat\beta} \frac{\mu_j}{T}\,.
\end{align}
Here we define
\begin{align}
	\Gamma_{\hat\alpha \hat\beta} \equiv \sum_{i,j} \bar{n}_i^{\hat\alpha} \Gamma_{ij} \bar{n}^{\hat\beta}_j = \sum_\gamma U_{\hat\alpha \gamma} \Gamma_\gamma U^T_{\gamma \hat\beta}\,,
	\quad
	S_{\hat\alpha \beta} \equiv \sum_i \bar{n}_i^{\hat\alpha} \Gamma_\beta n_i^\beta 
	= U_{\hat\alpha \beta} \Gamma_\beta\,.
	\label{eq:transport_matrix}
\end{align}
with
\begin{align}
	U_{\hat\alpha \beta} \equiv \sum_i \bar{n}^{\hat\alpha}_i n^\beta_i\,.
	\label{eq:conversion}
\end{align}
Note that, if all the vectors $\{ n_i^\alpha \}$ are linearly independent, \textit{i.e.}, $\hat \alpha = \alpha$, the matrix becomes diagonal, $\Gamma_{\alpha \beta} = \Gamma_\alpha \delta_{\alpha \beta}$, $S_{\alpha \beta} = \Gamma_\alpha \delta_{\alpha \beta}$, because $U_{\alpha \beta} = \sum_i \bar{n}_i^\alpha n_i^\beta = \delta_{\alpha \beta}$.

Since the matrix $\Gamma_{\hat\alpha \hat\beta}$ is invertible,\footnote{
	Suppose that a vector $v^{\hat{\alpha}}$ satisfies $0 = \sum_{\hat\alpha}v^{\hat{\alpha}}\Gamma_{\hat{\alpha}\hat\beta}$.
	The positivity of $\Gamma_\alpha$ implies that $0 = \sum_{\hat \alpha}v^{\hat\alpha}U_{\hat\alpha \beta}$.
	By restricting $\beta$ to $\hat{\beta}$, the matrix $U_{\hat\alpha\hat\beta}$ is invertible
	and hence $v^{\hat\alpha} = 0$. It follows that $\Gamma_{\hat\alpha \hat\beta}$ is invertible.
}
we obtain the following equation in  matrix notation [together with the conservation equations \eqref{eq:conservation}], which determines the equilibrium solution:
\begin{align}
	\left( M_{Xi} \right)
	\Bigg( \frac{\mu_{i}}{T} \Bigg)
	=
	\begin{pmatrix}
	\sum_{\hat \beta, \gamma} \Gamma^{-1}_{\hat \alpha \hat \beta} S_{\hat \beta \gamma} n_S^\gamma \frac{\dot a / f}{T} \\
	c_{A}
	\end{pmatrix}
	\,, \quad
	\left( M_{Xi} \right) \equiv
	\begin{pmatrix}
	\left( n^{\hat \alpha}_i \right) \\
	\left( g_i n^A_i \right)
	\end{pmatrix}\,.
	\label{eq:matrix-equil}
\end{align}
Here $\hat\alpha = 1, \cdots, N_{\hat\alpha}$, $A = 1, \cdots, N_A$, $i = 1, \cdots, N$, and $X$ runs through $\hat \alpha$ and $A$. $c_A$ and $\mu_i$ represent the $N_A$ and $N$ dimensional vectors, respectively. 
$(n^{\hat\alpha}_i)$ is an $N_{\hat\alpha} \times N$ matrix, a $(g_i n^A_i)$ is $N_A \times N$ matrix, and hence $M_{Xi}$ is an $N \times N$  matrix.
Multiplying an inverse matrix from the left,\footnote{
	We provide a proof that the $N \times N$ matrix $M_{Xi}$ is invertible.
	Suppose that a vector $v^i$ satisfies $0 = \sum_i M_{Xi} v^i$.
	By definition, we have $\sum_i M_{\hat \alpha i} \bar{n}_i^{\hat \beta} = \delta_{\hat \alpha \hat \beta}$ which is non-zero. Hence, $v^i$ can be expressed as a linear combination of $n_i^A$, \textit{i.e.}, $v^i = \sum_{A} n_i^A x_A$.
	Now, $0 = \sum_i M_{Xi} v^i$ is rewritten as $0= \sum_A x_A (\sum_i n_i^A g_i n_i^{A'} )$.
	The positivity of $g_i$ implies that $0 = \sum_A x_A n^A_i$.
	By multiplying $\bar{n}_i^{A'}$ and summing over $i$, we find $x_A = 0$ for all $A$. It follows that $M_{Xi}$ is invertible.
}
we obtain the equilibrium solution:
\begin{align}
	\Bigg( \frac{\mu_{i}}{T} \Bigg)_\text{eq}
	=
	\left( M^{-1}_{iX} \right)
	\begin{pmatrix}
	\sum_{\hat \beta, \gamma} \Gamma^{-1}_{\hat \alpha \hat \beta} S_{\hat \beta \gamma} n_S^\gamma \frac{\dot a / f}{T} \\
	c_{A}
	\end{pmatrix}\,.
	\label{eq:matrix-equil_sol}
\end{align}
where $M^{-1}_{iX}$ is the inverse matrix of $M_{Xi}$ defined in Eq.~\eqref{eq:matrix-equil}.
Note that, if $n_S^\gamma = c_{A} = 0$ for all $A$ and $\gamma$, the solution is a trivial one, $\mu_{i} = 0$ for all $i$. 
This formula is useful when we calculate, \textit{e.g.}, the resulting present-day baryon asymmetry.

\paragraph{Asymmetry generation.}

Here we derive the condition to produce an asymmetry in a specific charge.
Suppose that we are interested in generation of a certain charge $q_C$,
whose effective chemical potential is given by
\begin{align}
	\mu_C = \sum_i g_i n^C_i \mu_i\,.
	\label{eq:mu_c}
\end{align}
Inserting Eq.~\eqref{eq:matrix-equil_sol}, we can estimate the charge $q_C$ in the presence of the source term $n_S^\alpha$ and non-vanishing conserved charges $c_A$:
\begin{align}
	\frac{\mu_C^\text{eq}}{T} = 
	\begin{pmatrix}
		\left( g_i n_i^C \right)^T
	\end{pmatrix}
	\left( M^{-1}_{iX} \right)
	\begin{pmatrix}
	\sum_{\hat \beta, \gamma} \Gamma^{-1}_{\hat \alpha \hat \beta} S_{\hat \beta \gamma} n_S^\gamma \frac{\dot a / f}{T} \\
	c_{A}
	\end{pmatrix}\,,
	\label{eq:equilibrium_solution}
\end{align}
where $(g_i n_i^C)^T$ is a $1 \times N$ matrix.
In other words, this gives the condition on $n_S^\alpha$ and $c_A$ to obtain non-vanishing $q_C$.
In particular, in the case with vanishing conserved charges $c_A = 0$ for all $A$, 
we get non-vanishing $q_C$ if the vector $n_S^\gamma$ fulfills
\begin{align}
	(n_S^\gamma) \not\perp 
	\Bigg( \sum_{i,\hat \alpha, \hat \beta} g_i n^C_i M^{-1}_{i\hat\alpha} 
	\Gamma^{-1}_{\hat\alpha \hat\beta} S_{\hat\beta \gamma} \Bigg) \equiv v_\gamma^C\,.
	\label{eq:cond_asym}
\end{align}
If all the vectors $n_i^\alpha$ are linearly independent, this condition is further simplified as
\begin{align}
	(n_S^\gamma) \not\perp  
	\Bigg( \sum_{i} g_i n^C_i M^{-1}_{i\gamma} \Bigg)\,.
	\label{eq:cond_asym_lind}
\end{align}
These general formulae will prove useful when we discuss the condition to generate the baryon asymmetry in Secs.~\ref{sec:b+l} and \ref{sec:b-l}.

The physical intuition behind this formula is the following: The $CPT$-violating motion of the axion biases the processes encoded in the source vector $n_S^\alpha$ such that they induce non-vanishing chemical potentials $\mu_i$ for the particles involved in these processes. Meanwhile other processes (encoded in $M_{i \hat \alpha}^{-1} \Gamma^{-1}_{\hat \alpha \hat \beta} S_{\hat \beta \gamma}$) try to wash-out these chemical potentials. This competition determines the equilibrium solution. In order to generate the charge $q_C$ (which could be \textit{e.g.}, baryon number), we need a $q_C$-violating operator. The only way to obtain a \textit{vanishing} $q_C$ in the equilibrium solution is by choosing specific couplings such that the $q_C$-violating operator is not involved in this equilibration process.
Eq.~\eqref{eq:cond_asym} or Eq.~\eqref{eq:cond_asym_lind} indicate this specific coupling. After the decoupling of the $q_C$-violating interactions, the non-zero value of $q_C$ freezes out and becomes a conserved charge.
From this it is clear that, for $C$-genesis, we in particular do not have to couple the axion to the $q_C$-violating operator directly. This opens up a variety of couplings successful in creating $q_c$.

\paragraph{Backreaction to the axion.}
So far, we have assumed that the production of asymmetries does not affect the dynamics of the axion.
Here we discuss the backreaction to the equation of motion for the axion, and derive the condition under which we can neglect it.
Since we have already proven the invariance under field rotations, the coupling to the current can be rewritten as the coupling to a linear combination of operators $O_\alpha$.
Hence, it is sufficient to discuss the case with $\mathcal{L}_a = - (a/f) \sum_\alpha n_S^\alpha O_\alpha$.
The equation of motion for the homogeneous mode of the axion then becomes
\begin{align}
	0 &= \ddot a + V'(a) + \frac{1}{f} \sum_\alpha n_S^\alpha \vev{ O_\alpha |_{a/f}  }= \ddot a + V'(a) +  \sum_\alpha n_S^\alpha\frac{\Gamma_\alpha}{fT} \left( n_S^\alpha \frac{\dot a}{f} - \sum_j n^\alpha_j \mu_j \right)\,,
	\label{eq:axion_eom}
\end{align}
where the axion potential is $V(a)$. 
In the second equality, we have used Eq.~\eqref{eq:axion-op-lr}.

Let us assume that the equilibration for the chemical potentials is much faster than the axion dynamics.
Under this approximation, we can insert the equilibrium solution given in Eq.~\eqref{eq:matrix-equil_sol} in the last term of Eq.~\eqref{eq:axion_eom}.
Throughout this paper, we are interested in the case where there are no primordial asymmetries for all the conserved quantities, \textit{i.e.}, $c_A = 0$ for all $A$.
In this way, we can evaluate the last term in Eq.~\eqref{eq:axion_eom}, which defines the effective dissipation rate for the axion as
\begin{align}
	\sum_\alpha n_S^\alpha \frac{\Gamma_\alpha}{fT} \left(n_S^\alpha \frac{\dot a}{f} - \sum_j n^\alpha_j \mu_j^\text{eq} \right)
	=: \sum_{\alpha, \beta} n_S^\alpha \gamma^{\text{eff}}_{a,\alpha \beta} n_S^\beta \, \dot a\,,
\end{align}
implying
\begin{align}
	\gamma^{\text{eff}}_{a,\alpha\beta} = \frac{\Gamma_\alpha}{f^2 T} 
	\left( \delta_{\alpha\beta} -  \sum_{i,\hat \gamma, \hat \rho} n_i^\alpha M^{-1}_{i \hat \gamma} \Gamma_{\hat \gamma \hat \rho}^{-1} S_{\hat\rho \beta}  \right)
	= \frac{1}{f^2 T} 
	\left( \Gamma_\alpha \delta_{\alpha\beta} -  \sum_{\hat \gamma, \hat \rho} S_{\alpha \hat \gamma}^T \Gamma_{\hat \gamma \hat \rho}^{-1} S_{\hat \rho \beta} \right)\,.
	\label{eq:axion_dissipation}
\end{align}
In the second equality, we have used $n_i^\alpha = \sum_{\hat \beta} U^T_{\alpha \hat \beta} n_i^{\hat \beta}$, $\sum_{i} n_i^{\hat\beta} M_{i \hat\gamma}^{-1} = \delta_{\hat \beta \hat \gamma}$, and the definition of $S_{\hat \gamma \alpha} = U_{\hat \gamma \alpha} \Gamma_\alpha$ in Eq.~\eqref{eq:transport_matrix}.
If this rate is much slower than the typical interaction rate for chemical equilibration processes, the assumption of fast equilibration is justified a posteriori.

We remark that there is a special case where the effective dissipation, $\sum_{ \alpha  \beta} n_S^{\alpha} n_S^{\beta} \gamma_{a,\alpha \beta}^\text{eff}$, vanishes identically. Let us see when this happens.
As shown in Appendix.~\ref{sec:br_pr}, the condition where the effective dissipation term vanishes is given by
\begin{align}
	\sum_{\alpha \beta} \gamma^{\text{eff}}_{a,  \alpha  \beta} n_S^{\alpha} n_S^{\beta} = 0 \quad \text{iff}~n_S^{\alpha_\Delta} = \sum_{\hat \alpha_\parallel} U^T_{\alpha_\Delta \hat\alpha_\parallel} n_S^{\hat\alpha_\parallel} \,,
	\label{eq:zero-friction}
\end{align}
where we use the classification of charge vectors into linearly (in)dependent vectors, denoted by the superscripts $\alpha_\Delta, \alpha_\parallel, \alpha_\perp$, as introduced around Eq.~\eqref{eq:orthogonality}.
For instance, if the axion only couples to the operators whose charge vectors are linearly independent with respect to all other interactions, \textit{i.e}, $n_S^\alpha \neq 0$ only if $\alpha \in \{\hat{\alpha}_\perp\}$, 
the right-hand condition is trivially fulfilled and hence the effective friction term vanishes.
The non-vanishing effective friction term arises only if the axion couples to an operator whose charge vector lies in the span of the charge vectors of other interactions,
\textit{i.e}, $n_S^\alpha \neq 0$ for $\alpha \in \{\hat{\alpha}_\parallel, \alpha_\Delta\}$. 
Still, in this case, we could have a cancellation among the source vectors because the corresponding charge vectors are linearly dependent, and if the cancellation occurs, the effective friction term vanishes.
The condition of $n_S^{\alpha_\Delta} = \sum_{\hat \alpha_\parallel} U^T_{\alpha_\Delta \hat\alpha_\parallel} n_S^{\hat\alpha_\parallel}$ takes into account when this cancellation happens.
If the condition \eqref{eq:zero-friction} is fulfilled, the constant motion of the axion is never stopped by the asymmetry generation. This means that a non-vanishing $\dot a$ together with $\mu_i^\text{eq}$ is a non-trivial equilibrium solution even after the inclusion of the backreaction.

We can roughly understand its physical reason as follows.
Let us take the limit of $V(a) \to 0$ to get insight into the nature of this property.
As we will see below, the above non-trivial equilibrium solution exists if we get a new conserved charge in the limit of $V(a) \to 0$.
By multiplying the current equation \eqref{eq:current-eq} by $\bar n_i^{\hat\alpha}$ and taking a summation over $i$, 
we obtain $\sum_{i}\bar{n}_i^{\hat \alpha} \partial_\mu J_i^\mu = \sum_{\beta} U_{\hat\alpha \beta} O_{\beta} = O_{\hat\alpha} + \sum_{\alpha_\Delta} U_{\hat\alpha \alpha_\Delta} O_{\alpha_\Delta} $.
This implies $O_{\hat\alpha} = \sum_i \bar{n}_i^{\hat\alpha} \partial_\mu J_i^\mu - \sum_{\alpha_\Delta} U_{\hat\alpha \alpha_\Delta} O_{\alpha_\Delta}$.
Using this equation, we can rewrite the equation of motion for the axion as
\begin{align}
	\frac{\dd}{\dd t} \left( f \dot a + \sum_{\hat\alpha ,i} n_S^{\hat\alpha} \bar{n}_i^{\hat\alpha} q_i \right)
	= \sum_{\alpha_\Delta} \left( \sum_{\hat\alpha_\parallel} n_S^{\hat\alpha_\parallel} U_{\hat\alpha_\parallel \alpha_\Delta} - n_S^{\alpha_\Delta} \right)
	\vev{O_{\alpha_\Delta} |_{a/f}} 
\end{align}
Now it is clear that, if the condition \eqref{eq:zero-friction} is satisfied, we have a new conserved charge that is a summation of the axion shift charge $f \dot a$ and $\sum_{\hat\alpha,i} n_S^{\hat \alpha}\bar{n}^{\hat\alpha}_i q_i$.
The presence of this new charge in principle allows an equilibrium solution with both charges, $f \dot a$ and $\sum_{\hat\alpha,i} n_S^{\hat \alpha}\bar{n}^{\hat\alpha}_i q_i$, non-vanishing, which implies $\dot a \neq 0$ in equilibrium.
However, if the condition \eqref{eq:zero-friction} is violated, this new charge should vanish in equilibrium.
This means that there must exist a process driving the axion velocity to zero, which is nothing but a non-zero effective friction term.

\section{Transport equation in the Standard Model}
\label{sec:SM-transport}

In this section, we review the transport equation within the SM in the symmetric phase, before discussing the coupling to the axion in the subsequent sections.

\subsection{Standard Model interactions and charge vectors}
\label{sec:SM-interactions}

Let us first specify the number of chemical potentials required to describe the system.
The SM consists of the right-handed lepton $e_f$, the left-handed lepton $L_f$, the right-handed up-type quark $u_f$, the right-handed down-type quark $d_f$, the left-handed quark $Q_f$, and the Higgs $H$, where the index $f$ runs from $1$ to $N_f$ with the number of flavors being $N_f = 3$.
The vector of chemical potentials hence has $5 N_f + 1$ components:
\begin{align}
	\left( \mu_i \right) = \left( \mu_{e_1}, \mu_{e_2}, \mu_{e_3}, \mu_{L_1}, \mu_{L_2}, \mu_{L_3}, \mu_{u_1}, \mu_{u_2}, \mu_{u_3},  \mu_{d_1} ,  \mu_{d_2} ,  \mu_{d_3}, \mu_{Q_1}, \mu_{Q_2}, \mu_{Q_3}, \mu_H \right)\,.
\end{align}
The SM transport equation is written by means of this chemical potential vector: 
\begin{align}
	\dot q_i = - \sum_\alpha \Gamma_\alpha n_i^\alpha \sum_j n_j^\alpha \frac{\mu_j}{T}\,.
\end{align}
Here $\alpha$ runs over the SM interactions relevant for the chemical equilibrium, which are the electroweak sphaleron, the strong sphaleron, the lepton Yukawa, the up-type quark Yukawa, and the down-type quark Yukawa.

As we are interested in the evolution of the chemical potential in the early Universe, 
we should take into account the effect of the expansion of the Universe. 
Denoting $H$ as the Hubble parameter, 
we rewrite 
the transport equation \eq{eq:transport-with-a} by the replacement of $\dot{q}_i \to \dot{q}_i + 3 H q_i$. 
Assuming the radiation-dominated era, 
we obtain $\dot{T} = - H T$
and 
\begin{align}
	H = \sqrt{\frac{g_* \pi^2}{90}} \frac{T^2}{M_{\rm pl}}, 
\end{align}
where $g_*$ ($= 106.75$) is the effective degrees of freedom of relativistic particles. 
The transport equation is now written as\footnote{
This is not the case before the reheating completes. 
We implicitly assume that the reheating temperature is much higher than $10^{13} \GeV$ throughout this paper for simplicity. 
}
\begin{align}
	- \frac{\dd}{\dd \ln T} \lmk \frac{\mu_i}{T} \rmk = -\frac{1}{g_i}\sum_\alpha n_{i}^{\alpha} \frac{\gamma_\alpha}{H}
	\left[\sum_{j}n_{j}^{\alpha} \lmk \frac{\mu_j}{T} \rmk - n_{S}^{\alpha} \lmk \frac{\dot{a}/f}{T} \rmk \right], 
	\label{eq:fulltransporteq}
\end{align}
where we have included an axion source term. 
When the prefactor in the right-hand side becomes larger than of order unity, 
the square bracket in the right-hand side is driven to be zero within of order one Hubble time. 
It is thus convenient to define an equilibration temperature, below which a given interaction is in equilibrium within the time-scale of the Hubble expansion.

Let us focus on an interaction $\beta$ in the right-hand side of \eq{eq:fulltransporteq}. 
Multiplying $n_i^\beta$ and taking a summation over $i$, 
we obtain 
\begin{align}
	- \frac{\dd}{\dd \ln T} \lmk \sum_i n_i^\beta \frac{\mu_i}{T} \rmk 
	= - \sum_i \frac{1}{g_i} \lmk n_{i}^{\beta} \rmk^2 \frac{\gamma_\beta}{H}
	\left[\sum_{j}n_{j}^{\beta} \lmk \frac{\mu_j}{T} \rmk - n_{S}^{\beta} \lmk \frac{\dot{a}/f}{T} \rmk \right] + \dots, 
\end{align}
where the dots represent the other interaction terms. 
Then the quantity $\sum_i n_i^\beta \mu_i / T$ does not change much by the interaction $\beta$ if 
\begin{align}
  \sum_i \frac{1}{g_i} \lmk n_{i}^{\beta} \rmk^2  \gamma_\beta < H. 
  \label{eq:T_alpha}
\end{align}
We define the equilibration temperature of the interaction $\beta$ by the threshold of this condition.\footnote{Ref.~\cite{Garbrecht:2014kda} defines the equilibration temperature of the weak and strong sphaleron processes 
by $6 \gamma_{WS} = H$ and $4 \gamma_{SS} = H$, respectively. 
These are equivalent to our definitions. However, they define those of Yukawa interactions by $\gamma_i /g_L = H$, where $g_L$ is the degrees of freedom of left-handed lepton (quark) for lepton (quark) Yukawa interaction. 
This is different from ours by a factor of $7/2$ for the lepton Yukawa 
and $18/4$ for the quark Yukawa. 
\label{footnote:equilibration_temp}
}

In the following, we give the rate per unit time-volume $\Gamma_\alpha$, the charge vector $n_i^\alpha$, and the equilibration temperature $T_\alpha$ for each interaction,
see Tab.~\ref{tab:equilibration_temperature}. 
It is important to include the renormalization group (RG) flow of the parameters to evaluate these quantities,
which we have done using \texttt{SARAH}~\cite{Staub:2013tta}. 

Before going to the details of the interactions, 
we comment on the differences of our calculation of the equilibration temperature with respect to Ref.~\cite{Garbrecht:2014kda}. 
As explained in footnote~\ref{footnote:equilibration_temp}, 
we include the factor of $\sum_i \lmk n_{i}^{\beta} \rmk^2 / g_i$ in the definition of the equilibration temperature of Yukawa interactions rather than $1/g_L$, the latter of which is used in Ref.~\cite{Garbrecht:2014kda}. 
We also take into account the renormalization-group running of the Yukawa (as well as gauge) couplings, 
not only for top Yukawa but also the other Yukawas. 
This is quite important especially for the bottom Yukawa, where $T_b$ decreases by an order of magnitude. 
We also use updated sphaleron rates following Ref.~\cite{Moore:2010jd}.

\begin{table}[t]
	\centering
	\begin{tabular}{c|c|c|c|c|c|c} \hline
  		Interaction & Weinberg & WS & SS & $Y_e$ & $Y_\mu$ & $Y_\tau$ 
		\\ \hline
  \rule[-10pt]{0pt}{25pt}
		$\Gamma_\alpha/T^4$ & $\kappa_\text{W} \frac{m_\nu^2 T^2  }{ v_{\rm EW}^4} $& $\frac{1}{2} \kappa_\text{WS} \alpha_2^5 $ & $\frac{1}{2} \kappa_\text{SS} \alpha_3^5$ 
		& $\kappa_{Y_e}\, y_{e}^2$ & $\kappa_{Y_e}\, y_{\mu}^2$ & $\kappa_{Y_e}\, y_{\tau}^2$
		\\ \hline
  \rule[-10pt]{0pt}{25pt}
		$T_\alpha\,[\mathrm{GeV}]$ & $6.0 \times 10^{12}$ & $2.5 \times 10^{12}$ & $2.8 \times 10^{13}$
		& $1.1 \times 10^{5}$ & $4.7 \times 10^{9}$ & $1.3 \times 10^{12}$ \\ \hline\hline
  		Interaction & $Y_u$ & $Y_c$ & $Y_t$ & $Y_d$ & $Y_s$ & $Y_b$ 
		\\ \hline
  \rule[-10pt]{0pt}{25pt}
		$\Gamma_\alpha/T^4$ & $ \kappa_{Y_u}\, y_{u}^2$ & $ \kappa_{Y_u}\, y_{c}^2$  & $ \kappa_{Y_t}\, y_{t}^2$
		& $ \kappa_{Y_d}\, y_{d}^2$ & $\kappa_{Y_d}\, y_{s}^2$ & $\kappa_{Y_d}\, y_{b}^2$
		\\ \hline
  \rule[-10pt]{0pt}{25pt}
		$T_\alpha\,[\mathrm{GeV}]$ & $1.0 \times 10^6$ & $1.2 \times 10^{11}$ & $4.7 \times 10^{15}$
		& $4.5 \times 10^{6}$ & $1.1 \times 10^{9}$ & $1.5\times 10^{12}$ \\ \hline
  	\end{tabular}
  	\caption{ 
	A summary of the rate per unit-time volume $\Gamma_\alpha$ and the corresponding equilibration temperature $T_\alpha$ for the SM interactions and $L$-violating interaction by the dimension five Weinberg operator [see \eq{Gamma_W}]. 
	See the main text for the explicit values of the numerical coefficients $\kappa_\alpha$. The differences with respect to  Ref.~\cite{Garbrecht:2014kda} are discussed in the main text. 
	}
  	\label{tab:equilibration_temperature}
\end{table}
\begin{figure}[t]
	\centering
 	\includegraphics[width=0.65\linewidth]{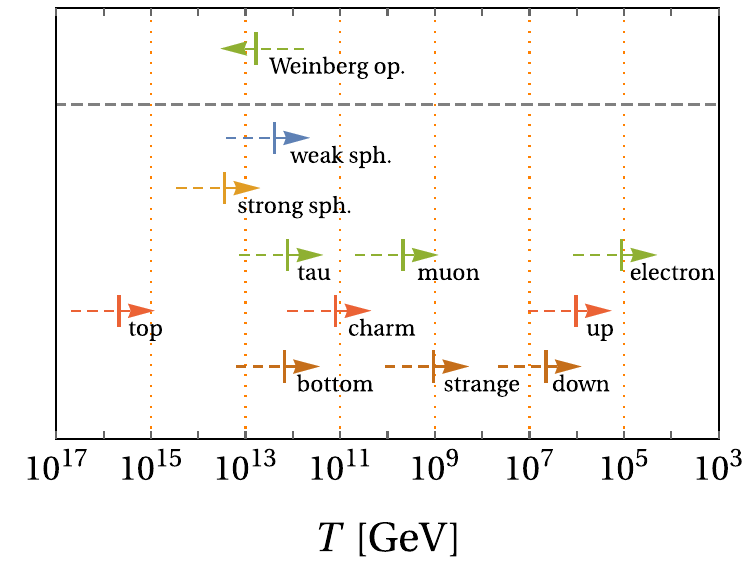}
	\caption{ 
	Equilibration temperatures for individual SM interactions, $T_\alpha$. 
	Each dashed line indicates the range from $10 T_\alpha$ to $T_\alpha$, within which one can expect non-trivial effects due to partial equilibration. The solid arrows (starting from the vertical lines) indicate that the interactions are in equilibrium for $T < T_\alpha$. 
	At the top of the figure, we also show the decoupling temperature of lepton number violating interaction via the dimension five Weinberg operator as a vertical line, above which it is in equilibrium [see Eq.~(\ref{eq:T_W}))]. The dashed line starts from $T_\alpha / 10$ in this case, as this interaction is weaker for lower temperature. 
	}
	\label{fig:equilibration_temp}
\end{figure}
\paragraph{Electroweak sphaleron.}

The electroweak sphaleron involves all the left-handed fermions, which are charged under SU$(2)_\text{W}$.
The corresponding charge vector, $n_i^\text{WS}$, is defined so that
\begin{align}
	\sum_i n_i^\text{WS} \mu_i = \sum_f \left( \mu_{L_f} + 3 \mu_{Q_f} \right)\,.
	\label{eq:ws}
\end{align}
The sphaleron rate per unit time-volume in SU($N_c$) gauge theory with $N_f$ vector fermions and $N_H$ complex scalars is  given by~\cite{Bodeker:1999gx, Arnold:1999ux, Arnold:1999uy, Moore:2000mx, Moore:2000ara, Moore:2010jd} 
\begin{align}
& 2 \Gamma_\text{sphal} \simeq 0.21 
 \lmk \frac{N_c g^2 T^2}{m_D^2} \rmk 
 \lmk \ln \frac{m_D}{\gamma} + 3.0410 \rmk 
 \frac{N_c^2 - 1}{N_c^2} \lmk N_c \alpha \rmk^5 T^4, 
 \label{eq:sphaleron_rate}
 \\
 &\gamma = N_c \alpha T \lmk \ln \frac{m_D}{\gamma} + 3.041 \rmk, 
 \\
 &m_D^2 = \frac{2 N_c + N_f + N_H}{6} g^2 T^2, 
\end{align}
where $g$ ($\equiv \sqrt{4 \pi \alpha}$) is a gauge coupling constant. 
Using $m_D^2 = (11/6) g_2^2 T^2$ in the SU(2) weak sector of the SM, 
we thus estimate the rate as 
\begin{align}
	\Gamma_\text{WS} = \frac{\kappa_\text{WS}}{2} \alpha_2^5 T^4\,,
\end{align}
where $\kappa_\text{WS} \simeq 24$ for $T = 10^{12}\,\mathrm{GeV}$.\footnote{
This sphaleron rate is about 1.3 times larger than the one reported in Ref.~\cite{DOnofrio:2014rug}. 
If one use the latter rate, $T_{\rm WS}$ is estimated as $1.9 \times 10^{12}\,\mathrm{GeV}$. 
}
Comparing the rate per unit time, $\gamma_\text{WS} \sum_i (n_i^{\rm WS})^2 / g_i = 36\Gamma_\text{WS} / T^3$,
to the Hubble parameter, one may estimate the equilibration temperature as
\begin{align}
	T_\text{WS} \simeq 2.5 \times 10^{12}\,\mathrm{GeV}\,.
\end{align}
\paragraph{Strong sphaleron.}

The strong sphaleron involves both left- and right-handed quarks, which are charged under SU$(3)_\text{C}$.
The charge vector, $n_i^\text{SS}$, is given so that
\begin{align}
	\sum_i n_i^\text{SS} \mu_i = \sum_f \left( 2 \mu_{Q_f} - \mu_{u_f} - \mu_{d_f} \right)\,.
	\label{eq:ss}
\end{align}
Substituting $m_D^2 = 2 g_3^2 T^2$ into \eq{eq:sphaleron_rate}, 
we can estimate the rate per unit time-volume as 
\begin{align}
	\Gamma_\text{SS} = \frac{\kappa_\text{SS}}{2} \alpha_3^5 T^4\,,
\end{align}
where $\kappa_\text{SS} \simeq 2.7 \times 10^2$ for $T = 10^{13}\,\mathrm{GeV}$.
Comparing the rate per unit time, $\gamma_\text{SS} \sum_i (n_i^{\rm SS})^2 / g_i = 24 \Gamma_\text{SS} / T^3$,  
to the Hubble parameter, we get the equilibration temperature:
\begin{align}
	T_\text{SS} \simeq 2.8 \times 10^{13}\,\mathrm{GeV}\,.
\end{align}
\paragraph{Lepton Yukawa.}
In general, the lepton Yukawa is an $N_f \times N_f$ matrix, $Y_e^{ff'}$.
If the effect of the neutrino mass can be neglected, one may redefine the leptons fields so that the lepton Yukawa becomes diagonal, \textit{i.e.}, 
$(Y_{e}^{f f'}) = \diag (y_e, y_\mu, y_\tau)$.
Let us take this field basis and denote the corresponding chemical potentials as $\mu_{e_f}$ and $\mu_{L_f}$.
The charge vector, $n_i^{Y_e^{ff}}$, is given so that
\begin{align}
	\sum_i n_i^{Y_e^{ff}} \mu_i = - \mu_{e_f} + \mu_{L_f} - \mu_H\,.
	\label{eq:Ye}
\end{align}
The rate per unit time-volume is estimated as
\begin{align}
	\Gamma_{Y_e^{ff}} = \kappa_{Y_e} (\alpha_2) \, y_{e_f}^2 T^4\,,
\end{align}
where we have made the dependence of $\kappa_{Y_e}$ on $\alpha_2$ explicit, 
which arises from taking into account $2 \leftrightarrow 2$ scattering processes with single gauge boson emission/absorption 
(among others).
The prefactor $\kappa_{Y_e}$ is estimated in Ref.~\cite{Garbrecht:2014kda} 
as $\kappa_{Y_e}(\alpha_2) \simeq 1.7\times 10^{-3}$. 
From this, we obtain the equilibration temperature of the lepton Yukawa for each flavor:
\begin{align}
	T_{y_e} \simeq 1.1 \times 10^{5}\,\mathrm{GeV}\,, 
	\quad 
	T_{y_\mu} \simeq 4.7 \times 10^{9}\,\mathrm{GeV} \,, 
	\quad 
	T_{y_\tau} \simeq 1.3 \times 10^{12}\,\mathrm{GeV} \,,
\end{align}
where we have used $\sum_i (n_i^{Y_e^{ff}})^2 / g_i = 7/4$.

\paragraph{Quark Yukawa.}
Since there exist two $N_f \times N_f$ matrices corresponding to the up-type and down-type quark Yukawas, we cannot diagonalize them simultaneously. 
This is the origin of the well-known CKM matrix which leads to flavor changing processes.
At very high temperature, only the top Yukawa is in equilibrium, and other quark Yukawa interactions start to become efficient as the Universe cools down.
As we discuss in the subsequent Sec.~\ref{sec:SM-charges}, special care about the quantum coherence of different flavors is required in order to describe this process properly.
As a result, we have to take an appropriate field basis in each temperature regime.
These effects have been investigated in the context of flavored leptogenesis~\cite{Abada:2006fw,Nardi:2006fx,Abada:2006ea,Dev:2017trv}.
Below, let us just neglect these subtleties for the moment, and estimate a typical size of transport coefficients.

The charge vector for the up-type quark Yukawa, $n_i^{Y_u^{ff'}}$, is given by:
\begin{align}
	\sum_i n_i^{Y_{u}^{ff'}} \mu_i = - \mu_{u_f} + \mu_{Q_{f'}} + \mu_H\,.
	\label{eq:Yu}
\end{align}
In an appropriate field basis of quarks, the transport coefficient is dominated by its diagonal part, which is estimated as
\begin{align}
	\Gamma_{Y_u^{ff}} = \kappa_{Y_u} (\alpha_2, \alpha_3) \, y_{u_f}^2 T^4\,,
\end{align}
where $\kappa_{Y_u}$ is again estimated in Ref.~\cite{Garbrecht:2014kda} as 
$\kappa_{Y_u}  (\alpha_2, \alpha_3) \simeq 1.0\times 10^{-2}$ for $T \simeq 10^{12}\,\mathrm{GeV}$, 
$1.2\times 10^{-2}$ for $T \simeq 10^9\,\mathrm{GeV}$,
and $1.5\times 10^{-2}$ for $T \simeq 10^6\, \mathrm{GeV}$, respectively.
We estimate it as $\kappa_{Y_u} \simeq 8.0\times 10^{-3}$ for $T \simeq 10^{15}\,\mathrm{GeV}$
from the running of $\alpha_3$.
Again the dependence of $\kappa_{Y_u}$ on $\alpha_2$ and $\alpha_3$ is made explicit.
As an indicator, let us estimate the corresponding equilibration temperature for the diagonal part:
\begin{align}
	T_{y_u} \simeq 1.0 \times 10^6\,\mathrm{GeV} \,, 
	\quad T_{y_c} \simeq 1.2 \times 10^{11}\,\mathrm{GeV} \,,
	\quad T_{y_t} \simeq 4.7 \times 10^{15}\,\mathrm{GeV} \,,
\end{align}
where we have used $\sum_i (n_i^{Y_u^{ff}})^2 / g_i = 3/4$. 
The equilibration temperature of the top Yukawa 
is comparable to the maximal temperature allowed by the constraints on the tensor-to-scalar ratio~\cite{Akrami:2018odb}.

The charge vector for the down-type quark Yukawa, $n_i^{Y_d^{ff'}}$, is given by:
\begin{align}
	\sum_i n_i^{Y_d^{ff'}} \mu_i = - \mu_{d_f} + \mu_{Q_{f'}} - \mu_H\,.
	\label{eq:Yd}
\end{align}
Again, in an appropriate field basis, the transport coefficient is dominated by its diagonal part, which is estimated as
\begin{align}
	\Gamma_{Y_d^{ff}} = \kappa_{Y_d} ( \alpha_2,  \alpha_3) \, y_{d_f}^2 T^4\,,
\end{align}
where $\kappa_{Y_d} \simeq \kappa_{Y_u}$~\cite{Garbrecht:2014kda}. 
As an indicator, we evaluate the equilibration temperature for the diagonal part:
\begin{align}
	T_{y_d} \simeq 4.5 \times 10^{6}\,\mathrm{GeV}\,, 
	\quad 
	T_{y_s} \simeq 1.1 \times 10^{9}\,\mathrm{GeV} \,, 
	\quad 
	T_{y_b} \simeq 1.5 \times 10^{12}\,\mathrm{GeV}\,,
\end{align}
where we have used $\sum_i (n_i^{Y_d^{ff}})^2 / g_i = 3/4$.

\subsection{Conserved quantities and decoupling}
\label{sec:SM-charges}

As we have seen in the previous section, some interactions in the SM may not be efficient in the early Universe.
Therefore, we expect that the number of (approximately) conserved quantities depends on the temperature of the ambient plasma.
In Secs.~\ref{sec:b+l} and \ref{sec:b-l}, we discuss the generation of baryon asymmetry around $T\gtrsim 10^2$\,GeV and $T \gtrsim 10^{13}\, \mathrm{GeV}$ respectively.
In the following, we summarize conserved quantities for these two cases.
We also mention the quantum coherence from different flavors.

\paragraph{$\bm{T \gtrsim 10^2}$\,GeV.}
At the temperature right before the electroweak phase transition, all the SM interactions are in equilibrium.
Without loss of generality, we can take a basis of chemical potentials so that the transport coefficients for the up-type quark Yukawa become diagonal $(\Gamma_{Y_u^{ff'}}) = \kappa_{Y_u} y_{u_f}^2 T^4 \delta_{ff'}$ while those for the down-type quark Yukawa have off-diagonal elements.
The unitarity of the CKM matrix implies $\Gamma_{Y_d^{f1}} + \Gamma_{Y_d^{f2}} + \Gamma_{Y_d^{f3}} = \kappa_{Y_d} y_{d_f}^2 T^4$.
As can be seen from Eqs.~\eqref{eq:ws}, \eqref{eq:ss}, \eqref{eq:Ye}, \eqref{eq:Yu},  and \eqref{eq:Yd}, we have $17$ charge vectors in this basis.
Out of $17$, $12$ charge vectors are linearly independent since we have the following $5$ relations among the charge vectors:
\begin{align}
	&n^\text{SS}_i = n^{Y_u^{11}}_i + n^{Y_u^{22}}_i + n^{Y_u^{33}}_i + n^{Y_d^{11}}_i + n^{Y_d^{22}}_i +  n^{Y_d^{33}}_i\,, \quad
	n^{Y_d^{11}}_i + n^{Y_d^{22}}_i +  n^{Y_d^{33}}_i = n^{Y_d^{31}}_i + n^{Y_d^{12}}_i +  n^{Y_d^{23}}_i\,, \nonumber \\
	&n^{Y_d^{11}}_i + n^{Y_d^{22}}_i = n^{Y_d^{12}}_i + n^{Y_d^{21}}_i\,, \quad
	n^{Y_d^{22}}_i + n^{Y_d^{33}}_i = n^{Y_d^{23}}_i + n^{Y_d^{32}}_i\,,  \quad
	n^{Y_d^{11}}_i + n^{Y_d^{33}}_i = n^{Y_d^{13}}_i + n^{Y_d^{31}}_i\,.
	\label{eq:relations-ew}
\end{align}
Therefore, the charge vectors of interactions span a $12$-dimensional subspace out of $16$, which indicates the presence of $4$ conserved quantities.
The $4$ charge vectors orthogonal to the charge vectors of interactions correspond to U$(1)_Y$, U$(1)_{B-L}$, U$(1)_{L_1 - L_2}$, U$(1)_{L_2 - L_3}$:
\begin{align}
	&( n_i^{Q_Y} ) = \left( -1,-1,-1,- \frac{1}{2},- \frac{1}{2},- \frac{1}{2},\frac{2}{3},\frac{2}{3},\frac{2}{3}, - \frac{1}{3}, - \frac{1}{3}, - \frac{1}{3}, \frac{1}{6}, \frac{1}{6}, \frac{1}{6}, \frac{1}{2} \right)\,, \nonumber \\
	&( n_i^{Q_{B-L}} ) = \left( -1,-1,-1,-1,-1,-1,\frac{1}{3},\frac{1}{3},\frac{1}{3},\frac{1}{3},\frac{1}{3},\frac{1}{3},\frac{1}{3},\frac{1}{3},\frac{1}{3},0 \right)\,, \nonumber\\
	&( n_i^{Q_{L_1 - L_2}} ) = ( 1,-1,0,1,-1,0,0,0,0,0,0,0,0,0,0,0 )\,, \nonumber \\
	&( n_i^{Q_{L_2 - L_3}} ) = ( 0, 1, -1, 0, 1, -1, 0,0,0,0,0,0,0,0,0,0 )\,.
	\label{eq:consv_ewscale}
\end{align}
The set of (linearly independent) $12$ charge vectors and $4$ conserved charge vectors \eqref{eq:consv_ewscale} forms a complete basis of the $16$-dimensional chemical potential space.

\paragraph{$\bm{T \gtrsim 10^{13}}$\,GeV.}
In Sec.~\ref{sec:b-l}, we discuss the $B-L$ asymmetry generation from the dimension five Weinberg operator, which gives the origin of the neutrino masses.
Since the Weinberg operator is efficient above $T \sim 10^{13}\, \mathrm{GeV}$, we are interested in the properties of the SM transport equation at this high temperature regime.
In this regime, many of the SM interactions are not efficient, and
only the following interactions are relevant:
the top Yukawa and the strong sphaleron are efficient;
the electroweak sphaleron, the bottom and tau Yukawa are marginal.

Let us briefly mention an appropriate field basis to treat the quantum coherence from different flavors.
For $T \sim 10^{13}\, \mathrm{GeV}$, the relevant quark Yukawa interactions are only the top and bottom Yukawas, and hence we can take a field basis of quarks which completely diagonalizes both the up/down-type Yukawa matrices: $y_t \overline{u}_3 Q_3 \cdot H$ and $y_b \overline{d}_3 Q_3 H^\dag$.
Aside from these top and bottom Yukawa interactions, no interactions distinguish different flavors.
Therefore, we expect that charges for $Q_3$, $u_3$, and $d_3$ in this field basis would differ from other quarks while the charges for the first and second generation quarks are the same.
We should take this field basis since otherwise we need to take into account the coherence of different flavors which is beyond the formalism developed in this paper as we see below.
The subtleties arise when one would like to use another field basis $d'_f$ which does not diagonalize the Yukawa interactions.
As an illustration, let us suppose that we take the field basis of the third generation down-type quark where the bottom Yukawa is not diagonal, \textit{i.e.}, $y_b \overline{d}_3 Q_3 H^\dag$ with $d_3 = \sum_f U_{3f} d_f'$.
As explained above, we expect a different charge density for a particular linear combination of $d_f'$, \textit{i.e.}, $d_3 = \sum_f U_{3f} d_f'$.
Therefore we need to describe the evolution of ``charges'' among different flavors for $d_f'$
because $Q_{d_3} = \int_{\bm{x}}\sum_{f,f'} U^\dag_{f3} U_{3f'} d_f^{\prime\dag} d_{f'}' \equiv \int_{\bm{x}} \sum_{f,f'} U^\dag_{f3} U_{3f'} Q_{d_{ff'}'}$.
Our transport equation is not applicable to this $d_f'$ basis because we assume that the charge densities do not develop coherence among different flavors.
A sophisticated formalism to deal with this quantum coherence has been developed in the context of flavored leptogenesis. See Refs.~\cite{Abada:2006fw,Nardi:2006fx,Abada:2006ea,Dev:2017trv} for more details.

Now we are ready to discuss conserved quantities.
As explained, in this temperature regime, the first and second generation left-handed leptons are indistinguishable. The same statement holds for the the first and second generation left-/right-handed quarks.
One may take common chemical potentials for them, \textit{i.e.}, $\mu_{L_1} = \mu_{L_2} = \mu_{L_{12}}$, $\mu_{Q_1} = \mu_{Q_2} = \mu_{Q_{12}}$, $\mu_{u_1} = \mu_{u_2} = \mu_{u_{12}}$, and $\mu_{d_1} = \mu_{d_2} = \mu_{d_{12}}$.
The first and second generation right-handed leptons are decoupled from all the interactions relevant for their asymmetry production, and hence their corresponding charges $Q_{e_f}$ with $f = 1,2$ become separately conserved quantities.
Therefore, we can focus on the chemical potentials of $10$ species, \textit{i.e.}, $\mu_i$ with $i = \tau, L_{12}, L_3, u_{12}, t, d_{12}, b, Q_{12}, Q_3, H$.
The multiplicity factor is given by $g_i = 1, 4, 2, 6, 3, 6, 3, 12, 6, 4$ respectively.
The charge vectors of each interaction in this basis are
\begin{align}
	&( n_i^\text{WS} ) = ( 0, 2, 1, 0,0,0,0, 6, 3, 0 )\,, \quad
	( n_i^\text{SS} ) = ( 0,0,0, -2, -1, -2, -1, 4, 2, 0 )\,, \quad
	( n_i^{Y_\tau} ) = ( -1,0,1,0,0,0,0,0,0,1 )\,,\nonumber\\
	&( n_i^{Y_t} )  = ( 0,0,0,0,-1,0,0,0,1,1 )\,, \quad
	( n_i^{Y_b} ) = ( 0,0,0,0,0,0, -1, 0, 1, -1 )\,.
	\label{eq:basis_WOscale}
\end{align}
These linearly independent vectors span a $5$-dimensional subspace out of $10$.
The remaining $5$ vectors orthogonal to Eq.~\eqref{eq:basis_WOscale} correspond to U$(1)_Y$, U$(1)_{B-L}$, U$(1)_{u_{12}-d_{12}}$, U$(1)_{L_{12}-2 L_3}$, and U$(1)_{B_{12}-2 B_3}$:
\begin{align}
	&( n_i^{Q_Y} ) = \left( -1, - \frac{1}{2}, - \frac{1}{2}, \frac{2}{3}, \frac{2}{3}, - \frac{1}{3}, - \frac{1}{3}, \frac{1}{6}, \frac{1}{6}, \frac{1}{2} \right)\,, \quad
	( n_i^{Q_{B-L}} ) = \left( -1, -1, -1, \frac{1}{3}, \frac{1}{3}, \frac{1}{3}, \frac{1}{3}, \frac{1}{3}, \frac{1}{3}, 0 \right)\,, \nonumber \\
	&( n_i^{Q_{u_{12} - d_{12}}} ) = ( 0,0,0,1,0,-1,0,0,0,0 )\,,\quad
	( n_i^{Q_{L_{12} - 2 L_3}} ) = ( -2, 1, -2, 0,0,0,0,0,0,0 )\,, \nonumber\\
	&( n_i^{Q_{B_{12} - 2 B_3}} ) = \left( 0,0,0, \frac{1}{3}, -\frac{2}{3}, \frac{1}{3}, -\frac{2}{3},\frac{1}{3}, -\frac{2}{3}, 0 \right)\,.
	\label{eq:consv_WOscale}
\end{align}
The set of $5$ charge vectors \eqref{eq:basis_WOscale} and conserved charge vectors \eqref{eq:consv_WOscale} forms a complete basis of the $10$-dimensional space of chemical potentials.

\section{Spontaneous $B+L$-genesis before the electroweak phase transition}
\label{sec:b+l}

Since the $B+L$ symmetry is violated by the electroweak sphaleron within the SM, it is tempting to discuss the possibility where the present-day baryon asymmetry is generated via this process.
At high temperature but below $T_\text{WS}$, the electroweak sphaleron is efficient and could source the $B+L$ asymmetry.
After the electroweak symmetry breaking, its rate per unit time is exponentially suppressed and $B+L$ becomes an approximately conserved quantity.
Therefore, if we could generate the $B+L$ asymmetry right before the electroweak phase transition, the resulting asymmetry can explain the present baryon density.
The minimal scenario in this context is electroweak baryogenesis, which is unfortunately excluded by the observed Higgs mass and the lack of the sufficient $CP$-violation in the CKM matrix.
However, as is known in the literature, the presence of an axion can reopen the possibility of baryogenesis at the electroweak phase transition~\cite{Servant:2014bla,Jeong:2018jqe,Co:2019wyp,Croon:2019ugf}.

In this section, we consider spontaneous $B+L$-genesis prior to the electroweak transition.
Suppose that the axion, which couples to the SM particles with (classically) shift symmetric couplings, has a non-vanishing velocity around the electroweak phase transition.
Though we do not specify the mechanism, one could for instance consider the coherent axion rotation initiated by a higher dimensional explicit breaking term~\cite{Co:2019jts,Co:2019wyp} or the onset of coherent axion oscillations.
As demonstrated in Sec.~\ref{sec:basis-indep}, the non-vanishing velocity of the axion biases the chemical potentials.
Consequently, the $B+L$ asymmetry is generated by the $B+L$-violating electroweak sphaleron, which can account for the present baryon density - even if the axion is not directly coupled to the electroweak sphaleron.
We clarify the condition of the coupling to the axion in order to generate the $B+L$ asymmetry.
We will see that couplings to the axion which have seemingly nothing to do with $B+L$ current, \textit{e.g.}, the coupling to the strong sphaleron $a G \tilde G$, can generate the sufficient $B+L$ asymmetry as shown in Ref.~\cite{Co:2019wyp}.

\subsection{Basic properties of the transport equation}

\paragraph{Reduction of chemical potentials.}

As discussed in Sec.~\ref{sec:SM-charges}, all SM interactions are in equilibrium around the electroweak phase transition.
This yields four independent conserved quantities, namely $Q_{Y}, Q_{B-L}$, $Q_{L_1 - L_2}$, and $Q_{L_2-L_3}$.
Since we are interested in a situation where they have no primordial asymmetries, we have $c_Y = c_{B-L} = c_{L_1 - L_2} = c_{L_2 - L_3} = 0$ [see Eq.~\eqref{eq:conservation}].
In order to reduce the number of species in the chemical potential vectors,
 it is convenient to implement the last two conditions from the beginning.
If we do not have primordial asymmetries in $Q_{L_1 - L_2}$ and $Q_{L_2-L_3}$, leptons in different flavors have the same properties.\footnote{
	For a lepton-flavor dependent axion coupling, this is not the case.
	We restrict ourselves to a lepton-flavor independent axion coupling throughout this paper for simplicity.
}
We can take common chemical potentials, \textit{i.e.}, $\mu_{e_f} = \mu_e$, $\mu_{L_f} = \mu_L$ for $f = 1, \cdots, N_f$.

As we have seen in Sec.~\ref{sec:SM-charges}, the charge vectors of the SM interactions involve $5$ non-trivial linearly dependent relations among the charge vectors \eqref{eq:relations-ew}.
If the axion couples to operators $O_{\hat\alpha_\perp}$, which are not involved in these relations,
we can simplify the equilibrium solution \eqref{eq:matrix-equil_sol} as
$\mu_i^\text{eq} = \sum_{\hat\alpha_\perp} M^{-1}_{i \hat\alpha_\perp} n_S^{\hat \alpha_\perp} \dot a / f$ where the actual value of the transport coefficients does not matter.
On the other hand, if the axion couples to operators involved in these relations $O_{\hat\alpha_\parallel}$ or $O_{\alpha_\Delta}$, the equilibrium solution cannot be simplified in this way, rather we have 
$\mu_i^\text{eq} = \sum_{\hat\alpha_\perp, \hat\beta_\parallel, \gamma} M^{-1}_{i \hat\alpha_\parallel} \Gamma^{-1}_{\hat\alpha_\parallel \hat\beta_\parallel} S_{\hat\beta_\parallel \gamma} n_S^{\gamma} \dot a / f$.
Here the matrix $\sum_{\hat\beta_\parallel}\Gamma^{-1}_{\hat\alpha_\parallel \hat\beta_\parallel} S_{\hat\beta_\parallel \gamma}$ does depend on the actual value of the transport coefficients.
This explicit dependence should be dominated by the smallest interaction among linearly dependent relations because we have $\sum_{\hat\beta_\parallel}\Gamma^{-1}_{\hat\alpha_\parallel \hat\beta_\parallel} S_{\hat\beta_\parallel \gamma} \to \delta_{\hat\alpha_\parallel \gamma}$ once one of them is switched off.
Therefore, while we need to keep a value of the smallest transport coefficient, we can take others to be infinite at the end of computations.
Since we restrict ourselves to a quark-flavor independent axion coupling, \textit{i.e.,} the axion can only couple to the entire up/down-type quark Yukawa, the relation among the strong Sphaleron and quark Yukawas in Eq.~\eqref{eq:relations-ew} is quite important.
The first generation up/down-type Yukawa interactions are the smallest couplings among them.
Hence, in order to estimate the equilibrium solution at leading order, we can take common chemical potentials for the second and third generation right-handed quarks, $\mu_{u_{23}} = \mu_{u_2} = \mu_{u_3}$ and $\mu_{d_{23}} = \mu_{d_2} = \mu_{d_3}$, while those for the first generation take different values.
Moreover, we can take $\mu_Q = \mu_{Q_1} = \mu_{Q_2} = \mu_{Q_3}$
since they are related by $\alpha = Y_d^{3f}$ and $Y_d^{2f}$ that are controlled by
the second and third generation down-type Yukawa couplings.

As a result, we can reduce the number of species in the chemical potential $\mu_i$ from $16$ to $8$ as $i = e, L, u_1, u_{23}, d_1, d_{23}, Q, H$.
The corresponding multiplicity factor is $g_i = 3, 6, 3,6, 3,6, 18, 4$ respectively.
One may readily read off the charge vectors in this basis from Eqs.~\eqref{eq:ws}, \eqref{eq:ss}, \eqref{eq:Ye}, \eqref{eq:Yu}, and \eqref{eq:Yd}:
\begin{align}
	&(n_i^\text{WS}) = ( 0, 3, 0, 0, 0, 0, 9, 0 )\,, \quad
	(n_i^\text{SS}) = (0, 0, -1,-2, -1,-2, 6, 0)\,, \quad
	(n_i^{Y_e}) = ( -1, 1, 0, 0, 0, 0, 0, -1)\,, \nonumber \\
	&(n_i^{Y_{u_1}}) = ( 0, 0, -1, 0, 0, 0, 1, 1 )\,, \quad
	(n_i^{Y_{u_{23}}}) = ( 0, 0, 0, -1, 0, 0, 1, 1 )\,, \nonumber \\
	&(n_i^{Y_{d_1}}) = ( 0, 0, 0, 0, -1, 0, 1, -1 )\,, \quad
	(n_i^{Y_{d_{23}}}) = ( 0, 0, 0, 0, 0, -1, 1, -1 )\,.
\end{align}
Here we have $n_i^\text{SS} = n_i^{Y_{u_1}} + 2 n_i^{Y_{u_{23}}} + n_i^{Y_{d_1}} + 2 n_i^{Y_{d_{23}}}$.
Two conserved quantities corresponding to $Q_Y$ and $Q_{B-L}$ provide
\begin{align}
	(n_i^{Q_Y}) = \left( -1, - \frac{1}{2}, \frac{2}{3}, \frac{2}{3}, - \frac{1}{3}, - \frac{1}{3}, \frac{1}{6}, \frac{1}{2} \right)\,, \quad
	(n_i^{Q_{B-L}}) = \left( -1, -1, \frac{1}{3}, \frac{1}{3}, \frac{1}{3}, \frac{1}{3}, \frac{1}{3}, 0 \right)\,.
\end{align}

\paragraph{Transport matrix.}
Here we provide explicit forms of matrices 
$\Gamma_{\hat\alpha \hat\beta}$ and $S_{\hat\alpha \beta}$ that are useful in obtaining equilibrium solution.
Throughout this section, we choose the complete basis of the charge vectors for interactions as $n_i^{\hat\alpha}$ with $\hat\alpha = \text{WS}, Y_e, \text{SS}, Y_{u_{23}}, Y_{d_{1}}, Y_{d_{23}}$.
Together with $n_i^A$ with $A = Q_Y, Q_{B-L}$, they form a complete basis, and
it is straightforward to compute its dual basis $\bar n_i^X$.
From this, we obtain the inverse matrix of $M_{Xi}$ in Eq.~\eqref{eq:matrix-equil} 
as\small
\begin{align}
	(M_{iX}^{-1}) = 
	\begin{pmatrix}
		\frac{22}{237} & -\frac{55}{79} & -\frac{13}{79} & 0 & \frac{15}{79} & \frac{30}{79} & -\frac{5}{79} & -\frac{3}{79} \\
 \frac{25}{237} & \frac{33}{158} & -\frac{4}{79} & 0 & -\frac{9}{158} & -\frac{9}{79} & \frac{3}{158} & -\frac{7}{79} \\
 \frac{7}{79} & -\frac{13}{79} & -\frac{206}{237} & 2 & \frac{61}{79} & \frac{122}{79} & \frac{6}{79} & -\frac{5}{237} \\
 \frac{7}{79} & -\frac{13}{79} & \frac{31}{237} & -1 & -\frac{18}{79} & -\frac{36}{79} & \frac{6}{79} & -\frac{5}{237} \\
 \frac{5}{79} & \frac{2}{79} & -\frac{23}{237} & 0 & -\frac{58}{79} & \frac{42}{79} & -\frac{7}{79} & \frac{19}{237} \\
 \frac{5}{79} & \frac{2}{79} & -\frac{23}{237} & 0 & \frac{21}{79} & -\frac{37}{79} & -\frac{7}{79} & \frac{19}{237} \\
 \frac{6}{79} & -\frac{11}{158} & \frac{4}{237} & 0 & \frac{3}{158} & \frac{3}{79} & -\frac{1}{158} & \frac{7}{237} \\
 \frac{1}{79} & -\frac{15}{158} & \frac{9}{79} & 0 & -\frac{39}{158} & -\frac{39}{79} & \frac{13}{158} & -\frac{4}{79}
	\end{pmatrix}\,,
	\label{eq:Minv_sec4}
\end{align}
\normalsize
and also transport matrices $\Gamma_{\hat\alpha \hat\beta}$ and $S_{\hat\alpha\beta}$ in Eq.~\eqref{eq:transport_matrix} as\tiny
\begin{align}	
	(\Gamma_{\hat\alpha \hat\beta}) = 
	\begin{pmatrix}
		\Gamma_\text{WS} & & & & & \\
 & \Gamma_{Y_e} & & & & \\
 & & \Gamma_\text{SS}+\Gamma_{Y_{u_1}}& -2 \Gamma_{Y_{u_1}}& -\Gamma_{Y_{u_1}}& -2 \Gamma_{Y_{u_1}}\\
 & & -2 \Gamma_{Y_{u_1}}& 4 \Gamma_{Y_{u_1}}+\Gamma_{Y_{u_{23}}} & 2 \Gamma_{Y_{u_1}}& 4 \Gamma_{Y_{u_1}}\\
 & & -\Gamma_{Y_{u_1}}& 2 \Gamma_{Y_{u_1}}& \Gamma_{Y_{d_1}}+\Gamma_{Y_{u_1}}& 2 \Gamma_{Y_{u_1}}\\
 & & -2 \Gamma_{Y_{u_1}}& 4 \Gamma_{Y_{u_1}}& 2 \Gamma_{Y_{u_1}}& \Gamma_{Y_{d_{23}}}+4 \Gamma_{Y_{u_1}}
	\end{pmatrix}\,, \quad	
	(S_{\hat\alpha \beta}) = 
	\begin{pmatrix}
		\Gamma_\text{WS} &  &  &  &  &  &  \\
  & \Gamma_{Y_e} &  &  &  &  &  \\
  &  & \Gamma_\text{SS} &  &  &  & \Gamma_{Y_{u_1}} \\
  &  &  & \Gamma_{Y_{u_{23}}} &  &  & -2 \Gamma_{Y_{u_1}} \\
  &  &  &  & \Gamma_{Y_{d_1}} &  & -\Gamma_{Y_{u_1}} \\
  &  &  &  &  & \Gamma_{Y_{d_{23}}} & -2 \Gamma_{Y_{u_1}}
	\end{pmatrix}\,.
	\label{eq:GammaS-sec4}
\end{align}
\normalsize
Here $\Gamma_{Y_{u_{23}}}$ and $\Gamma_{Y_{d_{23}}}$ in this matrix may be expressed as functions of $\Gamma_{Y_{u_f}}$ and $\Gamma_{Y_{d_f}}$ with $f=2,3$ because we have taken common chemical potentials for the second and third generation right-handed quarks.
As  explained, the actual values of the transport coefficients only matter if the axion couples to an operator whose charge vector belongs to the set of linearly dependent charge vectors.
Moreover, to evaluate the equilibrium solution at leading order in this case, we only need to keep the smallest interactions to be finite while taking the others to infinity at the end of the computation.
Therefore, the precise values of $\Gamma_{Y_{u_{23}}}$ and $\Gamma_{Y_{d_{23}}}$ are not important as long as $\Gamma_{Y_{u_{1}}}, \Gamma_{Y_{d_{1}}} \ll \Gamma_{Y_{u_{23}}},\Gamma_{Y_{d_{23}}}$, which is always fulfilled in our case because of $y_{u_1}, y_{d_1} \ll y_{u_2}, y_{u_3}, y_{d_2}, y_{u_3}$.
One can check this explicitly starting from the full $16 \times 16$ matrices and taking $y_{u_1}, y_{d_1} \ll y_{u_2}, y_{u_3}, y_{d_2}, y_{u_3}$ at the end of the computation.

\subsection{Equilibrium solution including the axion}
Now we are ready to discuss the equilibrium solution for the chemical potentials $\mu_i$ in the presence of an axion with non-vanishing $\dot a$.
From this, we get the condition of the axion coupling in order to generate a baryon asymmetry.
We also discuss the condition so that the axion is not stopped by the backreaction.

\paragraph{Condition for baryogenesis.}
The $B+L$ asymmetry is given by
\begin{align}
 	q_{B+L} = \mu_{B+L} \frac{T^2}{6}  \qquad \text{with} \quad \mu_{B+L} 
 	= 3\left(\mu_e + 2\mu_L\right) + \mu_{u_1} + \mu_{d_1} + 2\left(\mu_{u_{23}} + \mu_{d_{23}}\right) + 6\mu_Q\,.
\end{align}
The equilibrium solution for the chemical potentials $\mu_i$ is given by Eq.~\eqref{eq:matrix-equil_sol}, with the matrices $M_{iX}$, $\Gamma_{\hat \alpha \hat \beta}$ and $S_{\hat \alpha \beta}$ given in Eqs.~\eqref{eq:Minv_sec4} and \eqref{eq:GammaS-sec4}. Let's suppose for simplicity that we do not have any primordial asymmetries for $q_y$ or $q_{B-L}$, \textit{i.e.}, $c_{Q_y} = c_{Q_{B-L}} = 0$. The baryon asymmetry can thus be expressed as a linear combination of the source terms appearing on the right-hand side of Eq.~\eqref{eq:matrix-equil}, incorporating the couplings to the axion. A non-zero baryon asymmetry is generated as long as the source vector is \textit{not} orthogonal to the direction in $\alpha$-space which is subject to baryon number changing interactions, as derived in Eq.~\eqref{eq:cond_asym}. 
As mentioned, we assume that the axion couples to the SM particles in a flavor independent way, which means that the source vectors fulfill $n_S^{Y_u} = n_S^{Y_{u_1}} = n_S^{Y_{u_{23}}}$ and $n_S^{Y_d} = n_S^{Y_{d_1}} = n_S^{Y_{d_{23}}}$.
Inserting the expressions in Eqs.~\eqref{eq:Minv_sec4} and \eqref{eq:GammaS-sec4} we obtain the condition for generating a $B+L$ asymmetry:
\begin{align}
 \left(n_S^\text{WS}, n_S^{Y_e}, n_S^\text{SS}, n_S^{Y_d}, n_S^{Y_u} \right) \not\perp  v_\gamma^{B+L}
\end{align}
with
\begin{align}
 v_\gamma^{B+L} 
 &\simeq
  \frac{6}{79} \left(24, -22,  \frac{- 3 (7 \Gamma_{Y_{d_1}} + 5 \Gamma_{Y_{u_1}})}{\Gamma_{Y_{u_1}} + \Gamma_{Y_{d_1}}}, \frac{18 \Gamma_{Y_{d_1}}}{\Gamma_{Y_{u_1}} + \Gamma_{Y_{d_1}}},\frac{-18 \Gamma_{Y_{u_1}}}{\Gamma_{Y_{
  u_1}} + \Gamma_{Y_{d_1}}}   \right) 
\end{align}
The appearance of the interaction rates for the strong sphaleron and up/down-type Yukawas in the last three entries in the first line is due to the linear dependence between the respective charge vectors, as discussed above.
Here we have used the fact that $\Gamma_{Y_{u_1}},\Gamma_{Y_{d_1}} \ll \Gamma_\text{SS}, \Gamma_{Y_{u_{23}}}, \Gamma_{Y_{d_{23}}}$.  From Eq.~\eqref{eq:equilibrium_solution}, the equilibrium solution for the $B+L$ asymmetry is now immediately obtained as
\begin{align}
 \mu_{B+L}^\text{eq} = \sum_\gamma v_\gamma^{B+L} n^\gamma_S \frac{\dot a}{f} \,.
\end{align}

To give some concrete examples, the coupling to the electroweak sphaleron, $(n_S^\alpha) = (1,0,0,0,0)$, or a direct coupling to the $B+L$ current [see below \eq{eq:sourcevec}], 
\begin{align}
 (n_S^{\alpha}) 
 &= \sum_i n_i^{Q_{B+L}} (n_i^\alpha) 
 \nonumber\\
 &= (n^\alpha_{e}) + (n^\alpha_L) + \frac{1}{3} ( n_u^\alpha + n_d^\alpha + n_Q^\alpha ) = (6,0,0,0,0)\,,
 \label{eq:nS_B+L}
\end{align}
clearly satisfy the condition for generating a baryon asymmetry. This is not surprising since both operators violate $B+L$. 
By performing a $B+L$ rotation of the SM fermions, the coupling to the electroweak sphaleron can be rewritten as the coupling to the $B+L$ current. The above two charge vectors $n_S^\alpha$ coincide up to an overall factor reflecting the invariance under this field rotation.

According to the condition above, a coupling to the strong sphaleron $(n_S^\alpha) = (0,0,1,0,0)$, the lepton Yukawa $(n_S^\alpha) = (0,1,0,0,0)$, and the up/down-type quark Yukawas $(n_S^\alpha) = (0,0,0,0,1), (0,0,0,1,0)$ will also generate a baryon asymmetry. The coupling to the strong sphaleron $a G \tilde G$ is particularly interesting because it is present in QCD axion models.
These examples are more surprising since these operators do not violate $B+L$.
However, they generate an asymmetry for the left-handed leptons/quarks, which can then be converted into a baryon asymmetry by the electroweak sphaleron.

More generally, this result explicitly demonstrates that a generic shift-symmetric coupling of an axion to SM particles typically generates a baryon asymmetry - in fact there is only one particular linear combinations of operators which, when coupled to the axion, does not source a baryon asymmetry.  This is because, unless we choose a very specific coupling such that the electroweak sphaleron is not involved in achieving the equilibrium with $\dot a \neq 0$, the baryon asymmetry is generated.
Since there is no reason for this specific coupling to be realized, we conclude that the generation of the baryon asymmetry is a generic consequence of the axion coupling to the SM particles if the homogeneous axion velocity is non-vanishing at the electroweak phase transition.

\paragraph{Backreaction to the axion.}
Let us briefly discuss the effective friction term \eqref{eq:axion_dissipation} for the axion.\footnote{
	Throughout this paper, we assume that the SM particles are in equilibrium. This, however, implicitly assumes that the tachyonic instability of the gauge field via the Chern-Simons coupling $a W \tilde W$ is suppressed.
	In our case, this assumption is fulfilled because the typical axion velocity we have in mind is small, $\dot a / f T \sim 10^{-10}$, and the non-abelian gauge field acquires the magnetic mass term from the ambient plasma~(see \textit{e.g.}\cite{Hook:2016mqo}).
}
As shown in Eq.~\eqref{eq:zero-friction}, the effective friction term vanishes identically if the axion couples to the the electroweak Chern-Simons term or the lepton Yukawa:
\begin{align}
	\gamma^\text{eff}_{a, \text{WS}} = \gamma^\text{eff}_{a, Y_e} = 0\,.
\end{align}
On the other hand, the charge vectors for the strong sphaleron and the up/down-type quark Yukawas are linearly dependent: $n_i^\text{SS} = n_i^{Y_{u_1}} + 2 n_i^{Y_{u_{23}}} + n_i^{Y_{d_1}} + 2 n_i^{Y_{d_{23}}}$.
Hence, if the axion couples to these operators, the effective friction term becomes non-zero (for $\gamma^\text{eff}_{a, \text{SS}} $ see also Ref.~\cite{McLerran:1990de,Co:2019wyp}):
\begin{align}
	\gamma^\text{eff}_{a, \text{SS}} 
	\simeq \frac{1}{f^2 T} \frac{1}{\Gamma_{Y_{u_1}}^{-1} + \Gamma_{Y_{d_1}}^{-1}} \,,\quad
	\gamma^\text{eff}_{a, Y_u} = \gamma^\text{eff}_{a, Y_d} 
	\simeq \frac{1}{f^2 T} \frac{9}{\Gamma_{Y_{u_1}}^{-1} + \Gamma_{Y_{d_1}}^{-1}}\,.
\end{align}
Here again we have used $\Gamma_{Y_{u_1}},\Gamma_{Y_{d_1}} \ll \Gamma_\text{SS}, \Gamma_{Y_{u_{23}}}, \Gamma_{Y_{d_{23}}}$.
One can see that all of them have a similar value, \textit{i.e.}, $\gamma^\text{eff}_{a, \text{SS}} \sim \gamma^\text{eff}_{a, Y_{u/d}} \sim \kappa_{Y_u} y_u^2 T^3 / f^2$.
By comparing it with the Hubble parameter, we get the following condition for neglecting the backreaction:
\begin{align}
	\frac{f^2}{T} \gtrsim 10^6 \GeV\,.
\end{align}
Restricting the discussion to below the Peccei-Quinn breaking scale, $T/f \lesssim 1$, this implies that the backreaction can be neglected for $f \gtrsim 10^6$~GeV.

\section{Spontaneous $B-L$-genesis around the reheating epoch}
\label{sec:b-l}
In this section, we consider an example of spontaneous baryogenesis
at $T \sim 10^{13}\,\mathrm{GeV}$, \textit{i.e.}, during a much earlier epoch than the previous example in Sec.~\ref{sec:b+l}.
It is well-known that the SM left-handed neutrinos are massive, which cannot be explained
within the dimension four operators of the SM.
A simple way to explain the neutrino masses is to introduce the dimension five Weinberg operator 
(suppressing species indices) as
\begin{align}
	\mathcal{L}_{\nu} = - \frac{m_{\nu}}{2v_{\text{EW}}^{2}} \left( L \cdot H \right)^{2} + \text{H.c.}\,,
	\label{eq:LHLH}
\end{align}
where $m_\nu$ is the mass of the left-handed neutrino and $v_\mathrm{EW} \simeq 174\,\mathrm{GeV}$ 
is the Higgs vacuum expectation value.
This operator provides effective masses for the left-handed neutrinos after the electroweak symmetry breaking,
and  may be obtained from integrating out heavy right-handed neutrinos. Being a dimension five operator, the Weinberg operator becomes more effective at high temperatures.
As it violates lepton number, 
it (with the help of an axion) 
can be a source of $B-L$ asymmetry in the early universe.

An overview of our $B-L$-genesis scenario in this section is as follows.
We introduce an axion and its shift symmetric coupling to the SM sector,
\textit{e.g.}, $a W\tilde{W}$ or $a G\tilde{G}$.
Suppose that the axion develops a non-vanishing velocity before the Weinberg operator decouples from equilibrium.
The chemical potentials for the SM particles are then biased toward nonzero values via the shift-symmetric couplings. 
As a result, a $B-L$ asymmetry is generated by the lepton number violating Weinberg operator. 
As we will see shortly, 
the lepton number violating interaction  decouples at the temperature of order $10^{13}$\,GeV. 
If the axion keeps moving until this moment,
the produced $B-L$ asymmetry is never washed out afterwards,
and is eventually converted to the baryon asymmetry of the present universe.

More explicitly, the baryon asymmetry in the present-day Universe, $Y_B$
(= $9 \times 10^{-11}$ from observation~\cite{Ade:2015xua}),
is given in terms of the final $B-L$ asymmetry as
\begin{align}
 Y_B = \frac{q_B}{s} =  \frac{T^3}{6 \, s} \frac{\mu_B}{T} = \frac{C_\text{sph} T^3}{6 \, s} \frac{\mu_{B-L}}{T} \simeq 10^{-3} \, \frac{\mu_{B-L}}{T}
 \label{eq:B-LtoB}
\end{align}
where $s= 2 \pi^2/45 g_* T^3$ denotes the entropy of the thermal bath with $g_{*,0} = 106.75$ counting the effective degrees of freedom, and $C_\text{sph} = 28/79$ indicates the sphaleron conversion factor translating the $B-L$ asymmetry into a baryon asymmetry at the electroweak phase transition.

In this section, we compute the resulting $B-L$ asymmetry well after the decoupling of the Weinberg operator.
We also clarify the condition of the coupling to the axion so that the $B-L$ asymmetry is generated.
We will see, for instance, the coupling to the strong sphaleron, which at first glance has nothing to do with $B-L$ or $B+L$ charges, 
can produce a sufficient $B-L$ asymmetry.

\subsection{Transport equation including the Weinberg operator}
\label{subsec:b-l_transport_eq}

\paragraph{Weinberg operator.}
Here we summarize the basic properties of the Weinberg operator~\eqref{eq:LHLH}.
We assume that it is flavor-universal for simplicity.
Then the rate per unit volume is also flavor-blind and is estimated as
\begin{align}
	\Gamma_\mathrm{W} = \kappa_{\rm W} \frac{m_{\nu}^2 T^6}{v_{\text{EW}}^{4}}. 
	\label{Gamma_W}
\end{align}
where $\kappa_{\rm W} \sim 3 \times 10^{-3}$. 
We define the decoupling temperature of the lepton number violating process mediated by the flavor-universal Weinberg operator, $T_{\rm W}$,
by looking at the coefficient of the transport equation for the total lepton number density: 
\begin{align}
	- \frac{\dd}{\dd \ln T} \lmk 2 \frac{\mu_{L_1} +\mu_{L_2} + \mu_{L_3}}{3T} - 2 \frac{\mu_H}{T} \rmk 
	= - \sum_i \frac{1}{g_i} \lmk n_{i}^{\rm W} \rmk^2 \frac{3 \gamma_{\rm W}}{H}
 \lmk 2 \frac{\mu_{L_1} +\mu_{L_2} + \mu_{L_3}}{3T} - 2 \frac{\mu_H}{T} \rmk 
+ \dots, 
\end{align}
We thus define the decoupling temperature by $5 \gamma_{W} = H$. 
It is calculated as 
\begin{align}
	T_\mathrm{W} \simeq 6\times 10^{12}\,\mathrm{GeV}\times \left(\frac{0.05\,\mathrm{eV}}{m_\nu}\right)^2. 
	\label{eq:T_W}
\end{align}
Note that the lepton number violating interaction is in thermal equilibrium when the temperature is {\it higher} than $T_{\rm W}$. 
On the other hand, the other (SM) interactions $\alpha$ (the sphalerons and the Yukawa interactions) are in thermal equilibrium when the temperature is {\it lower} than $T_\alpha$.
This is the reason why we refer to $T_\mathrm{W}$ as the decoupling temperature as opposed to the term equilibration temperature used for the other interactions.

\paragraph{Transport equation.}
We are interested in the transport equation around the temperature of $T \sim T_W \sim 10^{13}\,\mathrm{GeV}$.
As we discussed in Sec.~\ref{sec:SM-charges}, 
we can focus on the chemical potentials of $10$ species
at such a high temperature,, \textit{i.e.}, $\mu_i$ with $i = \tau, L_{12}, L_3, u_{12}, t, d_{12}, b, Q_{12}, Q_3, H$.
We further assume that there is no initial charge asymmetry between $u_{12}$ and $d_{12}$, or $c_{u_{12}-d_{12}} = 0$,
in this section.
It allows us to combine $u_{12}$ and $d_{12}$ as $q_{12}$.
In summary, the chemical potentials of our interest are $\mu_i$ with
\begin{align}
	i = \tau, ~L_{12}, ~L_3, ~q_{12}, ~t, ~b, ~Q_{12}, ~Q_3, ~H,
\end{align}
and the multiplicity factor is $g_i = 1, 4, 2, 12, 3, 3, 12, 6, 4$ respectively.
The charge vectors of the relevant interactions are\footnote{
We should note that there are three lepton number violating interactions 
though we combine two of them into a single charge vector $n_i^{W_{12}}$. 
The interaction rate should be then given by $\Gamma_{{\rm W}_{12}} = 2 \Gamma_{{\rm W}_3} = 2 \Gamma_{\rm W}$. 
}
\begin{align}
	&( n_i^\text{WS} ) = ( 0, 2, 1, 0,0,0, 6, 3, 0 )\,, \quad
	( n_i^\text{SS} ) = ( 0,0,0, -4, -1, -1, 4, 2, 0 )\,, \quad
	( n_i^{Y_\tau} ) = ( -1,0,1,0,0,0,0,0,1 )\,,\nonumber\\
	&( n_i^{Y_t} )  = ( 0,0,0,0,-1,0,0,1,1 )\,, \quad
	( n_i^{Y_b} ) = ( 0,0,0,0,0, -1, 0, 1, -1 )\,, \nonumber\\
	&( n_i^{W_{12}} )  = (0,2,0,0,0,0,0,0,2)\,, \quad
	( n_i^{W_3} ) = (0,0,2,0,0,0,0,0,2).
	\label{eq:b-l_int_vectors}
\end{align}
These linearly independent vectors span a $7$-dimensional subspace out of $9$.
Note that all the charge vectors $n_i^\alpha$ are linearly independent,
and hence the axion does not have any friction term in equilibrium.
The remaining $2$ vectors orthogonal to Eq.~\eqref{eq:b-l_int_vectors} correspond 
to U$(1)_Y$ and U$(1)_{B_{12}-2 B_3}$:
\begin{align}
	&( n_i^{Q_Y} ) = \left( -1, - \frac{1}{2}, - \frac{1}{2}, \frac{1}{6}, \frac{2}{3}, - \frac{1}{3}, \frac{1}{6}, \frac{1}{6}, \frac{1}{2} \right)\,, \quad
	( n_i^{Q_{B_{12} - 2 B_3}} ) = \left( 0,0,0, \frac{1}{3}, -\frac{2}{3}, -\frac{2}{3},\frac{1}{3}, -\frac{2}{3}, 0 \right)\,.
\end{align}
These vectors form a complete basis of the $9$-dimensional chemical potential space.
Here that $U(1)_{B-L}$ is no longer a conserved charge because of the Weinberg operator.
The transport equation of our system is given by Eq.~\eqref{eq:fulltransporteq}, which we show here again for reader's convenience:
\begin{align}
	- \frac{\dd}{\dd \ln T} \lmk \frac{\mu_i}{T} \rmk = -\frac{1}{g_i}\sum_\alpha n_{i}^{\alpha} \frac{\gamma_\alpha}{H}
	\left[\sum_{j}n_{j}^{\alpha} \lmk \frac{\mu_j}{T} \rmk - n_{S}^{\alpha} \lmk \frac{\dot{a}/f}{T} \rmk \right],
	\label{eq:transport_eq_sec5}
\end{align}
with the charge vectors $n_i^\alpha$ defined above.

Since the bottom/tau Yukawa couplings and the electroweak sphaleron are only marginally relevant at $T \sim 10^{13}\,\mathrm{GeV}$,
we may further ignore them when we discuss the equilibrium solutions in Sec.~\ref{subsec:b-l_equilibrium}.
These interactions are however fully included in our numerical results in Secs.~\ref{subsec:b-l_numerics} and~\ref{subsec:axion_model}.

\subsection{Equilibrium solution including the axion}
\label{subsec:b-l_equilibrium}

In this subsection, we discuss the equilibrium solution
to get a rough idea of the $B-L$ asymmetry generation in our system.
Our primary goal here is to derive a condition for the axion source vector $n_S^\alpha$
to obtain a non-zero $B-L$ asymmetry in equilibrium.

In this subsection, we ignore the bottom and tau Yukawa interactions in order to simplify our analysis.
The right-handed tau lepton $\tau$ then plays no role and hence we omit it.
The right-handed bottom quark $b$ can be combined with $q_{12}$ (we denote them as $q$) 
by assuming that there is no initial asymmetry between $b$ and $q_{12}$.
We can also combine $L_{12}$ and $L_3$ as $L$ by again assuming that there is no initial asymmetry between them,
since we take the lepton number violating process as flavor-universal.
Thus, the chemical potentials of our interest reduce to $\mu_i$ with
\begin{align}
	i = L, ~q, ~t, ~Q_{12}, ~Q_3, ~H,
\end{align}
and the multiplicity factors are $g_i = 6, 15, 3, 12, 6, 4$ respectively.
The charge vectors of the relevant interactions are
\begin{align}
	&( n_i^\text{WS} ) = (3, 0, 0, 6, 3, 0 )\,, \quad
	( n_i^\text{SS} ) = ( 0,-5, -1, 4, 2, 0 )\,, \nonumber\\
	&( n_i^{Y_t} )  = ( 0,0,-1,0,1,1 )\,, \quad
	( n_i^{W} ) = (2,0,0,0,0,2),
\end{align}
and the conserved charges are  $Q_Y$ and  $Q_{B_{12} - 2 B_3}$ with their charge vectors
\begin{align}
	&( n_i^{Q_Y} ) = \left(-\frac{1}{2}, \frac{1}{15}, \frac{2}{3}, \frac{1}{6}, \frac{1}{6}, \frac{1}{2} \right)\,, \quad
	( n_i^{Q_{B_{12}-2B_3}} ) = \left(0, \frac{2}{15}, -\frac{2}{3}, \frac{1}{3}, -\frac{2}{3}, 0 \right)\,.
\end{align}
As the electroweak sphaleron is only marginally relevant, we may further ignore it.
In such a case the baryon number $Q_B$ is also conserved,
whose charge vector is
\begin{align}
	n_{i}^{Q_{B}} = \left(0, \frac{1}{3}, \frac{1}{3}, \frac{1}{3}, \frac{1}{3}, 0\right).
\end{align}
The $B-L$ charge vector in this basis is expressed as 
\begin{align}
	n_{i}^{Q_{B-L}} = \left(-1, \frac{1}{3}, \frac{1}{3}, \frac{1}{3}, \frac{1}{3}, 0\right).
\end{align}
In this case, all the charge vectors of the interactions are linearly independent,
and hence we can directly apply Eq.~\eqref{eq:cond_asym_lind} 
as a condition for the source vector $n_S^\alpha$
to generate a non-zero $B-L$ asymmetry.
The condition reads
\begin{align}
	(n_S^{\mathrm{WS}}, n_S^{\mathrm{SS}}, n_S^{Y_t}, n_S^{W})
	\not\perp
	\frac{1}{174} (92, -114, 270, -345),
	\label{eq:b-l_with_ws}
\end{align}
if the electroweak sphaleron is in equilibrium, and
\begin{align}
	(n_S^{\mathrm{SS}}, n_S^{Y_t}, n_S^{W})
	\not\perp
	\frac{3}{44} (-3, 18, -23),
	\label{eq:b-l_without_ws}
\end{align}
if the electroweak sphaleron is decoupled, respectively. 
Accordingly, the $B-L$ asymmetry is given by
\begin{align}
	\frac{\mu_{B-L}^\mathrm{eq}}{T} 
	=
	\left(\frac{46}{87}n_S^{\mathrm{WS}}
	-\frac{19}{29}n_S^{\mathrm{SS}} + \frac{45}{29}n_S^{Y_t} - \frac{115}{58}n_S^{W}  \right)\frac{\dot{a}/f}{T} ,
\end{align}
if the electroweak sphaleron is in equilibrium, and

\begin{align}
	\frac{\mu_{B-L}^\mathrm{eq}}{T}
	=
	\left(-\frac{9}{44}n_S^{\mathrm{SS}} + \frac{27}{22}n_S^{Y_t} - \frac{69}{44}n_S^{W}  \right)\frac{\dot{a}/f}{T},
\end{align}
if the electroweak sphaleron is out of equilibrium, respectively.
Here we have assumed $c_Y = c_{B_{12}-2B_3} = 0$ for the former case
and $c_Y = c_{B_{12}-2B_3} = c_{B} = 0$ for the latter case.

The conditions~\eqref{eq:b-l_with_ws} and~\eqref{eq:b-l_without_ws}
tell us that, in the presence of the Weinberg operator, it is difficult \textit{not} to produce the $B-L$ asymmetry
once the axion has shift-symmetric couplings to the SM particles
which are relevant at that temperature.
In order not to produce the $B-L$ asymmetry,
the axion has to couple to the operators in a specific form such that
its source vector is orthogonal to the right hand side of Eq.~\eqref{eq:b-l_with_ws} or \eqref{eq:b-l_without_ws}.
There is no reason for this to be the case, 
and hence we conclude that the generation of the $B-L$ asymmetry is a rather generic consequence
of the axion shift-symmetric couplings to the SM particles 
if the homogeneous axion velocity is non-vanishing around $10^{13} \GeV$.

So far we have studied the equilibrium solutions.
In the next section, 
we study three concrete scenarios numerically, without assuming equilibrium.
First, we study the scenario that the axion couples to the divergence of the $B-L$ current,
a scenario often considered in the context of  spontaneous baryogenesis.\footnote{
	Here we consider the $B-L$ current, not the lepton current, to match with Ref.~\cite{Kusenko:2014uta},
	which does not incorporate the electroweak sphaleron in the transport equation.
	We have numerically checked, however, that the final $B-L$ asymmetry is almost the same 
	for these two cases (the lepton current case tends to be slightly more suppressed).
	This is because the axion directly couples to the Weinberg operator in both cases
	which gives the dominant source of the $B-L$ asymmetry generation.
}
Second, we study the coupling $a W \tilde{W}$,
which is also studied in Refs.~\cite{Kusenko:2014uta,Takahashi:2015waa,Bae:2018mlv}.
As one can see from Eq.~\eqref{eq:b-l_with_ws}, 
it can produce the $B-L$ asymmetry if the electroweak sphaleron is efficient enough.
In reality, however, the electroweak sphaleron is only marginally relevant when the Weinberg operator is efficient 
(or $T \gtrsim 10^{13}\,\mathrm{GeV}$).
Thus, the resultant $B-L$ asymmetry is expected to be suppressed compared to the above estimation 
based on the full equilibration of the electroweak sphaleron.
We will study this suppression factor numerically below.
We also clarify an issue in Refs.~\cite{Kusenko:2014uta,Takahashi:2015waa} and its relation to the basis independence.
Finally, we study the coupling $a G\tilde{G}$,
which might be the most non-trivial scenario.
We can see from Eqs.~\eqref{eq:b-l_with_ws} and~\eqref{eq:b-l_without_ws}
that a nonzero $B-L$ asymmetry is generated
even if the axion couples only to the strong sphaleron or the top Yukawa coupling, 
which by them self cannot generate baryon nor lepton asymmetry. 
Below we numerically confirm that it is also the case without assuming equilibrium.

\subsection{Numerical results}
\label{subsec:b-l_numerics}
Now we study the $B-L$-genesis at $T \sim 10^{13}\,\mathrm{GeV}$
by solving the full transport equation~\eqref{eq:transport_eq_sec5} numerically.
Although we have ignored the bottom and tau Yukawa interactions in the previous Sec.~\ref{subsec:b-l_equilibrium},
we fully take them into account in our numerical code.
Thus the chemical potentials of our interest are $\mu_i$ with
\begin{align}
	i = \tau, ~L_{12}, ~L_3, ~q_{12}, ~t, ~b, ~Q_{12}, ~Q_3, ~H,
\end{align}
and we have solved the transport equation~\eqref{eq:transport_eq_sec5} for them
by assuming that there is no asymmetry at the end of the reheating,
\begin{align}
	\mu_i(T = T_R) = 0,
\end{align}
where $T_R$ is the reheating temperature.

The axion acts as an external force in Eq.~\eqref{eq:transport_eq_sec5}.
We consider two types of the axion dynamics.
For the first case, we simply take
\begin{align}
	\frac{\dot{a}/f}{T} = \eta_0,
\end{align}
with $\eta_0$ being a constant.
We also consider a more realistic case
that the axion starts to oscillate harmonically around its potential minimum 
at $T = T_\mathrm{osc}$, and decays at $T = T_\mathrm{dec}$.
An oscillating scalar field scales as
\begin{align}
	\dot{\phi} = v(t) \sin\left(m_\phi t\right),
	\quad
	\dot{v} = -\frac{3H}{2}v.
\end{align}
Therefore, we parametrize the axion dynamics assuming radiation domination as
\begin{align}
	\frac{\dot{a}/f}{T} = \eta_0 \left(\frac{T}{T_\mathrm{osc}}\right)^{1/2}
	\sin\left[\left(\frac{T_\mathrm{osc}}{T}\right)^2 - 1\right]
	\Theta\left[\left(T_\mathrm{osc}-T\right)\left(T-T_\mathrm{dec}\right)\right],
\end{align}
where we have taken the axion mass as $m_a = 2 H(T=T_\mathrm{osc})$ and $\Theta$ is the Heaviside theta function.
Here $\eta_0$ parametrizes the initial velocity of the axion.
The final $B-L$ asymmetry is proportional to $\eta_0$ since the transport equation is linear.
Note that $T_\mathrm{osc} \gtrsim T_W \gtrsim T_\mathrm{dec}$ is needed for the $B-L$-genesis
since otherwise either the produced asymmetry is washed out after the axion decay (for $T_\mathrm{dec} \gg T_W$),
or no asymmetry is produced (for $T_\mathrm{osc} \ll T_W$).

Below we show our numerical results of the resulting $B-L$ asymmetry
for three shift-symmetric couplings: $a \partial_\mu J^\mu_{B-L}$ where $J_{B-L}^\mu$ is the $B-L$ current,
$a W\tilde{W}$ and $a G \tilde{G}$.
Since the lepton number violating process is well-decoupled at the end of our numerical computation
(that is $T = 10^{10}\,\mathrm{GeV}$),
it can be directly translated to the baryon asymmetry in the present universe.
We fix $T_R$ and $\eta_0$ as
\begin{align}
	T_R = 10^{15}\,\mathrm{GeV}\,,
	\quad
	\eta_0 = 10^{-9}\,,
	\label{eq:initial_cond}
\end{align}
and the SM parameters as
\begin{align}
	g_2 = 0.55\,,
	\quad
	g_3 = 0.60\,,
	\quad
	y_\tau = 1.0\times 10^{-2}\,
	\quad
	y_t = 0.49\,,
	\quad
	y_b = 6.8\times 10^{-3}\,,
	\quad
	m_\nu = 0.05\,\mathrm{eV}\,,
\end{align}
in our numerical results below.
For the oscillating axion case, we fix the model parameters as
\begin{align}
	T_\mathrm{osc} = 10^{13}\,\mathrm{GeV},
	\quad
	T_\mathrm{dec} = 10^{11}\,\mathrm{GeV},
\end{align}
in this subsection. 
The dependence of the final $B-L$ asymmetry on these parameters is studied in the next subsection.

\paragraph{$\bm{B-L}$ current.}

First, we consider the shift-symmetric coupling to the $B-L$ current: $(a/f) \partial_\mu J^\mu_{B-L}$.
This type of coupling is probably most common in the context of the spontaneous baryogenesis,
since it can be understood as a pure shift of the chemical potential of the lepton number charge 
as we saw in Sec.~\ref{sec:basis-indep}.
The purpose to study this coupling here is two-fold. 
First, we demonstrate how our formalism applies to this most common example.
Second, we highlight a difference between this coupling and the coupling to the electroweak sphaleron $a W\tilde{W}$,
which we study next.

Since this coupling shifts the chemical potential of the quarks and leptons, 
the axion source vector is given by
\begin{align}
	n_S^\alpha 
	= \sum_i n_i^{Q_{B-L}} n_i^\alpha = -n_{\tau}^{\alpha} - n_{L_{12}}^{\alpha} - n_{L_3}^{\alpha}
	+\frac{1}{3}\left(n_{q_{12}}^{\alpha} + n_{t}^{\alpha} + n_{b}^{\alpha} + n_{Q_{12}}^{\alpha} + n_{Q_3}^{\alpha}\right).
\end{align}
From Eq.~\eqref{eq:b-l_int_vectors}, it is given as
\begin{align}
	( n_S^{\alpha} ) = (0,0,0,0,0,-2,-2),
\end{align}
where the ordering of the interactions is $\alpha = \mathrm{WS}, \mathrm{SS}, Y_\tau, Y_t, Y_b, W_{12}, W_{3}$.
Note that it has non-zero entries only for the Weinberg operators.
This is due to the fact that they are the interactions that violate the $B-L$ symmetry, 
and hence enter into the $B-L$ current equation.
\begin{figure}[t]
	\centering
 	\includegraphics[width=0.49\linewidth]{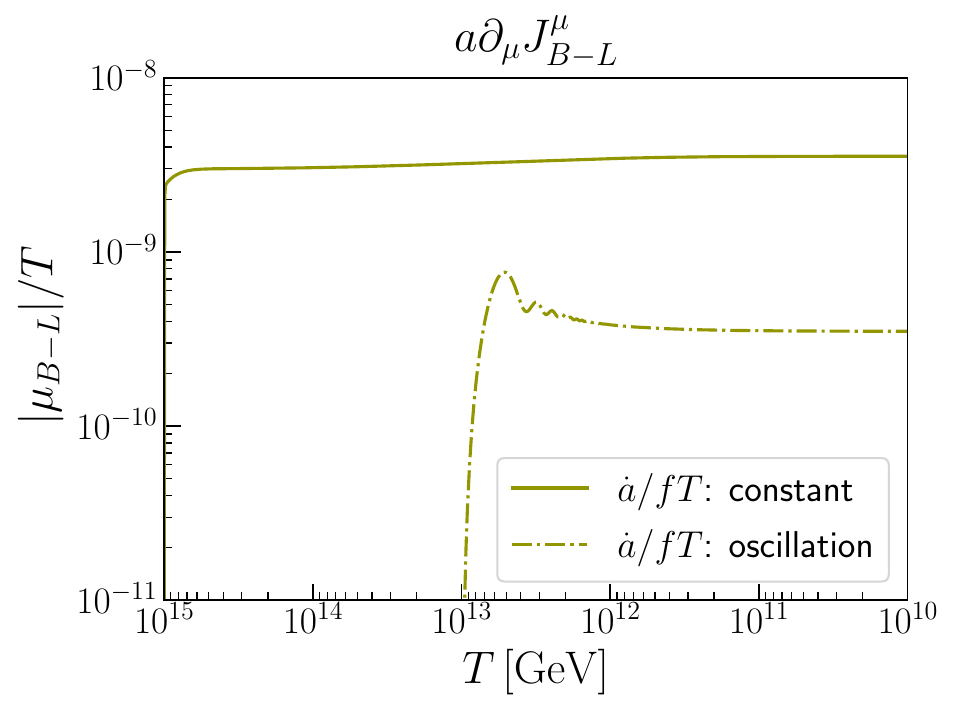} \hfill
 	\includegraphics[width=0.49\linewidth]{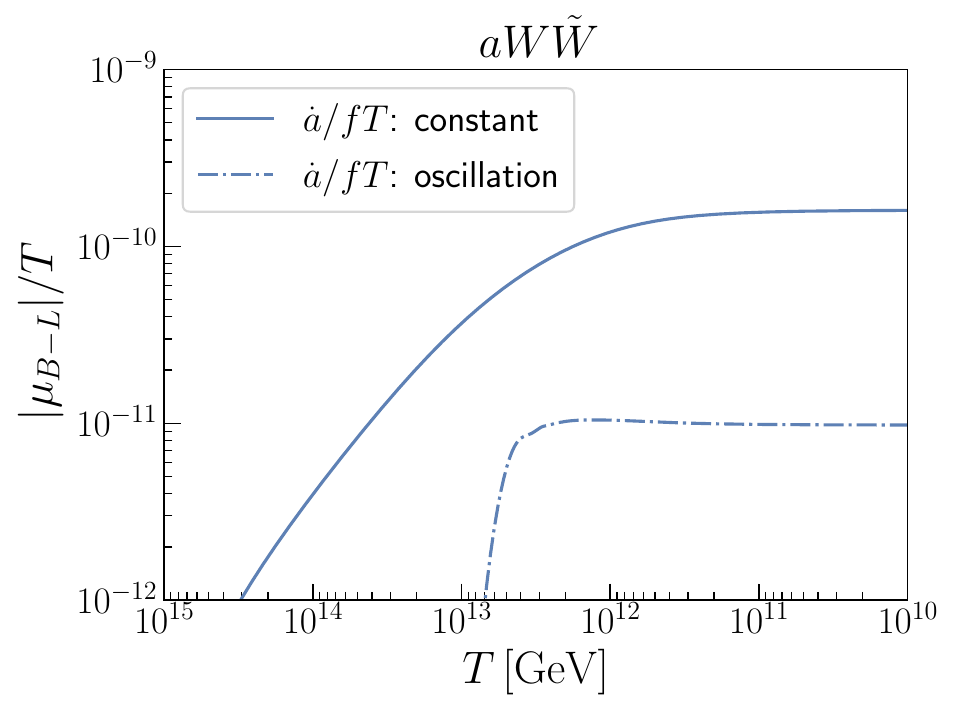}
	\caption{ 
	The time evolution of the $B-L$ asymmetry  produced from the shift-symmetric coupling $(a/f) \partial_\mu J^\mu_{B-L}$ (left panel) and  $(a/f) W \tilde{W}$ (right panel) for constant $\dot a/(f T)$ (solid) and oscillating $\dot a/(f T)$ (dashed).
	}
	\label{fig:b-l_aJBL}
\end{figure}
With this information, we can solve Eq.~\eqref{eq:transport_eq_sec5} numerically.
The results are shown in the left panel of Fig.~\ref{fig:b-l_aJBL}.
We can see from Eq.~\eqref{eq:B-LtoB} that for parameters in the ball-park of Eq.~\eqref{eq:initial_cond},  a sufficient amount of the $B-L$ asymmetry is produced from this coupling.

\paragraph{Electroweak sphaleron.}

Next, we consider the shift-symmetric coupling to the electroweak sphaleron: $(a/f) W\tilde{W}$.
The axion source vector in this case is given by
\begin{align}
	n_S^{\alpha} = (1,0,0,0,0,0,0),
	\label{eq:source_vector_WS}
\end{align}
where the ordering of the interactions is $\alpha = \mathrm{WS}, \mathrm{SS}, Y_\tau, Y_t, Y_b, W_{12}, W_{3}$.

We show our numerical result in the right panel of Fig.~\ref{fig:b-l_aJBL}.
It can be seen that, although this coupling can produce the $B-L$ asymmetry, 
the amount of the $B-L$ asymmetry is quite different from the coupling to the $B-L$ current.
In particular, the final $B-L$ asymmetry is suppressed
by $\mathcal{O}(10)$
(notice the different $y$-axis normalizations n the two panels of Fig.~\ref{fig:b-l_aJBL})
for both the constant case and the oscillation case 
with $T_\mathrm{osc} = 10^{13}\,\mathrm{GeV}$ and $T_\mathrm{dec} = 10^{11}\,\mathrm{GeV}$.
This suppression can be understood as follows.
The Weinberg operator is only the source of the $B-L$ violation in our scenario,
and hence it has to be effective to produce the $B-L$ asymmetry.
At the same time, the axion source term which in the current case is the electroweak sphaleron
has to be effective to produce the $B-L$ asymmetry.
As we saw in Secs.~\ref{sec:SM-interactions} and~\ref{subsec:b-l_transport_eq}, however,
the latter is at most only marginally relevant when the former is effective and vice versa, 
resulting in the suppression of the resulting $B-L$ asymmetry.

Here we comment on Ref.~\cite{Kusenko:2014uta}.
They started from the same coupling $(a/f) W\tilde{W}$ as we do.
They performed a chiral rotation of the leptons to remove this anomalous coupling,
and wrote down the Boltzmann equation
by assuming that the chemical potential of the lepton number charge 
is biased by the axion in the rotated basis.
This treatment is, however, not entirely correct in the presence of the Weinberg operator,
since the operators $W\tilde{W}$ and the divergence of the lepton current are equivalent only when
there is no additional source of the lepton number violation.\footnote{This was also noted in Ref.~\cite{Shi:2015zwa}, based on explicitly examining the Boltzmann equations in these two particular field bases. In our formalism, this invariance is automatic for any basis transformations by definition as we have shown.}
In other words, once one performs a chiral rotation to remove the anomalous coupling,
the axion couples both to the lepton current and the Weinberg operator.
Its couplings are such that the final expression of the source vector is still Eq.~\eqref{eq:source_vector_WS}, \textit{i.e.}, the same as the original coupling $(a/f) W\tilde{W}$,
which follows from our general proof of the basis independence in Sec.~\ref{sec:basis-indep}.
Thus, the coupling $(a/f) W\tilde{W}$ should not be interpreted as a pure shift of the chemical potential of the lepton number charge.
This subtlety is of phenomenological importance since the final $B-L$ asymmetry
can be quite different in the case of $(a/f)W\tilde{W}$ 
compared to, \textit{e.g.}, $(a/f)\partial_\mu J^\mu_{B-L}$,
particularly for the case in which the weak sphaleron is only marginally relevant at the decoupling of the $B-L$ violating process
as we saw above.

In a similar spirit, it was noted in Ref.~\cite{Takahashi:2015waa} that there can be 
a strong suppression in baryon asymmetry for the case in which the weak sphaleron is not efficient at the decoupling of the $B-L$ violating process.
By using the same chiral rotation as Ref.~\cite{Kusenko:2014uta} and discussing spontaneous baryogenesis, it was argued that this chiral rotation should not be performed if the weak sphaleron is not efficient.
Here we emphasize that one can however always 
perform the chiral rotation without specifying a state with which one takes an expectation value.
As the transport equation is basis independent, a non-vanishing velocity of the axion just biases the weak sphaleron after we perform the chiral rotation completely.
To understand whether this bias on the weak sphaleron in the $B+L$ current is transferred to the $B-L$ asymmetry, we need to know how all the relevant SM interactions are involved in attaining equilibrium with $\dot a \neq 0$,
and hence the chiral rotation, which leaves the transport equation unchanged, does not help us to understand this property.

\paragraph{Strong sphaleron.}

Finally we consider the axion coupling to the strong sphaleron: $(a/f) G \tilde{G}$.
The axion source vector in this case is given by
\begin{align}
	n_S^{\alpha} = (0,1,0,0,0,0,0),
\end{align}
where the ordering of the interactions is $\alpha = \mathrm{WS}, \mathrm{SS}, Y_\tau, Y_t, Y_b, W_{12}, W_{3}$.

\begin{figure}[t]
	\centering
 	\includegraphics[width=0.5\linewidth]{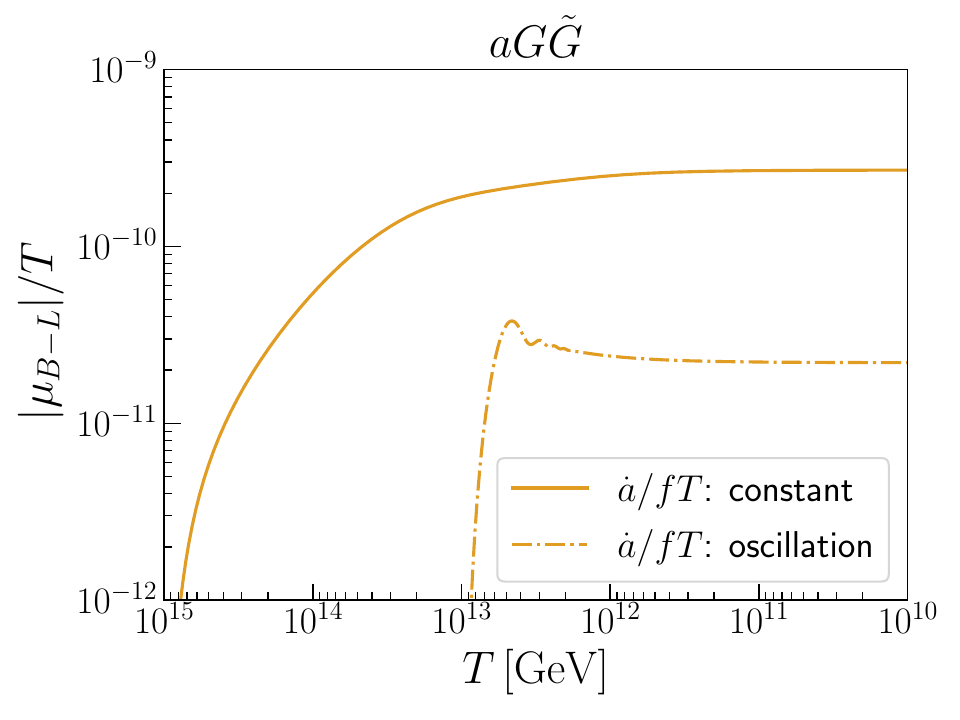}
	\caption{
	The time evolution of the $B-L$ asymmetry produced from the shift-symmetric coupling $(a/f) G \tilde{G}$
	for constant $\dot a/(f T)$ (solid) and oscillating $\dot a/(f T)$ (dashed).
	}
	\label{fig:b-l_aGG}
\end{figure}

In Fig.~\ref{fig:b-l_aGG}, we show our numerical result.
A sizable amount of the $B-L$ asymmetry can be produced from the coupling to the strong sphaleron.
At first sight, it might be surprising since the strong sphaleron has nothing to do with the $B-L$ nor $B+L$ symmetry.
It is nevertheless easily understood as follows.
First of all, we have to use the Weinberg operator to create the $B-L$ asymmetry since it is the only source of $B-L$ violation.
Since the Higgs and the leptons are involved in the Weinberg operator, 
the chemical potentials of the Higgs and/or 
the leptons have to be biased to create the $B-L$ asymmetry. 
In our case, the axion coupling $(a/f) G\tilde{G}$ 
first introduces a bias to the chemical potentials of the quarks.
This bias in the quark sector can be transferred into the Higgs sector by, \textit{e.g.}, the top and bottom Yukawa couplings, 
and the lepton sector by, \textit{e.g.} the electroweak sphaleron.
Once the Higgs and/or the leptons have a bias in their chemical potentials, 
the $B-L$ asymmetry is created through the lepton number violating process.
In short, a bias in a certain sector is eventually transferred to all the other sectors
once we have a sufficient variety of the interactions.
It is essentially what we have seen in Sec.~\ref{subsec:b-l_equilibrium}.

\subsection{Dependence on axion model parameters}
\label{subsec:axion_model}

\begin{figure}[t]
	\begin{minipage}{0.5\linewidth}
	\centering
 	\includegraphics[width=\linewidth]{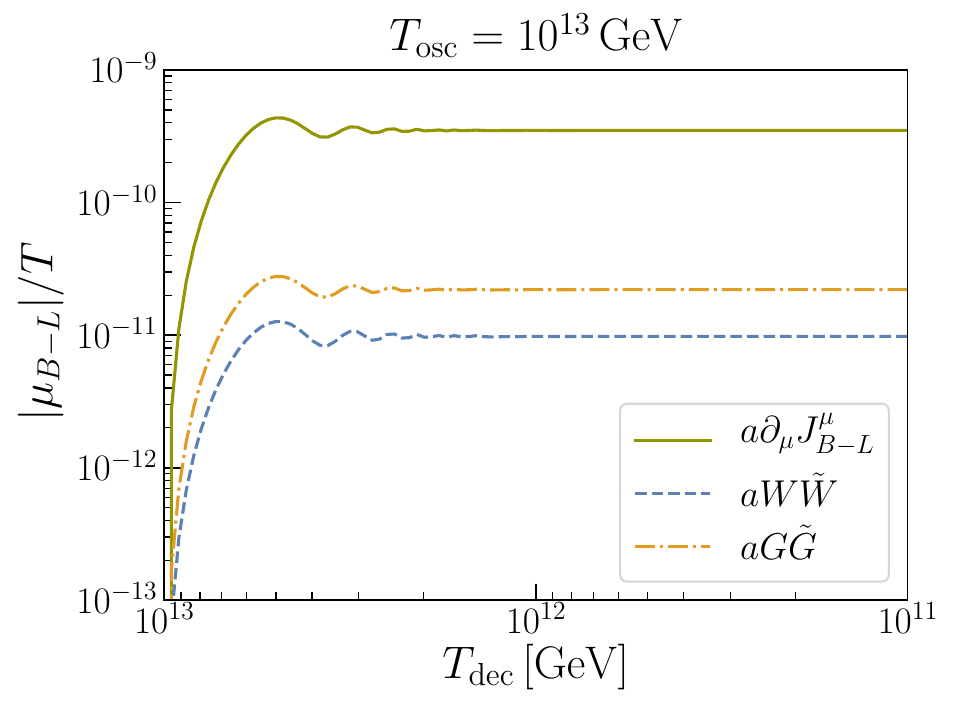}
	\end{minipage}
	\begin{minipage}{0.5\linewidth}
	\centering
 	\includegraphics[width=\linewidth]{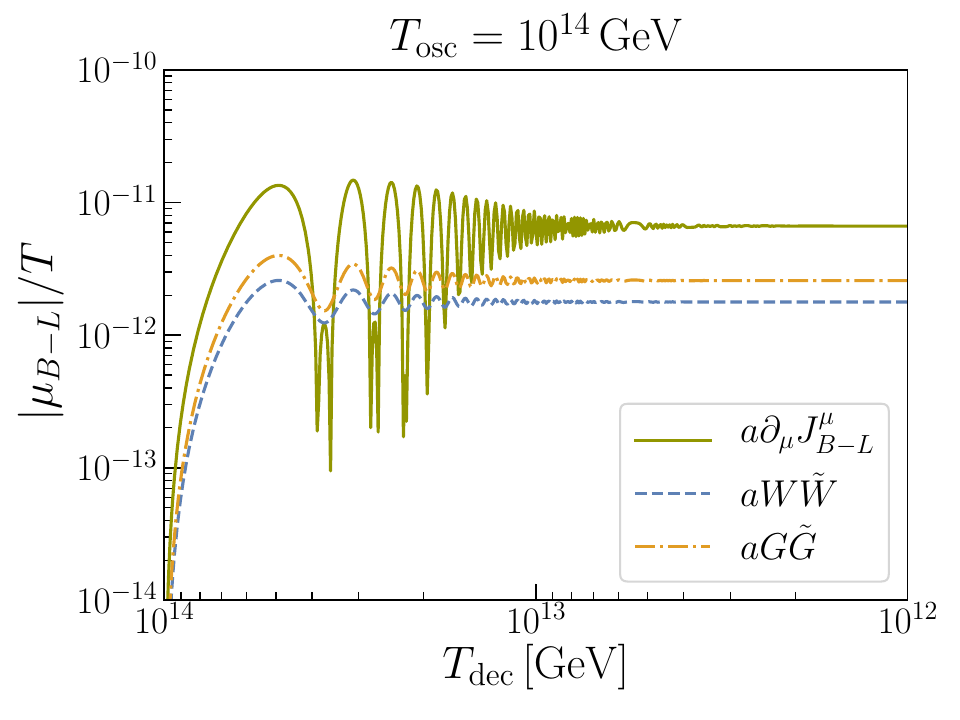}
	\end{minipage}
	\caption{
	The final $B-L$ asymmetry produced for different values of the axion decay temperature $T_\mathrm{dec}$.
	The axion oscillation temperature is taken as $T_\mathrm{osc} = 10^{13}\,\mathrm{GeV}$
	in the left panel, and $T_\mathrm{osc} = 10^{14}\,\mathrm{GeV}$ in the right panel.
	}
	\label{fig:b-l_TdecVsMuBL}
\end{figure}

In the previous Sec.~\ref{subsec:b-l_numerics}, we have fixed the axion model parameters as
$T_\mathrm{osc} = 10^{13}\,\mathrm{GeV}$ and $T_\mathrm{dec} = 10^{11}\,\mathrm{GeV}$.
In this subsection, we briefly discuss the dependence of the final $B-L$ asymmetry on these parameters.

\paragraph{Dependence on axion decay temperature.}

First we study the dependence of the final $B-L$ asymmetry on the axion decay temperature $T_\mathrm{dec}$.
In Fig.~\ref{fig:b-l_TdecVsMuBL}, we plot the final $B-L$ asymmetry for different values of $T_\mathrm{dec}$.
The axion oscillation temperature is $T_\mathrm{osc} = 10^{13}\,\mathrm{GeV}$ in the left panel,
and $T_\mathrm{osc} = 10^{14}\,\mathrm{GeV}$ in the right panel, respectively.

As is clear from the figure, the final $B-L$ asymmetry does not depend on $T_\mathrm{dec}$
for $T_\mathrm{dec} \lesssim 10^{13}\,\mathrm{GeV}$.
This is reasonable since the lepton number violating process decouples around this temperature,
and the $B-L$ asymmetry is conserved irrespective of the axion dynamics afterwards.
For $T_\mathrm{dec} \gtrsim 10^{13}\,\mathrm{GeV}$, the final $B-L$ asymmetry is an oscillating function
of $T_\mathrm{dec}$, following the axion oscillation.
In particular, not only the first oscillation but also the later oscillations affect the final $B-L$ asymmetry,
especially for the coupling  $a \partial_\mu J^{\mu}_{B-L}$ with $T_\mathrm{osc} = 10^{14}\,\mathrm{GeV}$.
This is because, in this case, 
the axion dynamics is directly coupled to the lepton number violating process that is quite effective at high temperatures
and hence the chemical potentials can track (part of) the axion oscillations.
Nevertheless, the final $B-L$ asymmetry on average is within roughly an order of magnitude from the asymptotic value 
for $T_\mathrm{dec} \ll 10^{13}\,\mathrm{GeV}$.

\paragraph{Dependence on axion oscillation temperature.}
\begin{figure}[t]
	\centering
 	\includegraphics[width=0.5\linewidth]{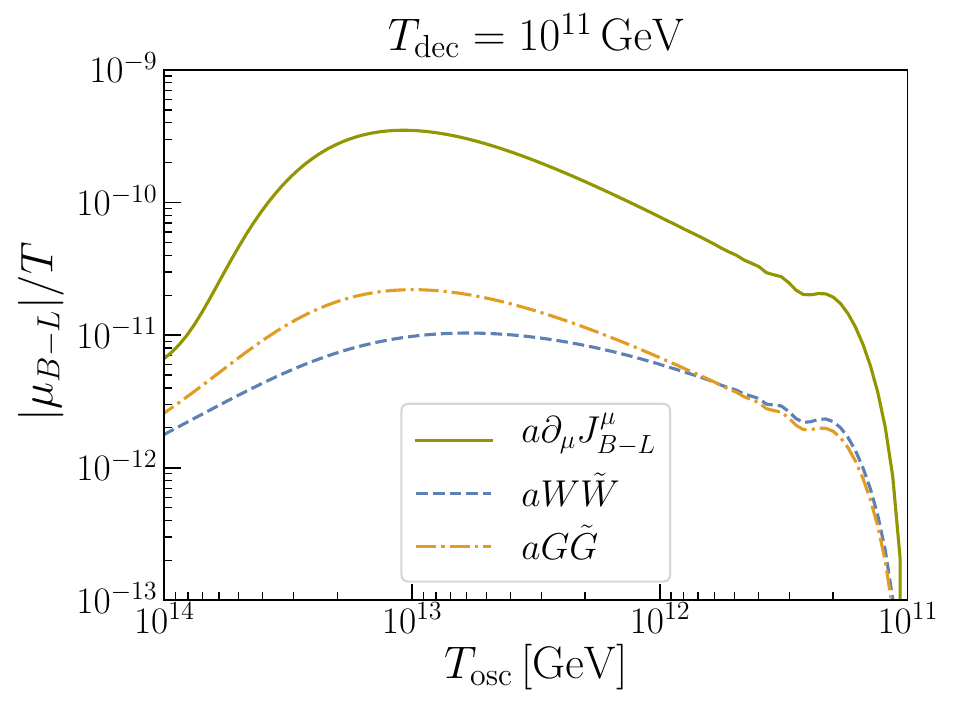}
	\caption{
	The final $B-L$ asymmetry produced for different values of the axion oscillation temperature $T_\mathrm{osc}$.
	The axion decay temperature is taken as $T_\mathrm{dec} = 10^{11}\,\mathrm{GeV}$.
	}
	\label{fig:b-l_ToscVsMuBL}
\end{figure}

Next we study the dependence of the final $B-L$ asymmetry on the axion oscillation temperature $T_\mathrm{osc}$.
In Fig.~\ref{fig:b-l_ToscVsMuBL}, we plot the final $B-L$ asymmetry for different values of $T_\mathrm{osc}$.
We focus on the asymptotic value of the final $B-L$ asymmetry for $T_\mathrm{dec} \ll 10^{13}\,\mathrm{GeV}$ here,
and hence the axion decay temperature is taken as $T_\mathrm{dec} = 10^{11}\,\mathrm{GeV}$.

We can roughly divide the parameter space into two regimes: $T_\mathrm{osc} \lesssim 10^{13}\,\mathrm{GeV}$
and $T_\mathrm{osc} \gtrsim 10^{13}\,\mathrm{GeV}$.
In the former regime, $T_\mathrm{osc} \lesssim 10^{13}\,\mathrm{GeV}$, 
the final $B-L$ asymmetry is an increasing function of $T_\mathrm{osc}$.
This is understood from the fact that the lepton number violating process decouples at $T \sim T_W \sim 10^{13}\,\mathrm{GeV}$,
and hence its effect is suppressed by $\gamma_W/H$ afterwards.
Indeed, the $B-L$ asymmetry depends roughly linearly on $T_\mathrm{osc}$ in this regime,
which is consistent with the above reasoning since $\gamma_W/H \propto T$.
In the latter regime, $T_\mathrm{osc} \gtrsim 10^{13}\,\mathrm{GeV}$, 
the final $B-L$ asymmetry is a decreasing function of $T_\mathrm{osc}$.
This property is easy to understand for the couplings $a W\tilde{W}$ and $a G\tilde{G}$
since these interactions are not in equilibrium,
and hence the produced $B-L$ asymmetry is suppressed by $\gamma_{\mathrm{WS}}/H$ and $\gamma_\mathrm{SS}/H$
for the first oscillation in this regime.
A larger value of $T_\mathrm{osc}$ (for fixed $\eta_0$) thus translates to a smaller value of the axion velocity when the axion couplings become effective.
The situation is more tricky for the coupling $a \partial_\mu J^\mu_{B-L}$.
In this case, the axion source term is effective even for the first oscillation 
since the axion directly couples to the lepton number violating process that is more effective for higher temperature.
Still, the final $B-L$ asymmetry is suppressed for a larger value of $T_\mathrm{osc}$. 
This is because the interaction is strong enough so that $\mu_{B-L}$ follows
(part of) the axion dynamics,
as one can also anticipate from the right panel of Fig.~\ref{fig:b-l_TdecVsMuBL}.
Since the axion oscillates a lot, the produced $B-L$ asymmetry is cancelled in the course of the oscillation,
resulting in the suppression shown in Fig.~\ref{fig:b-l_ToscVsMuBL}.

\section{Conclusion}
\label{sec:conc}
Axion-like particles not only solves the strong $CP$ problem but also has an ability to account for several cosmological issues
such as inflation, the dark matter, and the baryon asymmetry of the universe.
In particular, the axion(-like particle) is likely to be in a motion in the early universe,
providing a source of the $CPT$ symmetry violation.
If the axion is coupled to the SM,
this $CPT$ violation is transferred to the SM sector and,
with the help of a baryon number violating process, 
can be the origin of the baryon asymmetry of the present universe,
referred to as spontaneous baryogenesis~\cite{Cohen:1987vi,Cohen:1988kt}.
In this paper, we have developed a formalism that systematically accounts for spontaneous baryogenesis 
by an axion with general (classically) shift-symmetric couplings to the SM sector.
It consists of charge vectors $n_i^\alpha$ that characterize charges of particles
that are involved in a given operator $O_\alpha$,
and a source vector $n_S^\alpha$ that encodes couplings of the axion
to the operators $O_\alpha$.
Assuming thermal equilibrium, the final baryon asymmetry is obtained 
by solving simple linear algebraic equations [see Eq.~\eqref{eq:equilibrium_solution}].
Our formalism is also ready for numerical implementation 
so that the final baryon asymmetry is easily computed even without assuming equilibrium [see Eq.~\eqref{eq:fulltransporteq}].
Equipped with this formalism, we have revealed several aspects of spontaneous baryogenesis 
on both the theoretical and the phenomenological side.

On the theoretical side, we have shown that the transport equation and hence the final baryon asymmetry are invariant 
under a field rotation involving the axion (see Sec~\ref{sec:basis-indep}).
The explicit form of the axion coupling depends on the choice of the field basis.
For instance, an anomalous coupling to the SU(2) Chern-Simons term, $a W \tilde{W}$, can be eliminated 
by a chiral rotation of the leptons.
The axion then couples to the divergence of the lepton current, $a \partial_\mu J^\mu_{L}$, 
and (if present) to other lepton number violating operators such as the dimension-five Weinberg operator $\left( L \cdot H \right)^{2}$.
Since the chiral rotation is merely a field redefinition, physical quantities should not depend on the choice of this field basis,
which is automatically satisfied in our formalism.
Here we emphasize that the basis independence is not just an academic exercise.
Without accounting for this properly,
one may be lead to a wrong estimation of the final baryon asymmetry.
For instance, one may be tempted to regard the coupling $a W\tilde{W}$
just as a chemical potential of lepton number by a chiral rotation.
This is, however, not appropriate in the presence of the Weinberg operator,
since the axion also couples to the Weinberg operator after the chiral rotation.
Taking into account all the axion couplings properly which appear after this chiral rotation, one ends up with exactly the same transport equation as originally obtained with just the $a W \tilde W$ coupling.
This demonstrates that the field redefinition never helps to understand the dynamical of spontaneous baryogenesis because it does not change the governing equation, namely transport equation.
As a result, we find the final baryon asymmetry originating from the coupling $a W\tilde{W}$ (in the presence of the Weinberg operator) to be an order of magnitude smaller than the baryon asymmetry obtained for a coupling to the lepton current 
if the weak sphaleron is only marginally efficient at the decoupling of the lepton number violating process (see Sec.~\ref{subsec:b-l_numerics}). 
Since our formalism is basis-independent, it automatically takes into account this sort of subtleties.

We have also discussed the backreaction of the SM processes to the dynamics of the axion.
The axion coupling to the SM operator may act as a friction term in the axion equation of motion,
slowing and eventually stopping the motion of the axion.
In Sec.~\ref{sec:asymmetry_generation}, 
we have derived a condition 
under which the axion friction term identically vanishes.
The condition essentially states that the friction term vanishes if one can define a new conserved charge from 
a combination of the axion shift symmetry and the fermion rotation [see Eq.~\eqref{eq:zero-friction} for its precise definition].
The parameter space of the axion to obtain the correct amount of the baryon asymmetry is less
restricted if this condition is met, 
although a non-zero friction term does not necessarily spoil the spontaneous baryogenesis.

On the phenomenological side,
we have derived a condition for the axion couplings
to produce the baryon asymmetry [see Sec.~\ref{sec:asymmetry_generation},
in particular Eqs.~\eqref{eq:cond_asym} and~\eqref{eq:cond_asym_lind}],
which is invariant under a field rotation involving the axion.
It turns out that, once the axion has shift-symmetric couplings to the SM sector,
it is rather difficult \textit{not} to produce the baryon asymmetry,
as long as we have a baryon number violating process.
In particular, the axion does not have to couple directly to the baryon number violating operator.
The physical intuition behind this is as follows.
The axion coupling to one specific operator generates a bias in the chemical potential of particles
that are involved in that operator.
This bias is in general transferred to other particles via other interactions 
and eventually to the baryon number violating process,
resulting in the production of the baryon asymmetry.
As concrete examples, we have considered baryogenesis at $T\gtrsim 10^{2}\,\mathrm{GeV}$ in Sec.~\ref{sec:b+l},
and $T\gtrsim 10^{13}\,\mathrm{GeV}$ in Sec.~\ref{sec:b-l}, respectively,
where the baryon number violation is sourced by the electroweak sphaleron in the former case,
and the electroweak sphaleron together with the Weinberg operator in the latter case.
We have derived a condition of the baryon asymmetry production for these specific cases,
and confirmed that the baryon asymmetry is indeed a generic outcome of the axion shift-symmetric couplings.
For instance, we have shown for both cases that an axion coupling to the SU(3) Chern-Simons term, $a G \tilde{G}$,
ultimately leads to the generation of a baryon asymmetry,
although this operator itself has nothing to do with the $\mathrm{U}(1)_{B-L}$- nor $\mathrm{U}(1)_{B+L}$-violation.
Our findings open up a variety of new possibilities to produce the baryon asymmetry of the universe from axion-like particles.

Along the way, we have summarized the basic properties
of the SM transport equation in Sec.~\ref{sec:SM-transport}
as they are required in Secs.~\ref{sec:b+l} and~\ref{sec:b-l}.
In particular, we have estimated the equilibration temperature of the SM processes,
\textit{i.e.}, the strong/electroweak sphaleron and Yukawa interactions,
below which they are effective
(see Tab.~\ref{tab:equilibration_temperature} and Fig.~\ref{fig:equilibration_temp}).
Our estimation improves Ref.~\cite{Garbrecht:2014kda} by including 
the RG running of the Yukawa couplings in addition to the gauge couplings.
It is important especially for the quark Yukawa couplings as the strong interaction drives them to smaller values at high energy.
This section may be useful not only for the spontaneous baryogenesis but also for other baryogenesis scenarios 
such as the flavored leptogenesis~\cite{Abada:2006fw,Nardi:2006fx,Abada:2006ea,Dev:2017trv}.

\section*{Acknowledgments}
It is a pleasure to thank Kai Schmitz and Fuminobu Takahashi for helpful discussions and comments on the manuscript.
This work was partially funded by the Deutsche Forschungsgemeinschaft under Germany's Excellence Strategy - EXC 2121 ``Quantum Universe'' - 390833306.
This work was also supported by the ERC Starting Grant ‘NewAve’ (638528).
M.~Y. was supported by Leading Initiative for Excellent Young Researchers, MEXT, Japan. 
M.~Y. thanks the hospitality during his stay at DESY. 

\appendix
\newpage
\section{Definitions of symbols and indices}
\label{sec:symbols}

We summarize the definitions of  symbols and indices in Tab.~\ref{table:symbol} and Tab.~\ref{table:index} for the reader's convenience.

\begin{table}[h!]
\begin{center}
\begin{tabular}{|p{1.5cm}|p{9.5cm}|p{1.5cm}|p{1.6cm}|p{1.5cm}|}
  \hline
  \rule[-5pt]{0pt}{15pt}
Symbol & \hfil Definition \hfil 
    & \hfil Size \hfil 
    & \hfil Equation \hfil \\
 \hline \hline
  \rule[-5pt]{0pt}{15pt}
  \hfil $J_{i}^{\mu}$ \hfil  & \hfil current corresponding to a particle species $i$  \hfil  & \hfil $N$ \hfil & \hfil \hfil  \\
  \hline
  \rule[-5pt]{0pt}{15pt}
  \hfil $O_{\alpha}$ \hfil  & \hfil operator for an interaction $\alpha$ \hfil  & \hfil $N_\alpha$ \hfil & \hfil \hfil  \\
  \hline
  \rule[-5pt]{0pt}{15pt}
  \hfil $n_{i}^{\alpha}$ \hfil  & \hfil vector that specifies the charge of each species $i$ involved in the process of $O_{\alpha}$  \hfil  & \hfil $N_\alpha \times N$   \hfil & \hfil \hfil  \\
  \hline
  \rule[-5pt]{0pt}{15pt}
  \hfil $n_{i}^A$ \hfil  & \hfil vector that specifies the conserved charge $A$  \hfil  & \hfil $N_A \times N$ \hfil & \hfil \hfil  \\
  \hline
  \rule[-5pt]{0pt}{15pt}
  \hfil $q_{i}$ \hfil  & \hfil charge density of species $i$  \hfil  & \hfil  $N$ \hfil & \hfil \eq{eq:q_i} \hfil  \\
  \hline
  \rule[-5pt]{0pt}{15pt}
  \hfil $\mu_{i}$ \hfil  & \hfil chemical potentials for each charge $q_{i}$  \hfil  & \hfil $N$ \hfil & \hfil \eq{eq:mu_i} \hfil  \\
  \hline
  \rule[-5pt]{0pt}{15pt}
  \hfil $g_{i}$  \hfil  & \hfil multiplicity or effective degrees of freedom \hfil  & \hfil $N$ \hfil & \hfil \eq{eq:mu_i} \hfil  \\
  \hline
  \rule[-5pt]{0pt}{15pt}
  \hfil $\Gamma_{\alpha}$ \hfil  & \hfil interaction rate per unit time and volume \hfil  & \hfil $N_\alpha$ \hfil & \hfil \eq{eq:Gamma_alpha} \hfil  \\
  \hline
  \rule[-5pt]{0pt}{15pt}
  \hfil $\gamma_{\alpha}$  \hfil  & \hfil interaction rate per unit time ($\equiv \Gamma_{\alpha}/(T^{3}/6)$) \hfil  & \hfil $N_\alpha$ \hfil & \hfil \eq{eq:gamma_alpha} \hfil  \\
  \hline
  \rule[-5pt]{0pt}{15pt}
  \hfil $\Gamma_{ij}$ \hfil  & \hfil $N \times N$ matrix form of interaction rates, ($\equiv \sum_{\alpha} \Gamma_{\alpha} n_{i}^{\alpha} n_{j}^{\alpha}$) \hfil  & \hfil $N\times N$ \hfil & \hfil \eq{eq:transport} \hfil  \\
  \hline
  \rule[-5pt]{0pt}{15pt}
  \hfil $q_{A}$ \hfil  & \hfil conserved charge density for $A$, ($\equiv \sum_{i} n^{A}_{i} q_{i}$) \hfil  & \hfil $N_A$ \hfil & \hfil \eq{eq:q_A} \hfil  \\
  \hline
  \rule[-5pt]{0pt}{15pt}
  \hfil $c_{A}$ \hfil  & \hfil conserved charge for $A$, 
($\equiv q_{A}/ (T^{3} / 6)$) \hfil  & \hfil $N_A$ \hfil & \hfil \eq{eq:conservation} \hfil  \\
  \hline
  \rule[-5pt]{0pt}{15pt}
  \hfil $\bar{n}_i^{\hat \alpha}$, $\bar{n}_i^A$ \hfil  & \hfil dual basis vectors \hfil  & \hfil $N_{\hat\alpha}$, $N_A$ \hfil & \hfil \eq{eq:orthogonality} \hfil  \\
  \hline
  \rule[-5pt]{0pt}{15pt}
  \hfil $a/f$ \hfil  & \hfil axion field $a(t)$ divided by its decay constant $f$ \hfil  & \hfil  \hfil & \hfil \eq{eq:current-coupling} \hfil  \\
  \hline
  \rule[-5pt]{0pt}{15pt}
  \hfil $S_{i}$ \hfil  & \hfil source term, ($\equiv \sum_\alpha \Gamma_\alpha n_i^\alpha n_S^\alpha$)  \hfil  & \hfil $N$ \hfil & \hfil \eq{eq:transport-with-a} \hfil  \\
  \hline
  \rule[-5pt]{0pt}{15pt}
  \hfil $n_S^\alpha$ \hfil  & \hfil source vector, or charge vector that specifies the charge of the axion involved in the process of $O_\alpha$ \hfil  & \hfil $N_\alpha$ \hfil & \hfil \eq{eq:sourcevec} \hfil  \\
  \hline
  \rule[-5pt]{0pt}{15pt}
  \hfil $\Gamma_{\hat\alpha \hat\beta}$ \hfil  & \hfil $N_{\hat\alpha} \times N_{\hat\alpha}$ matrix form of interaction rates, ($\equiv \sum_{i,j} \bar{n}_i^{\hat\alpha} \Gamma_{ij} \bar{n}^{\hat\beta}_j$) \hfil  & \hfil $N_{\hat\alpha} \times N_{\hat\alpha}$ \hfil & \hfil \eq{eq:transport_matrix} \hfil  \\
  \hline
  \rule[-5pt]{0pt}{15pt}
  \hfil $S_{\hat\alpha \beta}$ \hfil  & \hfil $N_{\hat\alpha} \times N_\alpha$ matrix form of interaction rates, ($\equiv \sum_i \bar{n}_i^{\hat\alpha} \Gamma_\beta n_i^\beta$) \hfil  & \hfil $N_{\hat\alpha} \times N_\alpha$ \hfil & \hfil \eq{eq:transport_matrix} \hfil  \\
  \hline
  \rule[-5pt]{0pt}{15pt}
  \hfil $U_{\hat\alpha \beta}$ \hfil  & \hfil matrix that represents linear dependence, 
  ($\equiv \sum_i \bar{n}^{\hat\alpha}_i n^\beta_i$) \hfil  & \hfil  $N_{\hat\alpha} \times N_\alpha$ \hfil & \hfil \eq{eq:conversion} \hfil  \\
  \hline
  \rule[-5pt]{0pt}{15pt}
  \hfil $M_{Xi}$ \hfil  & \hfil matrix constructed from complete sets, ($\equiv ( n_i^{\hat\alpha}, g_i n_i^A )^T$) \hfil  & \hfil $N\times N$  \hfil & \hfil \eq{eq:matrix-equil} \hfil  \\
  \hline
  \rule[-5pt]{0pt}{15pt}
  \hfil $q_C$ \hfil  & \hfil a certain charge specified by a charge vector $n^C_i$ \hfil  & \hfil  \hfil & \hfil \eq{eq:mu_c} \hfil  \\
  \hline
  \rule[-5pt]{0pt}{15pt}
  \hfil $v_\alpha^C$ \hfil  & \hfil direction of source vector that results in $q_C = 0$ \hfil  & \hfil $N_\alpha$  \hfil & \hfil \eq{eq:cond_asym} \hfil  \\
  \hline
  \rule[-5pt]{0pt}{15pt}
  \hfil $\gamma^{\text{eff}}_{a,\alpha \beta}$ \hfil  & \hfil matrix that represents the backreaction to the axion \hfil  & \hfil $N_\alpha \times N_\alpha$ \hfil & \hfil \eq{eq:axion_dissipation} \hfil  \\
  \hline
  \rule[-5pt]{0pt}{15pt}
  \hfil $\kappa_\alpha$ \hfil  & \hfil numerical coefficient of interaction rate for $\alpha$ \hfil  & \hfil $N_\alpha$  \hfil & \hfil Tab.~\ref{tab:equilibration_temperature} \hfil  \\
  \hline
  \rule[-5pt]{0pt}{15pt}
  \hfil $T_\alpha$ \hfil  & \hfil equilibration temperature for an interaction $\alpha$  \hfil  & \hfil $N_\alpha$ \hfil & \hfil \eq{eq:T_alpha} \hfil  \\
  \hline
  \rule[-5pt]{0pt}{15pt}
  \hfil $\alpha_2$, $\alpha_3$ \hfil  & \hfil fine structure constants for SU(2) and SU(3) \hfil  & \hfil  \hfil & \hfil \hfil  \\
  \hline
\end{tabular}\end{center}
\caption{Definitions of symbols
\label{table:symbol}}
\end{table}

\begin{table}\begin{center}
\begin{tabular}{|p{1.5cm}|p{7cm}|p{1.5cm}|p{3.5cm}|}
  \hline
  \rule[-5pt]{0pt}{15pt}
Index & \hfil Definition \hfil 
    & \hfil Length \hfil 
    & \hfil Example \hfil \\
 \hline \hline
  \rule[-5pt]{0pt}{15pt}
  \hfil $\mu$ \hfil  & \hfil Lorentz index, sometimes omitted  \hfil  & \hfil 4 \hfil & \hfil $W\tilde{W} \equiv W_{\mu \nu} \tilde{W}^{\mu\nu}$ \hfil  \\
  \hline
  \rule[-5pt]{0pt}{15pt}
  \hfil $f$ \hfil  & \hfil flavor index \hfil  & \hfil $N_f$ $(=3)$ \hfil & \hfil  \hfil  \\
\hline
  \rule[-5pt]{0pt}{15pt}
  \hfil $i$ \hfil  & \hfil index for particle species \hfil  & \hfil $N$ \hfil & \hfil $i = e$, $L$, $q$, $Q$, \dots  \hfil  \\
  \hline
  \rule[-5pt]{0pt}{15pt}
  \hfil $\alpha$ \hfil  & \hfil index for interactions  \hfil  & \hfil $N_\alpha$ \hfil & \hfil $\alpha = {\rm WS}$, ${\rm SS}$, $Y_t$, \dots \hfil  \\
\hline
  \rule[-5pt]{0pt}{15pt}
  \hfil $A$ \hfil  & \hfil index for conserved charges  \hfil  & \hfil $N_{A}$ \hfil & \hfil $A = Q_Y$, $Q_{B-L}$, \dots \hfil  \\
\hline
  \rule[-5pt]{0pt}{15pt}
  \hfil $X$ \hfil  & \hfil collective index for $\hat\alpha$ and $A$ \hfil  & \hfil $N_{\hat\alpha} + N_A$ ($=N$) \hfil & \hfil  \hfil  \\
\hline
  \rule[-5pt]{0pt}{15pt}
  \hfil $\hat\alpha$  \hfil  & \hfil  index for a complete set of \newline \hspace*{1cm} linearly independent vectors $n^{\alpha}_{i}$ 
 \hfil  & \hfil  $N_{\hat\alpha}$ \hfil & \hfil  \hfil  \\
\hline
  \rule[-5pt]{0pt}{15pt}
  \hfil $\alpha_\Delta$  \hfil  & \hfil index for $\alpha$ other than $\hat\alpha$ \hfil  & \hfil $N_\alpha- N_{\hat\alpha}$ \hfil & \hfil  \hfil  \\
\hline
  \rule[-5pt]{0pt}{15pt}
  \hfil $\hat \alpha_\perp$ \hfil  & \hfil index for a set of vectors $n_i^{\hat\alpha}$ \newline \hspace*{5mm} that are orthogonal to $n_i^{\alpha_\Delta}$ for all $\alpha_\Delta$ 
\hfil  & \hfil \hfil & \hfil  \hfil  \\
\hline
  \rule[-5pt]{0pt}{15pt}
  \hfil  $\hat \alpha_\parallel$ \hfil  & \hfil  index for $\hat\alpha$ other than $\hat\alpha_\perp$ 
  \hfil  & \hfil \hfil & \hfil  \hfil  \\
\hline
\end{tabular}\end{center}
\caption{Definitions of indices
\label{table:index}}
\end{table}

\section{Derivation of transport equation}
\label{sec:derivation}

Here we provide an explicit derivation of the transport equation in the linear response for the sake of completeness, in particular deriving Eqs.~\eqref{eq:transport-coef} and \eqref{eq:axion-op-lr}.
We first derive the transport coefficients without the coupling to an axion and then discuss how this coupling sources the bias in the transport equation.

\subsection{Transport equation without axion}

We derive Eqs.~\eqref{eq:transport-coef} and \eqref{eq:transport} in the linear response, starting from the current equation of
\begin{align}
	\partial \cdot J_{i} = \epsilon_O \sum_{\alpha} n_{i}^{\alpha} O_{\alpha}\,,
	\label{eq:current-ap}
\end{align}
without the shift-symmetric coupling to an axion such as Eqs.~\eqref{eq:current-coupling} and \eqref{eq:operator-coupling}.
We introduce a formal parameter $\epsilon$ so as to make the ``slow'' relaxation explicit, which it a basic assumption of our derivation.
The following derivation is a slight extension of Ref.~\cite{Hosoya:1983id}.

\paragraph{Density operator.}
In the following derivation, we adopt the Heisenberg picture where the density operator $\rho$ does not evolve with time.
To derive the transport equation, we would like to perturb the system with chemical potentials and see how the system relaxes to equilibrium.
However, it is not easy to spell out such a non-equilibrium state directly. 

Instead, let us construct an equilibrium state that allows nonzero charge densities for broken symmetry.
Suppose that we couple the system with external chemical potentials, \textit{i.e.,} $\mu_i^{\rm (ext)} Q_i$.
We expect that the system attains a certain equilibrium state after a sufficiently long time, where the production from the external chemical potential is balanced by the decay.
Our first task here is to find out such an equilibrium state in terms of operators.
Instead of $\sum_{i} \mu_{i}^{\rm (ext)} Q_{i}$, we introduce an operator $\int \dd^3x\, \mathcal{X}$ which is time independent in a certain limit and is identical to $\sum_i \mu_i^{\rm (ext)} Q_i$ in the limit that the charges are conserved [\textit{i.e.}, $\epsilon_O \to 0$ in \eq{eq:current-ap}]:
\begin{align}
	\mathcal{X} (t, \bm{x}) &\equiv \epsilon \int^{t}_{-\infty}  \dd t' \, e^{\epsilon (t' - t)}
	\sum_{i} \mu_{i}^{\rm (ext)} J^{0}_{i} (t', \bm{x})\,.
\end{align}
Note that we should take $\epsilon \searrow 0$ in the end of computation.
Differentiating it with respect to $t$, one finds its time-independence in the limit of $\epsilon \searrow 0$:
\begin{align}
	\frac{\dd}{\dd t} \mathcal{X} (t,\bm{x}) = \epsilon \sum_{i} \mu_{i}^{\rm (ext)} J^{0}_{i} (t, \bm{x}) - \epsilon^{2} \int^{t}_{-\infty} \dd t' \, e^{\epsilon (t' - t)} \sum_{i} \mu_{i}^{\rm (ext)} J_{i}^{0} (t', \bm{x}) \to 0  \quad \text{for} \quad \epsilon \searrow 0\,.
\end{align}
We take the following density operator as a functional of $\mathcal{X}$, which is time independent and hence commutes with the Hamiltonian: 
\begin{align}
	\rho &\equiv \frac{e^{- \frac{1}{T} ( H - \int \dd^3 x \,  \mathcal{X} )}}{\Tr \left[ e^{- \frac{1}{T} ( H - \int \dd^3 x \,  \mathcal{X}) } \right]}
	\nonumber\\
	&= \frac{e^{- \frac{1}{T} ( H - \sum_{i} \mu_{i}^{\rm (ext)} Q_{i} ) +\epsilon_O X}}{\Tr \left[ e^{- \frac{1}{T} ( H - \sum_{i} \mu_{i}^{\rm (ext)} Q_{i} ) + \epsilon_O X } \right]}\,,
	\label{eq:density}
\end{align}
where in the second line we use
\begin{align}
	\int \dd^3 x\, \mathcal{X} (t,\bm{x}) = \sum_{i} \mu_{i}^{\rm (ext)} \int \dd^3 x J^{0}_{i} (t, \bm{x}) - \int^{t}_{-\infty} \dd t'   \, e^{\epsilon (t' - t)} \sum_{i} \mu_{i}^{\rm (ext)} \sum_{\alpha} n^{\alpha}_{i} \int \dd^3 x \,\epsilon_O O_{\alpha} (t', \bm{x})\,,
\end{align}
and define 
\begin{align}
	X \equiv -\sum_{i, \alpha} \frac{\mu_{i}}{T} n^{\alpha}_{i}  \int_{-\infty}^{t} \dd^{4} x'\, e^{\epsilon (t' - t)}  O_{\alpha} (x') \,,
\end{align}
with
\begin{align}
\label{eq:equilibriumc-ond}
	\mu_i = \mu_{i}^{\rm (ext)}\,.
\end{align}
Here we introduced a new variable $\mu_i$ for later use. 

Note that, at the leading order for
$\epsilon_O$, 
the charge densities can be expressed as 
\begin{align}
 q_i = \mu_i^{\rm (ext)} \frac{g_i T^2 }{ 6} = \mu_i \frac{g_i T^2 }{ 6}, 
\end{align}
for $\mu_i/T \ll 1$ with $g_i$ being the multiplicity factor.

In the Heisenberg picture, time evolution of any operator is given by the commutator with the Hamiltonian, which is also true for the current $\partial_{t} J_{i}^{0} = i [H, J_{i}^{0}]$.
An immediate consequence is that any operator evaluated with $\rho$ given in Eq.~\eqref{eq:density} does not evolve in time because the density operator commutes with the Hamiltonian, which also holds for the current, $\vev{ \dot J_{i}^{0} } = 0$. 
Note that we introduce $\mu_i^{\rm (ext)}$ and
the system is coupled to an external device that keeps producing $q_i$.
This observation implies that our state $\rho$ describes an equilibrium state where the production is balanced by the intrinsic decay of $q_i$ driven by the breaking operator $O_\alpha$.
This structure can be easily seen if we expand the $\rho$ with respect to $\epsilon_O$ as follows:
\begin{align}
	\rho \simeq  
 \rho_{\text{GC}} + 
 \left[ T \int^{\frac{1}{T}}_{0} \dd \tau\, e^{- H \tau} \epsilon_O X e^{H  \tau} - \vev{X}_{\text{C}} \right] \rho_\text{C}
	\,,
 \label{eq:rho-GC}
\end{align}
where we also drop the higher-order terms such as $\epsilon_O^n (\mu_i / T)^m $ with $n \geq 1$, $m \geq 2$. 
An expectation value with 
the subscripts `GC' is taken by the grand canonical ensemble, $\langle O_\alpha \rangle_\text{GC} = \Tr (\rho_\text{GC} O_\alpha)$, and that with `C' is for the canonical ensemble, 
$\langle O_\alpha \rangle_\text{C} = \Tr (\rho_\text{C} O_\alpha)$, where 
the grand canonical and canonical ensembles are defined as
\begin{align}
	\rho_{\text{GC}} (t) = \frac{e^{-  ( H - \sum_{i} \mu_{i}^{\rm (ext)} Q_{i} (t) )/ T}}{\Tr \left[ e^{-  ( H - \sum_{i} \mu_{i}^{\rm (ext)} Q_{i} (t) )/ T} \right]}\,,
	\qquad
	\rho_{\text{C}} = \frac{e^{- H/T}}{\Tr \left[ e^{- H/T} \right]}\,.
	\label{eq:grand-canonical_canonical}
\end{align}
The first and second terms in \eq{eq:rho-GC} correspond to the production by the external device and the intrinsic decay respectively.

Now we decouple the system from the external device that is responsible for the production term. 
Since we assume that the relaxation processes are slow, the charge densities can be still described by just replacing the chemical potentials $\mu_i$ with the time dependent one, \textit{i.e.,} $\mu_i \to \mu_i(t)$.
That is, we take $\mu_i^{\rm (ext)}/T \to 0$ with $\mu_i$ (contained in $X$ and $q_i$) kept nonzero values. 
Then the system is no longer the equilibrium and tends to relax into another equilibrium state via a non-trivial transport equation.
The underlying assumption of the transport equation is that the typical time scale of chemical equilibration is much slower than other reactions.
It is also assumed that the deviation from thermal equilibrium is weak. 
This motivates us to expand $\rho$ in a series of $\epsilon_O$: 
\begin{align}
	\rho \simeq  
 \rho_{\text{C}} + 
 \left[ T \int^{\frac{1}{T}}_{0} \dd \tau\, e^{- H \tau} \epsilon_O X e^{H  \tau} - \vev{X}_{\text{C}} \right] \rho_\text{C}
	\,.
	\label{eq:density-pert}
\end{align}

\paragraph{Transport coefficients.}
By using the perturbative expansion \eqref{eq:density-pert}, we obtain 
\begin{align}
	\vev{O_{\alpha}} - \vev{O_{\alpha}}_C
 &\simeq - \epsilon_O \sum_{j, \beta}n^{\beta}_{j}\frac{\mu_{j}}{T}
	\int^{t}_{-\infty} \dd^{4} x'\, e^{\epsilon (t' - t)} T \int^{\frac{1}{T}}_{0} \dd \tau\,
	\vev{O_{\alpha} (x)
		\left(e^{- H \tau} O_{\beta} (x') e^{H \tau}  - \vev{O_{\beta} (x')}_{\text{C}}
		\right)
	}_{\text{C}} \nonumber \\
	& = 
	- \epsilon_O\sum_{j, \beta}n^{\beta}_{j}\frac{\mu_{j}}{T}
	\int^{t}_{-\infty} \dd^{4} x'\, e^{\epsilon (t' - t)} T \int^{\frac{1}{T}}_{0} \dd \tau\,
	\vev{O_{\alpha} (x)
		\left( O_{\beta} (t'+i \tau, \bm{x}')  - \vev{O_{\beta} (x')}_{\text{C}}
		\right)
	}_{\text{C}} \,.
\end{align}
In the second line, we have 
used the fact that $e^{ H \tau}$ can be regarded as a complex time-evolution operator.
Assuming that the correlation drops for $t' \to - \infty$, \textit{i.e.}, $\langle O_{\alpha}(x) O_{\beta}(x')\rangle \to \langle O_{\alpha} (x) \rangle \langle O_{\beta} (x')\rangle$, the integrand can be expressed as
\begin{align}
	\vev{O_{\alpha} (x)
		\left( O_{\beta} (t'+i \tau, \bm{x}')  - \vev{O_{\beta} (x')}_{\text{C}}
		\right)
	}_{\text{C}}
	= \int^{t'}_{-\infty} \dd t''\, \vev{ O_{\alpha} (x) \frac{\dd}{\dd t''} O_{\beta} (t'' + i \tau, \bm{x}') }_{\text{C}}\,.
\end{align}
Rewriting the differentiation with respect to $t''$ as $i \tau$, one may perform the $\tau$ integration explicitly, which results in
\begin{align}
	\vev{O_{\alpha}}
 - \vev{O_{\alpha}}_C
 \simeq - \epsilon_O \sum_{j, \beta} n^{\beta}_{j} \mu_{j} 
	\int^{t}_{-\infty} \dd^{4} x'\, e^{\epsilon (t' - t)} \int^{t'}_{-\infty} \dd t'' \, 
	i \vev{ \left[ O_{\alpha} (x), O_{\beta} (t'', \bm{x}') \right] }_{\text{C}}\,.
	\label{eq:coll-temp}
\end{align}
Here we have used the Kubo-Martin-Schwinger relation:
\begin{align}
	\vev{O_{\alpha}(x) O_{\beta} (t'' + i/T, \bm{x}')}_{\text{C}} 
	= \Tr \left[ \rho_{\text{C}} O_{\alpha} (x) e^{-H/T} O_{\beta}(t'', \bm{x}') e^{H/T} \right]
	= \vev{ O_{\beta} (t'', \bm{x}') O_{\alpha} (x) }_{\text{C}}\,.
\end{align}

Now we are ready to evaluate the transport coefficient at the leading order in the interactions and $\mu_{i} / T$.
For later convenience, we define the spectral function for $O_{\alpha}$ by
\begin{align}
	G_{\alpha \beta}^{\rho} (x - x') \equiv \vev{ \left[ O_{\alpha} (x), O_{\beta}(x') \right] }_{\text{C}}\,, \quad
	G_{\alpha \beta}^{\rho} (\omega, \bm{p}) \equiv \int \dd^{4} x\, e^{i \omega t - i \bm{p} \cdot \bm{x}} G_{\alpha \beta}^{\rho} (x)\,.
	\label{eq:spectral}
\end{align}
Inserting Eq.~\eqref{eq:spectral} into Eq.~\eqref{eq:coll-temp}, we arrive at the following expression
\begin{align}
	\epsilon_O \lmk \vev{O_{\alpha}} - \vev{O_{\alpha}}_C \rmk &= - \epsilon_O^2\sum_{j, \beta} n^{\beta}_{j} \mu_{j} \,\lim_{\epsilon \searrow 0} \int \frac{\dd \omega}{2 \pi} \frac{1}{\omega - i \epsilon} \frac{1}{i \omega} 
	G_{\alpha \beta}^{\rho} (\omega, \bm{0} ) |_{\epsilon_O = 0} + \mathcal{O} (\epsilon_O^3) \\
	&\simeq - \epsilon_O^2 \sum_{j} n^{\alpha}_{j}\frac{\mu_{j}}{T} \, \lim_{\omega \searrow 0} \lim_{\epsilon_O \searrow 0} \frac{T G_{\alpha}^{\rho} (\omega, \bm{0})}{2 \omega}\,.
\end{align}
Here we keep the leading order expansion in $\epsilon_O$, and
we utilize the fact that the spectral function is an odd function in $\omega$, \textit{i.e.}, $G_{\alpha \beta}^{\rho} (\omega, \bm{0}) = - G_{\alpha \beta}^{\rho} (- \omega, \bm{0})$.
In the second equality, we also use $G^{\rho}_{\alpha \beta} (\omega, \bm{0}) / \omega = \delta_{\alpha \beta} G^{\rho}_{\alpha} (\omega, \bm{0}) / \omega$.
Therefore, we arrive at the following expression for the transport equation:
\begin{align}
	\dot q_{i}(t) = - \sum_{j} \sum_{\alpha} \Gamma_{\alpha} n^{\alpha}_{i} n^{\alpha}_{j} \frac{\mu_{j}(t)}{T}\,, \quad
		\Gamma_{\alpha} \equiv \epsilon_O^2 \lim_{\omega \searrow 0} \lim_{\epsilon_O \searrow 0} \frac{T G_{\alpha}^{\rho} (\omega, \bm{0})}{2 \omega}\,, 
	\label{eq:transport-coef-ap}
\end{align}
where we use $\epsilon_O \sum_\alpha n_i^\alpha \vev{O_{\alpha}}_C = \vev{\del \cdot {J}_i}_C = 0$.
One can calculate $\Gamma_\alpha$ [or $G_{\alpha}^{\rho} (\omega, \bm{0})$] by a diagrammatic calculation in a finite-temperature field theory. 

\subsection{Source term from the axion}

Now we turn on the shift-symmetry couplings with an axion.
As discussed in the main text, the coupling given in Eq.~\eqref{eq:current-coupling}  just shifts the chemical potential as $\mu_{k} \mapsto \mu_{k} - \dot a / f$.
A more non-trivial one is a direct coupling with the operator $O_{\beta}$
\begin{align}
	\mathcal{L}_{\text{int},\beta} = - \frac{a}{f} O_{\beta}\,.
	\label{eq:ext}
\end{align}
In the following, we derive Eq.~\eqref{eq:axion-op-lr} by regarding $a/f$ as an external time-dependent field in the linear response.

\paragraph{Linear response.}
Let us first recall the basic formula of the linear response theory.
Suppose that we turn on Eq.~\eqref{eq:ext} at $t_{\text{ini}}$.
After $t_{\text{ini}}$, all the fields obey the Hamiltonian with an explicit time dependence on the axion:
\begin{align}
	\bar H (t) = H + \frac{a(t)}{f} \int \dd^{3} x\, O_{\beta} (t, \bm{x})
	\quad \text{for} \quad t > t_{\text{ini}}\,.
\end{align}
Integrating the Heisenberg equation for $O_{\alpha}$, we obtain
\begin{align}
	O_{\alpha} (t, \bm{x}) |_{a/f} - O_{\alpha} (t_{\text{ini}}, \bm{x}) = 
	i \int^{t}_{t_{\text{ini}}} \dd t'\, \left[ \bar H (t'), O_{\alpha} (t', \bm{x}) |_{a/f} \right]\,,
\end{align}
where the operator with a subscript $a/f$ implies that it evolves under $\bar H (t)$.
Now we take an expectation value with respect to \eqref{eq:density-pert}:
\begin{align}
	&\vev{O_{\alpha} (x)|_{a/f} } - \vev{ O_{\alpha} (t_{\text{ini}}, \bm{x}) }
	\nonumber\\
	&= 
	i \int^{t}_{t_{\text{ini}}} \dd t'\, \vev{ \left[ H (t'), O_{\alpha} (t',\bm{x}) |_{a/f} \right] }
	+	\frac{a(t')}{f} i \int^{t}_{t_{\text{ini}}} \dd^4 x'\, \vev{ \left[ O_\beta(t',x'), O_{\alpha} (t',\bm{x}) |_{a/f} \right] }. 
	\label{eq:grand-canonical2}
\end{align} 
Here, we note that the canonical ensemble commutes with Hamiltonian and hence 
\begin{align}
 i \int^{t}_{t_{\text{ini}}} \dd t'\, \vev{ \left[ H (t'), O_{\alpha} (t',\bm{x}) |_{a/f} \right] }_{\text{C}} = 0\,. 
\end{align}
This implies that the first term in the right-hand side of \eq{eq:grand-canonical2} is $\mathcal{O}(\epsilon_O)$ can be rewritten as 
\begin{align}
 i \int^{t}_{t_{\text{ini}}} \dd t'\, \vev{ \left[ H (t'), O_{\alpha} (t',\bm{x}) |_{a/f} \right] }
 &\simeq
 i \int^{t}_{t_{\text{ini}}} \dd t'\, \vev{ \left[ H (t'), O_{\alpha} (t',\bm{x}) \right] } 
  \nonumber\\
 &= 
 \vev{O_{\alpha} (x) } - \vev{ O_{\alpha} (t_{\text{ini}}, \bm{x}) } \,,
\end{align}
where 
we drop the cross term between $a/f$ and $\epsilon_O$ and keep the term at the linear order in $a/f$ and 
$\epsilon_O$. 
Substituting this into \eq{eq:grand-canonical2}, we obtain the well-known formula of the linear response theory:
\begin{align}
	\vev{O_{\alpha} (x)|_{a/f} } - \vev{ O_{\alpha} (x) }
	&\simeq 
	- i \int^{t}_{-\infty} \dd^{4} x' \, \vev{ \left[ O_{\alpha} (x), O_{\beta} (x') \right] }_{\text{C}} \frac{a(t')}{f} \nonumber \\
	& = - i \int^{t}_{-\infty} \dd^{4} x'  G_{\alpha \beta}^{\rho} (x-x') \frac{a(t')}{f} \,,
	\label{eq:linear-ap}
\end{align}
where 
we keep the term at the linear order in $a/f$ and $\mu_{i}/T$. 
We also send the initial time $t_{\text{ini}}$ to $-\infty$.
In the previous section, we have estimated $\langle O_{\alpha} (t, \bm{x}) \rangle$.
This equation indicates how the axion coupling changes the expectation value of $O_{\alpha}$ at the linear order in $a/f$.

\paragraph{Source term from the axion.}
In order to derive the source term in the transport equation, we assume the time evolution of axion is so slow that one may perform the gradient expansion
$a (t') \simeq a_0 - \dot a (t - t')$ with $a_0$ and $\dot a$  constant.
Roughly speaking, the axion mass is assumed to be much smaller than a typical interaction rate in thermal plasma.
Let us discuss the contributions from $a$ and $\dot a$ in the right-hand side of Eq.~\eqref{eq:linear-ap} separately. 

We start with the time independent part of the axion field.
One may express the right-hand side of Eq.~\eqref{eq:linear-ap} as
\begin{align}
	- i  \frac{a_0}{f} \int^{t}_{-\infty} \dd^{4} x'  G_{\alpha \beta}^{\rho} (x-x')
	= - \frac{a_0}{f}\, \operatorname{Pv} \int \frac{\dd \omega}{2 \pi} \frac{1}{\omega} G^{\rho}_{\alpha \beta} (\omega, \bm{0})\,,
	\label{eq:axion-mass}
\end{align}
where `Pv' represents the Cauchy principal value.
The behavior of Eq.~\eqref{eq:axion-mass} is related to the mass of the axion which depends on the structure of the current equations.
Suppose that the charge vector $n^{\beta}_{i}$ associated with the coupling $a_0 O_{\beta}$ is not in the span of the charge vectors of all other operators.
In this case, one may rewrite this coupling as $a_0 \sum_{i} \bar n^{\beta}_{i} \partial \cdot J_{i}$.
Since the axion field is now constant, this coupling vanishes after integration by parts.
This observation implies that Eq.~\eqref{eq:axion-mass} becomes zero in this case.
On the other hand, if the charge vector $n^{\beta}_{i}$ can be expressed by a linear combination of other charge vectors, one cannot rotate out the constant axion field $a_0$.
In this case, Eq.~\eqref{eq:axion-mass} is non-zero in general.
This implies a non-zero mass of the axion and hence breaking of its shift symmetry
because the expectation value of $O_{\beta}$ enters in the equation of motion for the axion.
In the following, we assume that the axion mass coming from this coupling is negligible for simplicity (see also footnote~\ref{fn:axion_mass}).

We move on to the contribution from the non-zero axion velocity, $\dot a \neq 0$.
The right-hand side of Eq.~\eqref{eq:linear-ap} can be expressed as\footnote{
	For notational brevity, here we do not introduce a formal parameter $\epsilon_O$. The definition of $\Gamma$ should be understood as Eq.~\eqref{eq:transport-coef-ap}.
}
\begin{align}
	i \frac{\dot a}{f} \int^{t}_{-\infty} \dd^{4} x'\, G^{\rho}_{\alpha \beta} (x - x') (t - t')
	&= \frac{\dot a}{f} \int^{t}_{-\infty} \dd t'\, \int \frac{\dd \omega}{2 \pi} e^{- i \omega (t - t')} \partial_{\omega}G^{\rho}_{\alpha \beta} (\omega, \bm{0}) \nonumber \\
	&= \frac{\dot a}{f} \lim_{\epsilon \searrow 0} \int \frac{\dd \omega}{2 \pi i} \frac{1}{\omega - i \epsilon}\partial_{\omega}G^{\rho}_{\alpha \beta} (\omega, \bm{0})
	\nonumber \\
	&= \frac{\dot a/ f}{T} \Gamma_{\alpha} \delta_{\alpha \beta}\,.
\end{align}
In the last line, we have used the definition of the transport coefficient in Eq.~\eqref{eq:transport-coef-ap}.
Summing up all the equations obtained so far, we can write down Eq.~\eqref{eq:linear-ap} as follows:
\begin{align}
	\vev{O_{\alpha} (x)|_{a/f} } = 
	- \Gamma_{\alpha} \sum_{j} n^{\alpha}_{j} \frac{\mu_{j}}{T} + \frac{\dot a / f}{T} \Gamma_{\alpha} \delta_{\alpha \beta}\,,
\end{align}
reproducing Eq.~\eqref{eq:axion-op-lr}.

\section{Proof of the condition for vanishing backreaction}
\label{sec:br_pr}
Here we provide an explicit proof of Eq.~\eqref{eq:zero-friction}.
Let us consider the axion coupling of $ (a/f) \sum_{\alpha} n_S^{\alpha}O_{\alpha}$.
The effective friction term \eqref{eq:axion_dissipation} is given as
\begin{align}
	\gamma^\text{eff}_{a, \alpha \beta} = 
	\frac{1}{f^2 T} 
	\left( \Gamma_{\alpha}\delta_{\alpha \beta} - \sum_{\hat \gamma, \hat \rho} S^T_{\alpha \hat \gamma} \Gamma^{-1}_{\hat \gamma \hat \rho} S_{\hat \rho \beta} \right)\,.
	\label{eq:eff-friction-app}
\end{align}
The inverse matrix, $\Gamma^{-1}_{\hat\alpha \hat\beta}$, should not be confused with $1/\Gamma_{\hat \alpha \hat \beta}$.
Let us recall that the interaction indices $\alpha$ is composed of these for the basis vectors $\hat \alpha$ and the rest $\alpha_\Delta$ whose charge vector can be expressed as a linear combination of the basis vectors.
Furthermore, the basis vectors are classified into $\{ n_i^{\hat\alpha_\perp} \} \equiv \{ n_i^{\hat\alpha} | \sum_i \bar{n}_i^{\hat \alpha} n_i^{\alpha_\Delta} = 0 ~\text{for all}~\alpha_\Delta\}$ and  $\{ n_i^{\hat\alpha_\parallel} \} \equiv \{ n_i^{\hat\alpha} | \sum_i \bar{n}_i^{\hat \alpha} n_i^{\alpha_\Delta} \neq 0 ~\text{for some}~\alpha_\Delta\}$.
By definition of $\Gamma_{\hat \alpha \hat \beta}$ and $S_{\hat \alpha \beta}$ given in Eq.~\eqref{eq:transport_matrix}, we have $S_{\hat \alpha \hat\beta_\perp} = \Gamma_{\hat \alpha} \delta_{\hat \alpha \hat \beta_\perp}$ and $\Gamma_{\hat \alpha \hat \beta_\perp} = \Gamma_{\hat\alpha} \delta_{\hat \alpha \hat \beta_\perp}$. 
Therefore the effective friction vanishes if $\alpha \to \hat\alpha_\perp$ or $\beta \to \hat\beta_\perp$, \textit{i.e.}, $\gamma^\text{eff}_{a, \hat\alpha_\perp \beta} = \gamma^\text{eff}_{a, \alpha \hat\beta_\perp} = 0$.

Now we move on to the converse statement of Eq.~\eqref{eq:zero-friction}.
One may express $\Gamma_{\hat\alpha \hat\beta}$ as follows:
\begin{align}
	\Gamma_{\hat \alpha \hat \beta} = \Gamma_{\hat \alpha} \delta_{\hat \alpha \hat \beta} + \sum_{\alpha_\Delta} U_{\hat \alpha \alpha_\Delta} \Gamma_{\alpha_\Delta} U^T_{\alpha_\Delta \hat \beta}\,,
\end{align}
where we use $U_{\hat \alpha \hat \beta} = \delta_{\hat \alpha \hat \beta}$.
Note that $U_{\hat \alpha \alpha_\Delta}$ is non-zero only if $\hat\alpha \to \hat\alpha_\parallel$ by definition.
In the limit of $\Gamma_{\alpha_\Delta} = 0$ for all $\alpha_\Delta$, one finds $\Gamma^{-1}_{\hat \alpha \hat \beta} = \Gamma_{\hat \alpha}^{-1} \delta_{\hat\alpha \hat\beta}$
and $S_{\hat\alpha \beta} = \Gamma_\beta \delta_{\hat \alpha \beta}$.
Therefore, we get $\gamma_{a, \alpha \beta}^\text{eff} = 0$ for $\Gamma_{\alpha_\Delta} = 0$.
We would like to understand how $\gamma_{a, \alpha \beta}^\text{eff}$ changes in the presence of non-vanishing $\Gamma_{\alpha_\Delta}$.
For this purpose, it is useful to consider the differential equation of $\gamma_{a, \alpha \beta}^\text{eff}$ with respect to $\Gamma_{\alpha_\Delta}$:
\begin{align}
	\sum_{\alpha, \beta} n_S^\alpha\frac{\partial \gamma_{a, \alpha \beta}^\text{eff}}{\partial \Gamma_{\alpha_\Delta}} n_S^\beta
	= \left[ n_S^{\alpha_\Delta} - \left( n_S \cdot S^T \cdot \Gamma^{-1} \cdot U \right)_{\alpha_\Delta} \right]^2 \geq 0\,,
	\label{eq:diff_pr}
\end{align}
where we use the following shorthanded notation $(n_S \cdot S^T \cdot \Gamma^{-1} \cdot U)_{\alpha_\Delta} = \sum_{\beta, \hat \gamma, \hat \rho} n_S^\beta S^T_{\beta \hat \gamma} \Gamma^{-1}_{\hat \gamma \hat \rho} U_{\hat \rho \alpha_\Delta}$.
We also use that $\Gamma_{\hat \alpha \hat \beta}^{-1}$ is a symmetric matrix.
One can see that the effective friction term in \eqref{eq:eff-friction-app} is a monotonically increasing function of $\Gamma_{\alpha_\Delta}$.

Since $\gamma_{a, \alpha \beta}^\text{eff} = 0$ at $\Gamma_{\alpha_\Delta} = 0$, the effective friction term also vanishes with a non-vanishing $\Gamma_{\alpha_\Delta}$ only if the right-hand side of Eq.~\eqref{eq:diff_pr} is saturated for any $\Gamma_{\alpha_\Delta}$.
Hence, our goal is to understand the condition of $n_S^{\alpha_\Delta} = (n_S \cdot S^T  \cdot \Gamma^{-1} \cdot U)_{\alpha_\Delta}$.
By definition, one can show that the sectors $\{\hat{\alpha}_\perp\}$ and $\{\hat{\alpha}_\parallel, \alpha_\Delta\}$
are completely decoupled, \textit{i.e.},
$\Gamma_{\hat{\alpha}_\perp \hat{\beta}}^{-1} = \Gamma_{\hat{\alpha}_\perp \hat{\beta}_\perp}^{-1} \delta_{\hat{\beta}_\perp \hat{\beta}}$
and $U_{\hat{\alpha}_\perp \alpha_\Delta} = 0$,
and hence this condition can be rewritten as
\begin{align}
	\sum_{\beta_\Delta} n_S^{\beta_\Delta}\left[
	\delta_{\beta_\Delta \alpha_\Delta} 
	- \sum_{\hat\alpha_\parallel, \hat\gamma_\parallel}
	\Gamma_{\beta_\Delta} U^{T}_{\beta_\Delta \hat{\alpha}_\parallel} \Gamma_{\hat{\alpha}_\parallel \hat{\gamma}_\parallel}^{-1}U_{\hat{\gamma}_\parallel \alpha_\Delta}
	\right]
	= \sum_{\hat\beta_\parallel, \hat\gamma_\parallel} n_S^{\hat{\beta}_\parallel}\Gamma_{\hat{\beta}_\parallel}\Gamma_{\hat{\beta}_\parallel\hat{\gamma}_\parallel}^{-1}
	U_{\hat{\gamma}_\parallel \alpha_\Delta}\,.
\end{align}
We can further use the identity
\begin{align}
	\Gamma_{\hat{\alpha}_\parallel}\delta_{\hat{\alpha}_\parallel\hat{\beta}_\parallel}
	= \Gamma_{\hat{\alpha}_\parallel\hat{\beta}_\parallel}
	- U_{\hat{\alpha}_\parallel \beta_\Delta}\Gamma_{\beta_\Delta} U^{T}_{\beta_\Delta \hat{\beta}_\parallel}
\end{align}
and finally obtain
\begin{align}
	\sum_{\beta_\Delta} \Bigg[n_S^{\beta_\Delta} - \sum_{\hat\beta_\parallel} n_S^{\hat{\beta}_\parallel}U_{\hat{\beta}_\parallel \beta_\Delta}\Bigg]
	\Bigg[
	\delta_{\beta_\Delta \alpha_\Delta} 
	- \sum_{\hat\alpha_\parallel, \hat \gamma_\parallel} 
	\Gamma_{\beta_\Delta} U^{T}_{\beta_\Delta \hat{\alpha}_\parallel}
	\Gamma_{\hat{\alpha}_\parallel \hat{\gamma}_\parallel}^{-1}
	U_{\hat{\gamma}_\parallel \alpha_\Delta}
	\Bigg]
	= 0\,. \label{eq:cond_2}
\end{align}
In order to have $n_S^{\alpha_\Delta} = (n_S \cdot S^T  \cdot \Gamma^{-1} \cdot U)_{\alpha_\Delta}$ for any $\Gamma_{\alpha_\Delta}$, 
we need to find a solution to Eq.~\eqref{eq:cond_2} which holds for any $\Gamma_{\alpha_\Delta}$.
Hence, the only possible solution is 
\begin{align}
	n_S^{\beta_\Delta} - \sum_{\hat\beta_\parallel} n_S^{\hat{\beta}_\parallel}U_{\hat{\beta}_\parallel \beta_\Delta} = 0,
\end{align}
because $\delta_{\beta_\Delta \alpha_\Delta} 
	- \sum_{\hat\alpha_\parallel, \hat \gamma_\parallel} 
	\Gamma_{\beta_\Delta} U^{T}_{\beta_\Delta \hat{\alpha}_\parallel}
	\Gamma_{\hat{\alpha}_\parallel \hat{\gamma}_\parallel}^{-1}
	U_{\hat{\gamma}_\parallel \alpha_\Delta} = \delta_{\beta_\Delta \alpha_\Delta}$ 
for $\Gamma_{\alpha_{\Delta}} = 0$, which is invertible.

This completes the proof of the following statement:
\begin{align}
	\sum_{\alpha, \beta} n_S^\alpha n_S^\beta \gamma^\text{eff}_{a, \alpha \beta} = 0 \quad 
	\text{iff}~n_S^{\alpha_\Delta} - \sum_{\hat\alpha_\parallel} n_S^{\hat{\alpha}_\parallel}U_{\hat{\alpha}_\parallel \alpha_\Delta} = 0\,.
\end{align}

\newpage
\small
\bibliographystyle{utphys}
\bibliography{refs}
  
\end{document}